\documentclass[aps,prb,twocolumn,showpacs,amsmath,amssymb,superscriptaddress,floatfix]{revtex4-2}

\usepackage[english]{babel}
\usepackage{blindtext}
\usepackage{latexsym}
\usepackage{amssymb}
\usepackage{physics}
\usepackage{amsmath}
\usepackage{bm}
\usepackage{amsfonts}
\usepackage{relsize}
\usepackage{bbold}
\usepackage{slashed}
\usepackage{appendix}
\usepackage{graphicx}
\usepackage[colorlinks = true,
            linkcolor = blue,
            urlcolor  = blue,
            citecolor = blue,
            anchorcolor = blue]{hyperref}
\usepackage{graphicx} 
\usepackage{dcolumn} 

\newcommand{\up}{\uparrow}
\newcommand{\down}{\downarrow}
\newcommand{\bs}[1]{\mathbf{#1}}

\newcommand{\bq}{\bs{q}}
\newcommand{\bQ}{\bs{Q}}
\newcommand{\bzero}{\bs{0}}
\newcommand{\bk}{\bs{k}}

\newcommand{\chit}{\widetilde{\chi}}
\newcommand{\Gammat}{\widetilde{\Gamma}}

\newcommand{\eps}{\epsilon}

\newcommand{\psibar}{\overline{\psi}}
\newcommand{\hgl}[1]{{\color{black}#1}}

\newcommand{\msml}[1]{{\mathsmaller{\mathsmaller{\mathsmaller{#1}}}}}
\newcommand{\smsqr}{\msml{\square}}
\begin{document}
%
\title{Local Ward identities for collective excitations in fermionic systems with spontaneously broken symmetries}
\author{Pietro M.~Bonetti}
\affiliation{Max Planck Institute for Solid State Research,
 D-70569 Stuttgart, Germany}
 \date{\today}
\begin{abstract}
We derive Ward identities for fermionic systems exhibiting a gauge symmetry that gets globally broken. In particular, we focus on the relation that connects the gauge field response functions to the transverse susceptibilities of the order parameter. We find that the long-wavelength and zero energy limit of the former are related to the coefficients of a low-energy expansion of the latter. We examine three physical cases: the superconductor, the N\'eel antiferromagnet and the spiral magnet. In the case of a metallic spiral magnet that completely breaks the SU(2) spin symmetry we explicitly show that the Ward identities are fulfilled within the random phase approximation. We subsequently derive microscopic expressions for the spin stiffnesses and spin wave velocities, which can be plugged into low energy models to study the effect of long-wavelength bosonic fluctuations on top of mean-field solutions.
\end{abstract}
\maketitle
\section{Introduction}
Symmetries play a prominent role in several branches of physics. In condensed matter systems, they often allow for making rigorous statements about the nature of a phase transition, or provide insightful information about entire phases. A renowned example is the Goldstone theorem~\cite{Goldstone1961}, which predicts the number of gapless modes when a \emph{global} continuous symmetry is spontaneously broken, irrespective of the microscopic details of the system. 
In some systems, a \emph{global} invariance is promoted to a \emph{local} one with the introduction of a \emph{gauge field}. 


\hgl{In this case, constraints for the correlation functions can be derived, corresponding to fundamental conservation laws. These exact relations, going under the name of Ward (or Ward-Takahashi) identities~\cite{Ward1950,Takahashi1957}, play a key role in the development of approximations to tackle the many body problem~\cite{Baym1961,Abrikosov1965,Toyoda1987,Revzen1989}, as it is desirable for a method to satisfy them. Many works have been produced with this aim, focusing, among others, on the consistency of many-body approximations~\cite{Bychkov1966,Hertz1973,Bickers1989,Katanin2004,Krien2017}, or on the gauge invariance of response functions~\cite{GiulianiBook,Rostami2017,Rostami2021}. Ward identities have also been analyzed in systems with broken symmetries, such as ferromagnets~\cite{Hertz1973,Edwards1973} or superconductors~\cite{SchriefferBook,Anderson1958II,Anderson1958,Nambu1960}. Some of the most considered Ward identities are those which impose constraints relating single particle properties (as the self-energy) to the two-particle correlators (the vertex functions), providing a benchmark on the consistency of the approximations employed to compute these quantities}

In this paper, we derive Ward identities descending from a local symmetry that gets globally broken in a fermionic system. 
We consider the gauge field as an external perturbation and probe the linear response of the system to it, thus preserving the massless character of the Goldstone modes. \hgl{We derive a functional identity from which a hierarchy of exact constraints for correlation functions can be extracted. We focus on the Ward identity connecting the linear response to the external gauge field and the transverse susceptibilities of the order parameter, containing information on the Goldstone gapless excitations. In particular, we show that the low frequency and momentum expansion of the inverse of the susceptibilities is related to precise limits of the gauge field response kernel.}
We discuss three examples: a superconductor breaking the U(1) charge symmetry, a N\'eel antiferromagnet breaking the SU(2) spin symmetry leaving a residual U(1), and a spiral magnet, completely breaking SU(2). \hgl{Similar relations can be found in the literature for special cases (see for example Ref.~\cite{Schulz1995}). Here, we provide a formal derivation within a formalism that allows to extend them to any system and any possible kind of long-range order that breaks a continuous symmetry.}
These identities are of fundamental relevance, as from the sole knowledge of the local symmetry group, one can infer the form of the transverse susceptibilities for small energies and long wavelengths. This is reminiscent of and intimately connected to hydrodynamic theories for systems exhibiting spontaneous symmetry breaking, such as superfluid helium~\cite{KhalatinkovBook} or magnets~\cite{Halperin1969,Halperin1977}. Moreover, when the gauge field is of "pure gauge", that is, it can be removed by means of a local transformation, the Ward identities are useful to derive a low-energy theory for the slow directional fluctuations of the order parameter in the fashion of \emph{nonlinear sigma models}~\cite{Haldane1983}. 

We focus particularly on the case of doped antiferromagnets, that host low-lying fermionic excitations, forming small Fermi surfaces. We show that from the Ward identities one can compute a dynamical transverse susceptibility $\chi^\perp_\mathrm{dyn}\equiv\lim_{\omega\to0}\chi^\perp(\bq=\bzero,\omega)$ and a spin stiffness $J$ and extract the spin wave velocity as $c_s=\sqrt{J/\chi^\perp_\mathrm{dyn}}$. This is in contrast with the hydrodynamic approach of Refs.~\cite{Halperin1969,Halperin1977}, which predicts $c_s=\sqrt{J/\chi^\perp}$, with $\chi^\perp\equiv\lim_{\bq\to\bzero}\chi^\perp(\bq,\omega=0)$ the uniform transverse susceptibility. As it was noticed in Ref.~\cite{Sachdev1995}, the equality $\chi^\perp=\chi^\perp_\mathrm{dyn}$ holds only for insulating antiferromagnets at low temperatures, as the spin systems considered in Refs.~\cite{Halperin1969,Halperin1977}. Furthermore, as a consequence of the doping, N\'eel antiferromagnetism is often destroyed, making way to a coplanar spiral magnet. In this state, the spins are no longer antiferromagnetically ordered but the magnetization rotates in a plane, completely breaking the SU(2) rotational symmetry and giving rise to three rather than two Goldstone modes~\cite{Rastelli1985,Chandra1990,Shraiman1992,Kampf1996}. Spiral states have been found to emerge in the two dimensional Hubbard model and in the $t$-$J$ model at moderate hole doping away from half filling, and they are possible candidates for the incommensurate magnetic correlations observed in the cuprate superconductors~\cite{Shraiman1989,Fresard1991,Chubukov1992,Chubukov1995,Kotov2004,Igoshev2010,Yamase2016,Eberlein2016,Mitscherling2018,Bonetti2020}. They have also been proposed in the context of frustrated antiferromagnets~\cite{Azaria1990,Azaria1992,Azaria1993,Azaria1993_PRL,Sachdev1994,Chubukov1994,Azaria1995}. We present a direct evaluation of the Ward identities for a metallic spiral magnet via the random phase approximation, and derive expressions for the spin stiffness and dynamical susceptibility of each gapless mode. \hgl{These formulas generalize to finite temperature and doping and to spiral magnetic states the previously derived expressions for spin stiffnesses in the ground state at half filling~\cite{Schulz1995,Borejsza2004}.} 
This formalism represents a convenient starting point to study long-wavelength fluctuation effects on top of mean-field solutions. 

The paper is organized as follows. In Sec.~\ref{sec: Ward Identities} we derive the anomalous Ward identities descending from a gauge symmetry that gets globally broken in a fermionic system. We consider three different systems: the superconductor, the antiferromagnet, and the spiral magnet. In Sec.~\ref{sec: Explicit calculation} we present an explicit and detailed evaluation of the Ward identities for a spiral magnet and derive approximate expressions for the spin stiffnesses and dynamical susceptibilities. A conclusion in Sec.~\ref{sec: conclusion} closes the presentation. 
\section{Ward Identities}
\label{sec: Ward Identities}
In this section we derive and discuss the Ward identities connected with a specific gauge symmetry which gets globally broken due to the onset of long-range order in the fermionic system. We focus on two specific symmetry groups: the (abelian) U(1) charge symmetry and the (nonabelian) SU(2) spin symmetry. All over the paper we employ Einstein's notation, that is, a sum over repeated indices is implicit. 
\subsection{U(1) symmetry}
\hgl{We now analyze the Ward identities in a superconductor. According to the Elitzur's theorem~\cite{Elitzur1975}, the breaking of the U(1) gauge symmetry is not possible, leading to the interpretation of the superconductor as a \emph{topologically ordered} state~\cite{Wen1989}, due to a subtle interplay of the dynamical gauge field and the order parameter~\cite{Hansson2004}. In this work, we treat the gauge field as a \emph{classical} external perturbation and measure the system's response to it, so that a discussion on the topological nature of the order is not required.}

We consider the generating functional of susceptibilities of the superconducting order parameter and gauge kernels, defined as:
\begin{equation} 
    \begin{split}
        &\mathcal{G}\left[A_\mu,J,J^*\right]=\\
        &\hskip5mm-\ln \int \!\mathcal{D}\psi\mathcal{D}\psibar e^{-\mathcal{S}\left[\psi,\psibar,A_\mu\right]+(J^*,\psi_\down\psi_\up)+(J,\psibar_\up\psibar_\down)},
    \end{split}
    \label{eq: G-functional U(1)}
\end{equation}
where $\psi=(\psi_\up,\psi_\down)$ ($\psibar=(\psibar_\up,\psibar_\down)$) are Grassmann spinor fields corresponding to the annihilation (creation) of a fermion, $A_\mu$ is the electromagnetic field, $J$ ($J^*$) is a source field that couples to the superconducting order parameter $\psibar_\up\psibar_\down$ ($\psi_\down\psi_\up$), and $\mathcal{S}[\psi,\psibar,A_\mu]$ is the action of the system. The index $\mu=0,1,\dots,d$, with $d$ the system dimensionality, runs over temporal ($\mu=0$) and spatial ($\mu=1,\dots,d$) components. In the above equation and from now on, the expression $(A,B)$ has to be intended as $\int_x A(x)B(x)$, where $x$ is a collective variable consisting of a spatial coordinate $\bs{x}$ (possibly discrete, for lattice systems), and an imaginary time coordinate $\tau$, and $\int_x$ is a shorthand for $\int d^d\bs{x}\int_0^\beta d\tau$, with $\beta$ the inverse temperature. Even in the case of a lattice system, we define the gauge field over a \emph{continuous} space time, so that expressions involving its gradients are well defined. 

We let the global U(1) charge symmetry be broken by an order parameter that, to make the treatment simpler, we assume to be local ($s$-wave)
\begin{equation}
    \langle \psi_\down(x)\psi_\up(x)\rangle 
    = \langle \psibar_\up(x)\psibar_\down(x)\rangle= \varphi_0,
\end{equation}
where the average is computed at zero source and gauge fields, and, without loss of generality, we choose $\varphi_0\in \mathbb{R}$. A generalization to systems with nonlocal order parameters, such as $d$-wave superconductors, is straightforward. 

The functional~\eqref{eq: G-functional U(1)} has been defined such that its second derivative with respect to $J$ and $J^*$ at zero $J$, $J^*$ and $A_\mu$ gives (minus) the susceptibility of the order parameter $\chi(x,x')$, while (minus) the gauge kernel $K_{\mu\nu}(x,x')$ can be extracted differentiating twice with respect to the gauge field. In formulas
\begin{subequations}
    \begin{align}
        &\chi(x,x')=-\frac{\delta^2\mathcal{G}}{\delta J(x)\delta J^*(x')}\bigg\rvert_{J=J^*=A_\mu=0},\\
        &K_{\mu\nu}(x,x')=-\frac{\delta^2\mathcal{G}}{\delta A_\mu(x)\delta A_\nu(x')}\bigg\rvert_{J=J^*=A_\mu=0}.
    \end{align}
\end{subequations}
Let us now consider the constraints that the U(1) gauge invariance imposes on the functional $\mathcal{G}$. Its action on the fermionic fields is
\begin{subequations}\label{eq: U(1) gauge transf on psis}
    \begin{align}
        &\psi(x) \to e^{i\theta(x)}\psi(x),\\
        &\psibar(x) \to e^{-i\theta(x)}\psibar(x),
    \end{align}
\end{subequations}
\noindent with $\theta(x)$ a generic function. Similarly, the external fields transform as
\begin{subequations}\label{eq: ext fields U(1) transformation}
    \begin{align}
        &J(x) \to J^\prime(x)= e^{2i\theta(x)}J(x),\\
        &J^*(x) \to [J^{\prime}(x)]^*= e^{-2i\theta(x)}J^*(x),\\
        & A_\mu(x) \to A_\mu^\prime(x)=A_\mu(x)-\partial_\mu \theta(x),
    \end{align}
\end{subequations}
where $\partial_\mu =(i\partial_\tau,\boldsymbol{\nabla})$. In Eqs.~\eqref{eq: U(1) gauge transf on psis} and \eqref{eq: ext fields U(1) transformation} the spatial coordinate $\bs{x}$ of the spinors $\psi$ and $\psibar$, as well as the sources $J$ and $J^*$ may be a lattice one, while the gauge field $A_\mu$ and the parameter $\theta$ are always defined over a continuous space. To keep the notation lighter, we always indicate the space-time coordinate as $x$, keeping in mind that its spatial component could have a different meaning depending on the field it refers to. 

For $\mathcal{G}$ to be invariant under a U(1) gauge transformation, it must not to depend on $\theta(x)$:
\begin{equation}
    \frac{\delta}{\delta \theta(x)}\mathcal{G}[A_\mu',J',(J^{\prime})^*]=0.
    \label{eq: dG/dtheta(x) U(1)}
\end{equation}
Considering an infinitesimal transformation, that is $|\theta(x)|\ll 1$, from Eqs.~\eqref{eq: ext fields U(1) transformation} and~\eqref{eq: dG/dtheta(x) U(1)}, we obtain 
\begin{equation}
    \partial_\mu \left(\frac{\delta\mathcal{G}}{\delta A_\mu(x)}\right)+2i\left[\frac{\delta\mathcal{G}}{J(x)}J(x)-\frac{\delta\mathcal{G}}{J^*(x)}J^*(x)\right]=0.
    \label{eq: Ward identity G}
\end{equation}
We now consider the change of variables 
\begin{subequations}
    \begin{align}
        & J(x)=J_{1}(x)+iJ_{2}(x),\\
        & J^*(x)=J_{1}(x)-iJ_{2}(x),
    \end{align}
\end{subequations}
such that $J_{1}(x)$ ($J_{2}(x)$) is a source field coupling to longitudinal (transverse) fluctuations of the order parameter, and the functional $\Gamma$, defined as the Legendre transform of $\mathcal{G}$,
\begin{equation}
    \begin{split}
        \Gamma[A_\mu,\phi_{1},\phi_{2}]=\sum_{a=1,2}&\int_x\phi_{a}(x) J_{a}(x)
        +\mathcal{G}[A_\mu,J_{1},J_{2}],
    \end{split}
\end{equation}
where $\phi_{a}(x)=\frac{\delta\mathcal{G}[A_\mu,J_{1},J_{2}]}{\delta J_{a}(x)}$. The gauge kernel can be computed from $\Gamma$ as well:
\begin{equation}
    K_{\mu\nu}(x,x')=-\frac{\delta^2 \Gamma}{\delta A_\mu(x) \delta A_\nu(x')}\Big\rvert_{\vec{\phi}=A_\mu=0},
\end{equation}
because, thanks to the Legendre transform properties, $\delta\Gamma/\delta A_\mu(x) = \delta\mathcal{G}/\delta A_\mu(x)$. Differently, differentiating $\Gamma$ twice with respect to the fields $\phi_{a}$ returns the inverse correlator
\begin{equation}
    C^{ab}(x,x')=-\frac{\delta^2\Gamma}{\delta\phi_{a}(x)\delta\phi_{b}(x')}\bigg\rvert_{\vec{\phi}=A_\mu=0},
\end{equation}
which obeys a reciprocity relation~\cite{NegeleOrland}
\begin{equation}
    \int_{x^{\prime\prime}}C^{ac}(x,x^{\prime\prime})\chi^{cb}(x^{\prime\prime},x')=\delta_{ab}\delta(x-x'),
    \label{eq: reciprocity relation}
\end{equation}
with the generalized susceptibility $\chi^{ab}(x,x')$, defined as
\begin{equation}
    \chi^{ab}(x,x')=-\frac{\delta^2\mathcal{G}}{\delta J_{a}(x)\delta J_{b}(x')}\bigg\rvert_{J_{a}=A_\mu=0}.
\end{equation}
Eq.~\eqref{eq: Ward identity G} can be expressed in terms of $\Gamma$ as
\begin{equation}
    \begin{split}
        &\partial_\mu \left(\frac{\delta\Gamma}{\delta A_\mu(x)}\right)
        -2\left[\frac{\delta\Gamma}{\delta\phi_{1}(x)}\phi_{2}(x)
        -\frac{\delta\Gamma}{\delta\phi_{2}(x)}\phi_{1}(x)\right]
        =0.
    \end{split}
    \label{eq: Ward identity Gamma}
\end{equation}
Eq.~\eqref{eq: Ward identity Gamma} is an identity for the generating functional $\Gamma$ stemming from U(1) gauge invariance of the theory. Taking derivatives with respect to the fields, one can derive an infinite set of Ward identities. 

We are interested in the relation between the gauge kernel and the transverse inverse susceptibility $C^{22}(x,x')$.
For this purpose, we differentiate Eq.~\eqref{eq: Ward identity Gamma} once with respect to $\phi_{2}(x')$ and once with respect to $A_\nu(x')$, and then set the fields to zero. We obtain the set of equations
\begin{subequations}
    \begin{align}
        &-\partial_\mu \mathcal{C}_\mu^2(x,x')=2\varphi_0\, C^{22}(x,x'),\label{eq: WI 1}\\ 
        &-\partial_\mu K_{\mu\nu}(x,x')=2\varphi_0\, \mathcal{C}_\nu^{2}(x,x'),
        \label{eq: WI 2}
    \end{align}
\end{subequations}
where $\varphi_0=\langle\phi(x)\rangle=\langle\phi_{1}(x)\rangle=\langle\psi_\down(x)\psi_\up(x)\rangle$, and we have defined the quantity
\begin{equation}
    \mathcal{C}_\mu^{a}(x,x')=-\frac{\delta^2\Gamma}{\delta A_\mu(x)\delta\phi_{a}(x')}\bigg\rvert_{\vec{\phi}=A_\mu=0}.
\end{equation}
Combining \eqref{eq: WI 1} and \eqref{eq: WI 2}, we obtain
\begin{equation}
    \partial_\mu\partial_\nu K_{\mu\nu}(x,x')=4\varphi_0^2\,C^{22}(x,x').
    \label{eq: WI for SC}
\end{equation}
Fourier transforming Eq.~\eqref{eq: WI for SC} and rotating to real frequencies, we have
\begin{equation}
    -q_\mu q_\nu K_{\mu\nu}(q)=4\varphi_0^2 C^{22}(q),
    \label{eq: WI for SC in q space}
\end{equation}
with $q=(\bq,\omega)$ a collective variable combining momentum and real frequency. 


We now define the superfluid stiffness $J_{\alpha\beta}$ and the uniform density-density susceptibility $\chi_n$~\cite{notedefstiffnesses} as
\begin{subequations}
    \begin{align}
        &J_{\alpha\beta}\equiv-\lim_{\bq\to\bzero}K_{\alpha\beta}(\bq,\omega=0) \label{eq: Jsc definition},\\
        &\chi_n\equiv\lim_{\omega\to 0}K_{00}(\bq=\bzero,\omega),\label{eq: den-den chi SC}
    \end{align}
\end{subequations}
where the minus sign in \eqref{eq: Jsc definition} has been introduced so that $J_{\alpha\beta}$ is positive definite. Notice that, even though the limits $\bq\to\bzero$ and $\omega\to 0$ in Eq.~\eqref{eq: den-den chi SC} have been taken in the opposite order compared to what is conventionally done, they commute in a $s$-wave superconductor because of the absence of gapless fermionic excitations. In the above equation and from now on, we employ the convention that the indices labeled as $\mu$, $\nu$ include temporal and spatial components, whereas $\alpha$ and $\beta$ only the latter. Taking the second derivative with respect to $q$ on both sides of \eqref{eq: WI for SC in q space}, we obtain
\begin{subequations}\label{eq: J and chi from K22 SC}
    \begin{align}
        &J_{\alpha\beta}=2\varphi_0^2  \partial^2_{q_\alpha q_\beta}C^{22}(\bq,\omega=0)\big\rvert_{\bq\to\bzero},\\
        &\chi_n=-2\varphi_0^2  \partial^2_{\omega}C^{22}(\bq=\bzero,\omega)\big\rvert_{\omega\to0},
    \end{align}
\end{subequations}
where $\partial^2_{q_\alpha q_\beta}$ and $\partial^2_\omega$ are shorthands for $\frac{\partial^2}{\partial q_\alpha q_\beta}$ and $\frac{\partial^2}{\partial\omega^2} $, respectively. Moreover, we have made use of the Goldstone theorem, reading $C^{22}(\bzero,0)=0$. To derive Eq.~\eqref{eq: J and chi from K22 SC} from \eqref{eq: WI for SC in q space} we have exploited the finiteness of the gauge kernel $K_{\mu\nu}(q)$ in the $\bq\to\bzero$ and $\omega\to 0$ limits. 
Eq.~\eqref{eq: J and chi from K22 SC} states that the superfluid stiffness and the uniform density-density correlation function are not only the zero momentum and frequency limit of the gauge kernel, but also the coefficients of the inverse transverse susceptibility when expanded for small $\bq$ and $\omega$, respectively. Inverting Eq.~\eqref{eq: reciprocity relation}, $C^{22}(q)$ can be expressed in terms of $\chi^{ab}(q)$ as 
\begin{equation}
    C^{22}(q)= \frac{1}{\chi^{22}(q)-\chi^{21}(q)\frac{1}{\chi^{11}(q)}\chi^{12}(q)}.
\end{equation}
In the limit $q\to 0=(\bzero,0)$, $\chi^{22}(q)$ diverges for the Goldstone theorem, while the second term in the denominator vanishes like some power of $q$. This implies that, for small $q$,
\begin{equation}
    C^{22}(q)\simeq \frac{1}{\chi^{22}(q)}. 
\end{equation}
From this consideration, together with \eqref{eq: J and chi from K22 SC}, we can deduce that the transverse susceptibility can be written as 
\begin{equation}
    \chi^{22}(\bq,\omega)\simeq \frac{4\varphi_0^2}{-\chi_n \omega^2+J_{\alpha\beta}q_\alpha q_\beta},
    \label{eq: chi22 SC small q}
\end{equation}
for small $\bq$ and $\omega$. 

The above form of the $\chi^{22}(q)$ can be also deduced from a low energy theory for the phase fluctuations of the superconducting order parameter. Setting $J$ and $J^*$ to zero in \eqref{eq: G-functional U(1)}, and integrating out the Grassmann fields, one obtains an effective action for the gauge fields. The quadratic contribution in $A_\mu$ is 
\begin{equation}
    \mathcal{S}_\mathrm{eff}^{(2)}[A_\mu]=-\frac{1}{2}\int_q K_{\mu\nu}(q) A_\mu(-q) A_\nu(q),
\end{equation}
where $\int_q$ is a shorthand for $\int \frac{d\omega}{2\pi}\int \frac{d^d \bq}{(2\pi)^d}$. Since we are focusing only on slow and long-wavelength fluctuations of $A_\mu$, we replace $K_{\mu\nu}(q)$ with $K_{\mu\nu}(0)$. Considering a pure gauge field, $A_\mu(x)=-\partial_\mu\theta(x)$, where $\theta(x)$ is (half) the phase of the superconducting order parameter ($\phi(x)=\varphi_0 e^{-2i\theta(x)}$), we obtain 
\begin{equation}
    \mathcal{S}_\mathrm{eff}[\theta]=\frac{1}{2}\int_x \left\{-\chi_n\left[\partial_t\theta(x)\right]^2+J_{\alpha\beta}\partial_\alpha\theta(x)\partial_\beta\theta(x)\right\},
    \label{eq: phase field action}
\end{equation}
with $\theta(x)\in[0,2\pi]$ a periodic field. The above action is well known to display a Berezinskii-Kosterlitz-Thouless (BKT) transition~\cite{Berezinskii1971,Kosterlitz1973} for $d=1$ (at $T=0$) and $d=2$ (at $T>0)$, while for $d=3$ ($T\geq 0$) or $d=2$ ($T=0$), it describes a gapless phase mode known as Anderson-Bogoliubov phonon~\cite{Anderson1958}.

From~\eqref{eq: phase field action}, we can extract the propagator of the field $\theta(x)$
\begin{equation}
    \langle \theta(-q)\theta(q)\rangle =  \frac{1}{-\chi_n \omega^2+J_{\alpha\beta}q_\alpha q_\beta},
\end{equation}
where we have neglected the fact that $\theta(x)$ is defined modulo $2\pi$. 
Writing $\phi_{2}(x)=(\phi(x)-\phi^*(x))/(2i)=-\varphi_0 \sin(2\theta(x))\simeq -2\varphi_0 \theta(x)$, $\chi^{22}(q)$ can be expressed as
\begin{equation}
    \begin{split}
        \chi^{22}(q)=&\langle \phi_{2}(-q)\phi_{2}(q)\rangle \simeq 4\varphi_0^2 \langle \theta(-q)\theta(q)\rangle \\
        =& \frac{4\varphi_0^2}{-\chi_n \omega^2+J_{\alpha\beta}q_\alpha q_\beta},
    \end{split}
\end{equation}
which is in agreement with Eq.~\eqref{eq: chi22 SC small q}. 
\subsection{SU(2) symmetry}
In this Section, we repeat the same procedure we have applied in the previous one to derive the Ward identities connected to a SU(2) gauge invariant system. We consider the functional
\begin{equation}
    \mathcal{G}[A_\mu,\vec{J}]=-\ln \int \!\mathcal{D}\psi\mathcal{D}\psibar e^{-\mathcal{S}\left[\psi,\psibar,A_\mu\right]+(\vec{J},\frac{1}{2}\psibar\vec{\sigma}\psi)},
\end{equation}
where $A_\mu(x)=A_\mu^a(x)\frac{\sigma^a}{2}$ is a SU(2) gauge field, $\vec{\sigma}$ are the Pauli matrices, and $\vec{J}(x)$ is a source field coupled to the fermion spin operator $\frac{1}{2}\psibar(x)\vec{\sigma}\psi(x)$. Similarly to the previous section, derivatives of $\mathcal{G}$ with respect to $A_\mu$ and $\vec{J}$ at zero external fields give minus the gauge kernels and spin susceptibilities, respectively. In formulas,
\begin{subequations}
    \begin{align}
        &\chi^{ab}(x,x')=-\frac{\delta^2\mathcal{G}}{\delta J_{a}(x)\delta J_{b}(x')}\bigg\rvert_{\vec{J}=A_\mu=0},\\
        &K_{\mu\nu}^{ab}(x,x')=-\frac{\delta^2\mathcal{G}}{\delta A^a_\mu(x)\delta A^b_\nu(x')}\bigg\rvert_{\vec{J}=A_\mu=0}.
    \end{align}
\end{subequations}
We let the SU(2) symmetry be broken by a (local) order parameter of the form 
\begin{equation}\label{eq: magnetic order parameter}
    \left\langle \frac{1}{2}\psibar(x)\vec{\sigma}\psi(x) \right\rangle = \varphi_0 \hat{v}(\bs{x}),
\end{equation}
with $\hat{v}(\bs{x})$ a position-dependent unit vector pointing along the local direction of the magnetization~\cite{NoteAFnoanomalousaverages}. 

A SU(2) gauge transformation on the fermionic fields reads
\begin{subequations}
    \begin{align}
        &\psi(x) \to R(x)\psi(x),\\
        &\psibar(x) \to \psibar(x)R^\dagger(x),
    \end{align}
    \label{eq: SU(2) gauge symm}
\end{subequations}
where $R(x)\in\mathrm{SU(2)}$ is a matrix acting on the spin indices of $\psi$ and $\psibar$. The external fields 
transform as
\begin{subequations}
    \begin{align}
        J_{a}(x) \to J^{\prime}_a(x)= &\mathcal{R}^{ab}(x)J_{b}(x),\\
        A_\mu(x) \to A_\mu^\prime(x)=&R^\dagger(x)A_\mu(x)R(x)\nonumber\\
        &+iR^\dagger(x)\partial_\mu R(x), \label{eq: SU(2) Amu transformation}
    \end{align}
\end{subequations}
where $\mathcal{R}(x)$ is the adjoint representation of $R(x)$
\begin{equation}
    \mathcal{R}^{ab}(x)\sigma^b = R(x)\sigma^a R^\dagger(x).
    \label{eq: mathcalR definition}
\end{equation}
The SU(2) gauge invariance of $\mathcal{G}$ can be expressed as
\begin{equation}
    \frac{\delta}{\delta R(x)}\mathcal{G}[A_\mu',\vec{J}^\prime]=0.
\end{equation}
Writing $R(x)=e^{i \theta_a(x)\frac{\sigma^a}{2}}$, $R^\dagger(x)=e^{-i \theta_a(x)\frac{\sigma^a}{2}}$, and considering an infinitesimal transformation $|\theta_a(x)|\ll1$, we obtain the functional identity
\begin{equation}
    \begin{split}
        \partial_\mu \left(\frac{\delta\Gamma}{\delta A_\mu^a(x)}\right)
        -\varepsilon^{a\ell m}\bigg[
        &\frac{\delta\Gamma}{\delta\phi^\ell(x)}\phi^m(x)\\
        &-\frac{\delta\Gamma}{\delta A_\mu^\ell(x)}A_\mu^m(x)
        \bigg]
        =0,
    \end{split}
    \label{eq: Ward identity SU(2)}
\end{equation}
where $\varepsilon^{abc}$ is the Levi-Civita tensor. $\Gamma[A_\mu,\vec{\phi}]$ is the Legendre transform of $\mathcal{G}$, defined as
\begin{equation}
    \Gamma[A_\mu,\vec{\phi}]=\int_x\vec{\phi}(x)\cdot\vec{J}(x) + \mathcal{G}[A_\mu,\vec{J}],
\end{equation}
with $\phi_{a}(x)=\frac{\delta\mathcal{G}[A_\mu,\vec{J}]}{\delta J_{a}(x)}$. The inverse susceptibilities $C^{ab}(x,x')$, defined as,
\begin{equation}
    C^{ab}(x,x')=-\frac{\delta^2\Gamma}{\delta\phi_{a}(x)\delta\phi_{b}(x')}\bigg\rvert_{\vec{\phi}=A_\mu=0},
\end{equation}
obey a reciprocity relation with the spin susceptibilities $\chi^{ab}(x,x')$ similar to \eqref{eq: reciprocity relation}.

Defining the quantities
\begin{subequations}
    \begin{align}
        &\mathcal{C}_\mu^{ab}(x,x')=-\frac{\delta^2\Gamma}{\delta A^a_\mu(x)\delta\phi_{b}(x')}\bigg\rvert_{\vec{\phi}=A_\mu=0},\\
        &\mathcal{B}_{\mu}^{a}(x)=-\frac{\delta\Gamma}{\delta A^a_\mu(x)}\bigg\rvert_{\vec{\phi}=A_\mu=0},
    \end{align}    
\end{subequations}
we obtain from~\eqref{eq: Ward identity SU(2)} the set of equations
\begin{subequations}
    \begin{align}
        -\partial_\mu  \mathcal{C}_\mu^{ab}(x,x')&=\varphi_0 \varepsilon^{a\ell m}C^{\ell b}(x,x')v_m(\bs{x}),\label{eq: WI SU(2) I}\\
        -\partial_\mu  K_{\mu\nu}^{ab}(x,x')&=\varphi_0 \varepsilon^{a\ell m}\mathcal{C}^{b\ell}_\nu(x,x')v_m(\bs{x})\nonumber\\
        &\phantom{=}-\varepsilon^{a\ell b}\mathcal{B}_\nu^\ell(x)\delta(x-x')\label{eq: WI SU(2) II},\\
        \partial_\mu  \mathcal{B}_{\mu}^{a}(x)&=0\label{eq: WI SU(2) III},
    \end{align}
\end{subequations}
where \eqref{eq: WI SU(2) I}, \eqref{eq: WI SU(2) II}, have been obtained differentiating \eqref{eq: Ward identity SU(2)} with respect to $\phi_b(x')$ and $A_\nu(x')$, respectively, and setting the fields to zero. Eq.~\eqref{eq: WI SU(2) III} simply comes from \eqref{eq: Ward identity SU(2)} computed at zero $A_\mu$, $\phi_{a}$. According to Eq.~\eqref{eq: magnetic order parameter}, the expectation value of $\vec{\phi}(x)$ takes the form $\langle \vec{\phi}(x)\rangle=\varphi_0 \hat{v}(\bs{x})$. Combining \eqref{eq: WI SU(2) I}, \eqref{eq: WI SU(2) II}, and \eqref{eq: WI SU(2) III}, we obtain the Ward identity
\begin{equation}
    \partial_\mu\partial_\nu K_{\mu\nu}^{ab}(x,x')=\varphi_0^2 \varepsilon^{a\ell m}\varepsilon^{bnp} v^\ell(
    \bs{x}) v^n(\bs{x}') C^{mp}(x,x'),
    \label{eq: WI gauge Kab SU(2)}
\end{equation}
which connects the gauge kernels with the inverse susceptibilities. 

In the following, we analyze two concrete examples where the above identity applies, namely the N\'eel antiferromagnet and the spiral magnet. We do not consider ferromagnets or, in general, systems with a net average magnetization, as in this case the divergence of the transverse components of the kernel $K_{00}^{ab}(q)$ for $q\to 0$ leads to changes in the form of the Ward identities. In this case, one can talk of type-II Goldstone bosons~\cite{Wilson2022}, characterized by a nonlinear dispersion. 
\subsubsection{N\'eel order}
We now consider the particular case of antiferromagnetic (or N\'eel) ordering for a system on a $d$-dimensional bipartite lattice. In this case $\hat{v}(\bs{x})$ takes the form $(-1)^\bs{x}\hat{v}$, with $(-1)^\bs{x}$ being 1 ($-1$) on the sites of sublattice A (B), and $\hat{v}$ a constant unit vector. In the following, without loss of generality, we consider $\hat{v}=(1,0,0)$. Considering only the diagonal ($a=b$) components of \eqref{eq: WI gauge Kab SU(2)}, we have
\begin{subequations}
    \begin{align}
        \partial_\mu\partial_\nu  K_{\mu\nu}^{11}(x,x')&=0,\label{eq: WI K11}\\
        \partial_\mu\partial_\nu  K_{\mu\nu}^{22}(x,x')&=\varphi_0^2 (-1)^{\bs{x}-\bs{x}'} C^{33}(x,x')\label{eq: WI K22},\\
        \partial_\mu\partial_\nu  K_{\mu\nu}^{33}(x,x')&=\varphi_0^2 (-1)^{\bs{x}-\bs{x}'} C^{22}(x,x')\label{eq: WI K33}.
    \end{align}
\end{subequations}
Despite N\'eel antiferromagnetism breaking the lattice translational symmetry, the components of the gauge Kernel considered above depend only on the difference of their arguments $x-x'$, and thus have a well-defined Fourier transform.
Eq.~\eqref{eq: WI K11} implies $q_\mu q_\nu K_{11}^{\mu\nu}(\bq,\omega)=0$, as expected due to the residual U(1) gauge invariance in the N\'eel state. In particular, one obtains $\lim_{\bq\to\bzero}K^{11}_{\alpha\beta}(\bq,0)=0$, and $\lim_{\omega\to 0}K^{11}_{00}(\bzero,\omega)=0$. Eqs.~\eqref{eq: WI K22} and \eqref{eq: WI K33} are the same equation as we have $K_{22}(x,x')=K_{33}(x,x')$, again because of the residual symmetry. If we rotate them onto the real time axis and perform the Fourier transform, we get
\begin{subequations}\label{eq: chi and J Neel}
    \begin{align}
        &J_{\alpha\beta}\equiv-\lim_{\bq\to\bzero}K^{22}_{\alpha\beta}(\bq,0)=\frac{1}{2}\varphi_0^2 \partial^2_{q_\alpha q_\beta} C^{33}(\bq,0)\Big\rvert_{\bq\to\bQ},\label{eq: spin stiffness Neel}\\
        &\chi_\mathrm{dyn}^\perp\equiv\lim_{\omega\to 0}K^{22}_{00}(\bzero,\omega)=-\frac{1}{2}\varphi_0^2 \partial^2_{\omega} C^{33}(\bQ,\omega)\Big\rvert_{\omega\to 0},\label{eq: chi perp Neel}
    \end{align}
\end{subequations}
where $J_{\alpha\beta}$ is the spin stiffness, $\bQ=(\pi/a_0,\dots,\pi/a_0)$, with $a_0$ the lattice spacing, and we name $\chi_\mathrm{dyn}^\perp$ as transverse dynamical susceptibility~\cite{notedefstiffnesses}. In the above equations we have made use of the Goldstone theorem, which in the present case reads
\begin{equation}\label{eq: Goldstone Neel}
    C^{22}(\bQ,0)=C^{33}(\bQ,0)=0.
\end{equation}
%
Furthermore, to derive Eq.~\eqref{eq: chi and J Neel} from \eqref{eq: WI K22}, we have used the finiteness of the $\bq\to\bzero$ and $\omega\to 0$ limits of the gauge kernels. 
Following the argument given in the previous section, for $q=(\bq,\omega)$ close to $Q=(\bQ,0)$, we can replace $C^{33}(q)$ by $1/\chi^{33}(q)$ in \eqref{eq: chi and J Neel}, implying
\begin{equation}
    \begin{split}
        \chi^{22}(q\simeq Q)&=\chi^{33}(q\simeq Q)\\
        &\simeq\frac{\varphi_0^2}{-\chi_\mathrm{dyn}^\perp \omega^2 + J_{\alpha\beta}(q-Q)_\alpha(q-Q)_\beta}.
    \end{split}
    \label{eq: chi22 and chi33 Neel small q}
\end{equation}
Notice that in Eq.~\eqref{eq: chi22 and chi33 Neel small q} we have neglected the imaginary parts of the susceptibilities, that, for doped antiferromagnets, can lead to \emph{Landau damping} of the Goldstone modes~\cite{Bonetti2022}.

Also for N\'eel ordering, form \eqref{eq: chi22 and chi33 Neel small q} of the transverse susceptibilities can be deduced from a low energy theory for the gauge field $A_\mu(x)$, that is,
\begin{equation}
    \begin{split}
        \mathcal{S}_\mathrm{eff}[A_\mu]=-\frac{1}{2}\int_q 
        \Big[
        &K_{00}^{ab}(\bzero,\omega\to0) A^a_0(-q)
        A^b_0(q)\\
        +&K_{\alpha\beta}^{ab}(\bq\to\bzero,0) A^a_\alpha(-q)
        A^b_\beta(q)
        \Big].
    \end{split}
\end{equation}
Considering a pure gauge field
\begin{equation}
    A_\mu(x)=iR^\dagger(x)\partial_\mu R(x),
    \label{eq: pure gauge Amu SU(2)}
\end{equation}
with $R(x)$ a SU(2) matrix, we obtain the action
\begin{equation}
    \mathcal{S}_\mathrm{eff}[\hat{n}]=\frac{1}{2}\int_x \left\{-\chi_\mathrm{dyn}^\perp \left|\partial_t\hat{n}(x)\right|^2+J_{\alpha\beta} \partial_\alpha\hat{n}(x)\cdot\partial_\beta\hat{n}(x)\right\},
    \label{eq: SU(2)/U(1) NLsM}
\end{equation}
where $\hat{n}(x)=(-1)^\bs{x}\mathcal{R}(x)\hat{v}(\bs{x})$, with $\mathcal{R}(x)$ defined as in Eq.~\eqref{eq: mathcalR definition}, and $|\hat{n}(x)|^2=1$. Eq.~\eqref{eq: SU(2)/U(1) NLsM} is the well-known $\mathrm{O(3)/O(2)}$ nonlinear sigma model (NL$\sigma$M) action, describing low-energy properties of quantum antiferromagnets~\cite{Haldane1983,AuerbachBook1994}.

Writing $R(x)=e^{i\theta_a(x)\frac{\sigma^a}{2}}$, and expanding to first order in $\theta_a(x)$, $\hat{n}(x)$ becomes $\hat{n}(x)\simeq(1,\theta_2(x),-\theta_3(x))$.
Considering the expression $\vec{\phi}(x)=(-1)^\bs{x} \varphi_0 \hat{n}(x)$ for the order parameter field, we see that small fluctuations in $\hat{n}(x)$ only affect the 2- and 3-components of $\vec{\phi}(x)$. The transverse susceptibilities can be therefore written as
\begin{equation}
    \begin{split}
        \chi^{22}(q)&=\chi^{33}(q)=\langle \phi_{2}(q)\phi_{2}(-q)\rangle\\
        & \simeq 
        \varphi_0^2 \langle n_2(q+Q) n_2(-q-Q)\rangle \\
        &= \frac{\varphi_0^2}{-\chi_\mathrm{dyn}^\perp  \omega^2+J_{\alpha\beta}(q-Q)_\alpha(q-Q)_\beta}, 
    \end{split}
    \label{eq: goldstone chis neel}
\end{equation}
which is the result of Eq.~\eqref{eq: chi22 and chi33 Neel small q}. In Eq.~\eqref{eq: goldstone chis neel} we have made use of the propagator of the $\hat{n}$-field dictated by the action of Eq.~\eqref{eq: SU(2)/U(1) NLsM}, that is, 
\begin{equation}
    \langle n_a(q) n_a(-q)\rangle = \frac{1}{-\chi_\mathrm{dyn}^\perp \omega^2+J_{\alpha\beta}q_\alpha q_\beta}.
\end{equation}
Eq.~\eqref{eq: goldstone chis neel} predicts two degenerate magnon branches with linear dispersion for small $\bq-\bQ$. In the case of an isotropic antiferromagnet ($J_{\alpha\beta}=J\delta_{\alpha\beta}$), we have $\omega_\bq = c_s |\bq|$, with the spin wave velocity given by $c_s=\sqrt{J/\chi_\mathrm{dyn}^\perp}$.
\subsubsection{Spiral magnetic order}
\label{subsec: WI spiral}
We now turn our attention to the case of spin spiral ordering, described by the magnetization direction 
\begin{equation}
    \hat{v}(\bs{x})=(\cos(\bQ\cdot\bs{x}),\sin(\bQ\cdot\bs{x}),0),
    \label{eq: spiral magnetization}
\end{equation}
where at least one component of $\bQ$ is neither 0 nor $\pi/a_0$. In this case, it is convenient to rotate the field $\vec{\phi}(x)$ to a basis in which $\hat{v}(\bs{x})$ is uniform. This is achieved by the transformation~\cite{Kampf1996}
\begin{equation}
    \vec{\phi}^\prime(x) = \mathcal{M}(x) \vec{\phi}(x),
    \label{eq: spiral rotation field}
\end{equation}
with
\begin{equation}
    \mathcal{M}(x)=\left(
    \begin{array}{ccc}
        \cos(\bQ\cdot\bs{x}) & \sin(\bQ\cdot\bs{x}) & 0 \\
        -\sin(\bQ\cdot\bs{x}) & \cos(\bQ\cdot\bs{x}) & 0 \\
        0 & 0 & 1
    \end{array}
    \right). 
    \label{eq: spiral rotation matrix}
\end{equation}
In this way, the inverse susceptibilities are transformed into
\begin{equation}
    \begin{split}
        \widetilde{C}^{ab}(x,x')=&- \frac{\delta^2\Gamma}{\delta \phi^{\prime}_a(x)\delta \phi^{\prime}_b(x')}\bigg\rvert_{\vec{\phi}'=A_\mu=0}\\
        =&[\mathcal{M}^{-1}(x)]^{ac}[\mathcal{M}^{-1}(x')]^{bd} C^{cd}(x,x').
    \end{split}
    \label{eq: rotated Ks spiral}
\end{equation}
If we now apply the Ward identity \eqref{eq: WI gauge Kab SU(2)}, we obtain
\begin{subequations}\label{eq: WI real space spiral}
    \begin{align}
        \partial_\mu\partial_\nu  K_{\mu\nu}^{11}(x,x')&=\varphi_0^2 \sin(\bQ\cdot\bs{x})\sin(\bQ\cdot\bs{x}')\nonumber \\ &\phantom{=}\times\widetilde{C}^{33}(x,x'),\\
        \partial_\mu\partial_\nu  K_{\mu\nu}^{22}(x,x')&=\varphi_0^2 \cos(\bQ\cdot\bs{x})\cos(\bQ\cdot\bs{x}')\nonumber\\ &\phantom{=}\times\widetilde{C}^{33}(x,x'),\\
        \partial_\mu\partial_\nu  K_{\mu\nu}^{33}(x,x')&=\varphi_0^2\widetilde{C}^{22}(x,x'),
    \end{align}
\end{subequations}
with 
\begin{subequations}
    \begin{align}
        \widetilde{C}^{33}(x,x')&=C^{33}(x,x'),\\
        \widetilde{C}^{22}(x,x')&=\sin(\bQ\cdot\bs{x})\sin(\bQ\cdot\bs{x}') C^{11}(x,x')\nonumber\\
        &\phantom{=}+\cos(\bQ\cdot\bs{x})\cos(\bQ\cdot\bs{x}') C^{22}(x,x')\nonumber\\
        &\phantom{=}-\sin(\bQ\cdot\bs{x})\cos(\bQ\cdot\bs{x}')C^{12}(x,x')\nonumber\\
        &\phantom{=}-\cos(\bQ\cdot\bs{x})\sin(\bQ\cdot\bs{x}')C^{21}(x,x').
    \end{align}
\end{subequations}
We remark that an order parameter of the type~\eqref{eq: spiral magnetization} completely breaks the SU(2) spin symmetry, which is why none of the right hand sides of the equations above vanishes. We have considered a \emph{coplanar} spiral magnetic order, that is, we have assumed all the spins to lie in the same plane, so that out of the three Goldstone modes, two are degenerate and correspond to out-of-plane fluctuations, and one to in-plane fluctuations of the spins.
Furthermore, translational invariance is broken, so the Fourier transforms of the gauge kernels $K_{\mu\nu}^{ab}(\bq,\bq',\omega)$ and inverse susceptibilities $C^{ab}(\bq,\bq',\omega)$ are nonzero not only for $\bq-\bq'=\bzero$ but also for $\bq-\bq'=\pm\bQ$ or $\pm2\bQ$. Time translation invariance is preserved, and the gauge kernels and the inverse susceptibilities depend on one single frequency. However, in the basis obtained with transformation~\eqref{eq: spiral rotation matrix}, translational invariance is restored, so that the Fourier transform of $\widetilde{C}^{ab}(x,x')$ only depends on one spatial momentum. With this in mind, we can extract expressions for the spin stiffnesses and dynamical susceptibilities from~\eqref{eq: WI real space spiral}. After rotating to real frequencies, and using the property that for a spiral magnet the gauge kernels are finite in the limits $\bq=\bq'\to\bzero$ and $\omega\to 0$, we obtain~\cite{notedefstiffnesses}
\begin{subequations}\label{eq: spin stiff spiral def}
    \begin{align}
        J^{\perp,1}_{\alpha\beta}\equiv&-\lim_{\bq\to\bzero}K^{11}_{\alpha\beta}(\bq,0)\nonumber\\
        =&\frac{1}{8}\varphi_0^2\partial^2_{q_\alpha q_\beta}\sum_{\eta=\pm}\widetilde{C}^{33}(\bq+\eta\bQ,0)\bigg\rvert_{\bq\to\bzero},\label{eq: expression J1 spiral}\\
        J^{\perp,2}_{\alpha\beta}\equiv&-\lim_{\bq\to\bzero}K^{22}_{\alpha\beta}(\bq,0)\nonumber\\
        =&\frac{1}{8}\varphi_0^2\partial^2_{q_\alpha q_\beta}\sum_{\eta=\pm}\widetilde{C}^{33}(\bq+\eta\bQ,0)\bigg\rvert_{\bq\to\bzero},\label{eq: expression J2 spiral}\\
        J^\smsqr_{\alpha\beta}\equiv&-\lim_{\bq\to\bzero}K^{33}_{\alpha\beta}(\bq,0)\nonumber\\
        =&\frac{1}{2}\varphi_0^2\partial^2_{q_\alpha q_\beta}\widetilde{C}^{22}(\bq,0)\bigg\rvert_{\bq\to\bzero},\label{eq: expression J3 spiral}
    \end{align}
\end{subequations}
and
\begin{subequations}\label{eq: Z factors spiral def}
    \begin{align}
        \chi_\mathrm{dyn}^{\perp,1}\equiv&\lim_{\omega\to 0}K^{11}_{00}(\bzero,\omega)\nonumber\\
        =&-\frac{1}{8}\varphi_0^2\partial^2_\omega \sum_{\eta=\pm}\widetilde{C}^{33}(\eta\bQ,\omega)\bigg\rvert_{\omega\to0},\label{eq: expression chi1 spiral}\\
        \chi_\mathrm{dyn}^{\perp,2}\equiv&\lim_{\omega\to 0}K^{22}_{00}(\bzero,\omega)\nonumber\\
        =&-\frac{1}{8}\varphi_0^2\partial^2_\omega \sum_{\eta=\pm}\widetilde{C}^{33}(\eta\bQ,\omega)\bigg\rvert_{\omega\to0},\label{eq: expression chi2 spiral}\\
        \chi_\mathrm{dyn}^\smsqr\equiv&\lim_{\omega\to 0}K^{33}_{00}(\bzero,\omega)\nonumber\\
        =&-\frac{1}{2}\varphi_0^2\partial^2_\omega \widetilde{C}^{22}(\bzero,\omega)\bigg\rvert_{\omega\to0}\label{eq: expression chi3 spiral},
    \end{align}
\end{subequations}
where the labels $\perp$ and $\smsqr$ denote out-of-plane and in-plane quantities, respectively. In the equations above, we have defined $K^{ab}_{\mu\nu}(\bq,\omega)$ as the prefactors of the components of the gauge kernels $K^{ab}_{\mu\nu}(\bq,\bq',\omega)$ which are proportional to $(2\pi)^d\delta(\bq-\bq')$.
From Eqs.~\eqref{eq: spin stiff spiral def} and \eqref{eq: Z factors spiral def} it immediately follows that $J^{\perp,1}_{\alpha\beta}=J^{\perp,2}_{\alpha\beta}\equiv J^{\perp}_{\alpha\beta}$, and $\chi_\mathrm{dyn}^{\perp,1}=\chi_\mathrm{dyn}^{\perp,2}\equiv \chi_\mathrm{dyn}^{\perp}$, as expected in the case of coplanar order~\cite{Azaria1990}. To derive the equations above, we have made use of the Goldstone theorem, which for spiral ordering reads (see for example Refs.~\cite{Chubukov1995,Kampf1996})
\begin{subequations}
    \begin{align}
        &\widetilde{C}^{33}(\pm\bQ,0)=0,\\
        &\widetilde{C}^{22}(\bzero,0)=0.
    \end{align}
\end{subequations}
Notice that the above relations can be also derived from a functional identity similar to~\eqref{eq: Ward identity SU(2)} but descending from the \emph{global} SU(2) symmetry.
Moreover, close to their respective Goldstone points ($(\bzero,0)$ for $\widetilde{C}^{22}$, and $(\pm\bQ,0)$ for $\widetilde{C}^{33}$), $\widetilde{C}^{22}(q)$ can be replaced with $1/\chit^{22}(q)$, and $\widetilde{C}^{33}(q)$ with $1/\chit^{33}(q)$, with the rotated susceptibilities defined analogously to~\eqref{eq: rotated Ks spiral}.   
If the spin spiral state occurs on a lattice that preserves parity, we have $\widetilde{C}^{aa}(\bq,\omega)=\widetilde{C}^{aa}(-\bq,\omega)$, from which we obtain
\begin{subequations}\label{eq: stiffenss and chi spiral final}
    \begin{align}
        &J_{\alpha\beta}^\perp=\frac{1}{4}\varphi_0^2 \partial^2_{q_\alpha q_\beta}\left( \frac{1}{\chit^{33}(\bq,0)}\right)\bigg\rvert_{\bq\to \pm \bQ},
        \label{eq: stiffenss and chi spiral final J2}\\
        &J^\smsqr_{\alpha\beta}=\frac{1}{2}\varphi_0^2 \partial^2_{q_\alpha q_\beta}\left( \frac{1}{\chit^{22}(\bq,0)}\right)\bigg\rvert_{\bq\to \bzero},
        \label{eq: stiffenss and chi spiral final J3}\\
        &\chi_\mathrm{dyn}^\perp=-\frac{1}{4}\varphi_0^2 \partial^2_{\omega}\left( \frac{1}{\chit^{33}(\pm\bQ,\omega)}\right)\bigg\rvert_{\omega\to 0},
        \label{eq: stiffenss and chi spiral final chi2}\\
        &\chi_\mathrm{dyn}^\smsqr=-\frac{1}{2}\varphi_0^2 \partial^2_{\omega}\left( \frac{1}{\chit^{22}(\bzero,\omega)}\right)\bigg\rvert_{\omega\to 0}.
        \label{eq: stiffenss and chi spiral final chi3}
    \end{align}
\end{subequations}
Neglecting the imaginary parts of the susceptibilities, giving rise to dampings of the Goldstone modes~\cite{Bonetti2022}, from Eq.~\eqref{eq: stiffenss and chi spiral final} we can obtain expressions for the susceptibilities near their Goldstone points
\begin{subequations}\label{eq: low energy chis spiral}
    \begin{align}
        &\chit^{22}(q\simeq(\bzero,0))\simeq \frac{\varphi_0^2}{-\chi^\smsqr_\mathrm{dyn}\omega^2+J^\smsqr_{\alpha\beta}q_\alpha q_\beta},\\
        &\chit^{33}(q\simeq(\pm\bQ,0))\simeq \frac{\varphi_0^2/2}{-\chi^\perp_\mathrm{dyn}\omega^2+J_{\alpha\beta}^\perp(q\mp Q)_\alpha (q\mp Q)_\beta}.
    \end{align}
\end{subequations}

Expressions \eqref{eq: low energy chis spiral} can be deduced from a low energy model also in the case of spin spiral ordering. Similarly to what we have done for the N\'eel case, we consider a pure gauge field, giving the nonlinear sigma model action
\begin{equation}
    \mathcal{S}_\mathrm{eff}[\mathcal{R}]=\frac{1}{2}\int_x \tr\left[\mathcal{P}_{\mu\nu}\partial_\mu \mathcal{R}(x)\partial_\nu \mathcal{R}^T(x)\right],
    \label{eq: O(3)xO(2)/O(2) NLsM}
\end{equation}
where $\mathcal{R}(x)\in\mathrm{SO(3)}$ is defined as in Eq.~\eqref{eq: mathcalR definition}, and now $\partial_\mu$ denotes $(-\partial_t,\vec{\nabla})$. The matrix $\mathcal{P}_{\mu\nu}$ is given by
\begin{equation}
    \mathcal{P}_{\mu\nu}=
    \left(
    \begin{array}{ccc}
        \frac{1}{2}J^\smsqr_{\mu\nu} & 0 & 0 \\
        0 & \frac{1}{2}J^\smsqr_{\mu\nu} & 0 \\
        0 & 0 & J^\perp_{\mu\nu}-\frac{1}{2}J^\smsqr_{\mu\nu}
    \end{array}
    \right),
\end{equation}
with 
\begin{equation}
    J^a_{\mu\nu}=
    \left(
    \begin{array}{c|c}
        -\chi_\mathrm{dyn}^a & 0 \\ \hline
        0 & J^a_{\alpha\beta}
    \end{array}
    \right),
\end{equation}
for $a\in\{\smsqr,\perp\}$.
Action~\eqref{eq: O(3)xO(2)/O(2) NLsM} is a NL$\sigma$M describing low energy fluctuations around a spiral magnetic ordered state. It has been introduced and studied in the context of frustrated antiferromagnets~\cite{Azaria1990,Azaria1992,Azaria1993,Azaria1993_PRL}.

We now write the field $\vec{\phi}^\prime(x)$ as $\vec{\phi}^\prime(x)=\varphi_0 \mathcal{M}(x)\mathcal{R}(x)\hat{v}(\bs{x})$, and consider an  $\mathcal{R}(x)$ stemming from a SU(2) matrix $R(x)=e^{i\theta_a(x)\frac{\sigma^a}{2}}$ with $\theta_a(x)$ infinitesimal, that is, 
\begin{equation}
    \mathcal{R}_{ab}(x)\simeq\delta_{ab}-\varepsilon^{abc}\theta_c(x),
    \label{eq: mathcal R small theta}
\end{equation}
we get
\begin{equation}
    \begin{split}
        \vec{\phi}^\prime(x)&\simeq\varphi_0 \mathcal{M}(x)[\hat{v}(\bs{x})-\hat{v}(\bs{x})\times\vec{\theta}(x)]\\
        &=\varphi_0[\hat{e}_1-\hat{e}_1\times\vec{\theta}^\prime(x)],
    \end{split}
\end{equation}
with $\hat{e}_1=(1,0,0)$, and $\vec{\theta}^\prime(x)=\mathcal{M}(x)\vec{\theta}(x)$. Inserting \eqref{eq: mathcal R small theta} into \eqref{eq: O(3)xO(2)/O(2) NLsM}, we obtain 
\begin{equation}
    \begin{split}
        \mathcal{S}_\mathrm{eff}[\vec{\theta}]=\frac{1}{2}\int_x \bigg\{&J^\perp_{\mu\nu}\sum_{a=1,2}\left[\partial_\mu \theta_a(x)\partial_\nu \theta_a(x)\right]\\
        &+J^\smsqr_{\mu\nu}\partial_\mu \theta_3(x)\partial_\nu \theta_3(x)\bigg\}.
    \end{split}
    \label{eq: NLsM spiral linearized}
\end{equation}
We are finally in the position to extract the form of the susceptibilities for small fluctuations
\begin{subequations}
    \begin{align}
        \chit^{22}(q)=&\langle\phi^{\prime}_2(q)\phi^{\prime}_2(-q)\rangle\simeq\varphi_0^2\langle\theta^{\prime}_3(q)\theta^{\prime}_3(-q)\rangle\nonumber\\
        =&\frac{\varphi_0^2}{-\chi_\mathrm{dyn}^\smsqr\omega^2+J^\smsqr_{\alpha\beta}q_\alpha q_\beta},\\
        \chit^{33}(q)=&\langle\phi^{\prime}_3(q)\phi^{\prime}_3(-q)\rangle\simeq\varphi_0^2\langle\theta^{\prime}_2(q)\theta^{\prime}_2(-q)\rangle\nonumber\\
        =&\sum_{\eta=\pm}\frac{\varphi_0^2/2}{-\chi_\mathrm{dyn}^\perp\omega^2+J_{\alpha\beta}^\perp(q-\eta Q)_\alpha (q-\eta Q)_\beta},
    \end{align}
\end{subequations}
which is the result of Eq.~\eqref{eq: low energy chis spiral}. 
In the above equations we have used the correlators of the $\theta$ field descending from action \eqref{eq: NLsM spiral linearized}. Form~\eqref{eq: low energy chis spiral} of the susceptibilities predicts three linearly dispersing Goldstone modes, two of which (the out-of-plane ones) are degenerate and propagate with velocities $c_\perp^{(n)} = \sqrt{\lambda_\perp^{(n)}/\chi^\perp_\mathrm{dyn}}$, where $\lambda^{(n)}_\perp$ are the eigenvalues of $J_{\alpha\beta}^\perp$ and $n=1,\dots,d$. Similarly, the in-plane mode velocity is given by $c_\smsqr^{(n)}=\sqrt{\lambda_\smsqr^{(n)}/\chi_\smsqr^\mathrm{dyn}}$, with $\lambda_\smsqr^{(n)}$ the eigenvalues of $J_{\alpha\beta}^\smsqr$.
\section{Explicit calculation for a spiral magnet}
\label{sec: Explicit calculation}
In this section, we present an explicit calculation for a spiral magnet, emerging from a Hubbard model, of the left and right hand sides of \eqref{eq: spin stiff spiral def} and \eqref{eq: Z factors spiral def}, and show that the Ward identities are fulfilled. We compute the spin susceptibilities and the gauge kernels within the random phase approximation (RPA), which is a conserving approximation in the sense of Baym and Kadanoff~\cite{Baym1961}, and it is therefore expected to fulfill the Ward identities.
\subsection{Mean-field approximation}
In a fermionic lattice system, the amplitude $\varphi_0$ of spin spiral order parameter
\begin{equation}
    \langle\vec{\phi}(x)\rangle=\varphi_0 (\cos(\bQ\cdot\bs{x}),\sin(\bQ\cdot\bs{x}),0),
\end{equation}
can be expressed as 
\begin{equation}
    \varphi_0 = \int_\bk \langle\psibar_{\bk,\up}\psi_{\bk+\bQ,\down} \rangle,
    \label{eq: spiral OP fermions}
\end{equation}
where $\int_\bk$ is a shorthand for an integral over the full $d$-dimensional Brillouin zone, that is $\int_\mathrm{\bk\in BZ}\frac{d^d\bk}{(2\pi)^d}$.
From~\eqref{eq: spiral OP fermions}, we deduce that spiral ordering couples only the electron states $(\bk,\up)$ and $(\bk+\bQ,\down)$, and the mean-field fermion Green's function can be expressed as a $2\times2$ matrix:
\begin{equation}
    \widetilde{\mathbf{G}}_\bk(i\nu_n) = 
    \left(
    \begin{array}{cc}
        i\nu_n-\xi_{\bk} & -\Delta \\
        -\Delta & i\nu-\xi_{\bk+\bQ}
    \end{array}\right)^{-1},
    \label{eq: spiral Gf}
\end{equation}
where $\nu_n=(2n+1)\pi T$ ($n\in\mathbb{Z}$) is a fermionic Matsubara frequency, $\xi_\bk=\eps_\bk-\mu$, with $\eps_\bk$ the single particle dispersion and $\mu$ the chemical potential, and $\Delta$ is the spiral order parameter. Diagonalizing~\eqref{eq: spiral Gf}, one obtains the quasiparticle dispersions
\begin{equation}
    E^\pm_\bk=g_\bk\pm \sqrt{h_\bk^2+\Delta^2},
\end{equation}
with $g_\bk=(\xi_\bk+\xi_{\bk+\bQ})/2$, and $h_\bk=(\xi_\bk-\xi_{\bk+\bQ})/2$. 

The Green's function can be conveniently written as a linear combination of quasiparticle poles
\begin{equation}
    \widetilde{\mathbf{G}}_\bk(i\nu_n) = \sum_{\ell=\pm} \frac{1}{2}u^\ell_{\bk} \frac{1}{i\nu_n-E^\ell_{\bk}},
    \label{eq: spiral Gf comf}
\end{equation}
where the coefficients $u^\ell_\bk$ are given by
\begin{equation}
    u^\ell_\bk=\sigma^0+\ell \frac{h_{\bk}}{e_{\bk}} \sigma^3 + \ell\frac{\Delta}{e_{\bk}}\sigma^1,
\end{equation}
with $\sigma^0=\mathbb{1}$, and $e_\bk=\sqrt{h_\bk^2+\Delta^2}$.
We assume the spiral states to emerge from a lattice model with onsite repulsive interactions (Hubbard model), with imaginary time action
\begin{equation}
    \begin{split}
        \mathcal{S}[\psi,\psibar]=\int_0^\beta\!d\tau\bigg\{&\sum_{j,j',\sigma}\psibar_{j,\sigma}\left[(\partial_\tau - \mu)\delta_{jj'} + t_{jj'}\right]\psi_{j',\sigma}\\
        &+ U\sum_{j}\psibar_{j,\up}\psibar_{j,\down}\psi_{j,\down}\psi_{j,\up}\bigg\},
    \end{split}
    \label{eq: Hubbard action}
\end{equation}
where $t_{jj'}$ describes the hopping amplitude between the lattice sites labeled by $j$ and $j'$ and $U$ is the Hubbard interaction. 
The Hartree-Fock or mean-field (MF) gap equation at temperature $T$ reads
\begin{equation}
    \Delta = -U \int_\bk T\sum_{\nu_n} \widetilde{G}^{\up\down}_\bk(i\nu_n)=U\int_\bk \frac{\Delta}{2e_\bk}\left[f(E^-_\bk)-f(E^+_\bk)\right],
    \label{eq: gap equation}
\end{equation}
where $\sum_{\nu_n}$ denotes a sum over the fermionic Matsubara frequencies, and $f(x)=1/(e^{x/T}+1)$ is the Fermi function. The bosonic order parameter $\varphi_0$ is related to $\Delta$ as $\Delta=U\varphi_0$. Finally, the optimal $\bQ$-vector is obtained minimizing the mean-field free energy
\begin{equation}
    \begin{split}
        \mathcal{F}_\mathrm{MF}(\bQ)=&-T\sum_{\nu_n}\int_\bk \Tr\ln \widetilde{\mathbf{G}}_\bk(i\nu_n)+\frac{\Delta^2}{U}+\mu n \\
        =&-T\int_\bk \sum_{\ell=\pm}\ln\left(1+e^{-E^\ell_\bk/T}\right)+\frac{\Delta^2}{U}+\mu n, 
    \end{split} 
    \label{eq: MF free energy}
\end{equation}
where $n$ is the fermion density.
%
%
%
\subsection{Random phase approximation}
\label{sec: RPA}
In this Section, we summarize the RPA for a spiral magnet, closely following Ref.~\cite{Bonetti2022}. Further details can be also found in Ref.~\cite{Kampf1996}.

The charge and spin susceptibilities are coupled in a spin spiral state. It is therefore convenient to consider the retarded real-time correlators~\cite{Bonetti2022}
\begin{equation}
    \chi^{ab}_{jj'}(t)=-i\Theta(t)\langle \left[S^a_j(t),S^b_{j'}(0)\right]\rangle,
\end{equation}
where $j$ and $j'$ label two lattice sites, $t$ is the time, $[\bullet,\bullet]$ denotes the commutator, $\Theta(t)$ is the Heaviside function, and $a$ and $b$ run from 0 to 3. The generalized spin operators are defined in terms of the fermionic fields as
\begin{equation}
    S^a_j=\frac{1}{2}\sum_{s,s'=\up,\down}\psibar_{j,s}\sigma^a_{ss'}\psi_{j,s'}.
\end{equation}
The $a=1,2,3$ components correspond to the usual spin operator, while $a=0$ gives half the fermion density $n_j =\psibar_{j,\up}\psi_{j,\up}+\psibar_{j,\down}\psi_{j,\down}$. 

It is convenient to work in the rotated basis~\cite{Kampf1996} introduced in Eqs.~\eqref{eq: spiral rotation field} and \eqref{eq: spiral rotation matrix}, the susceptibilities transform as
\begin{equation}
    \chit_{jj'}(t)=\mathcal{M}_j \chi_{jj'}(t) \mathcal{M}^T_{j'},
    \label{eq: rotated chis spiral RPA}
\end{equation}
with 
\begin{equation}
    \mathcal{M}_j=\left(
    \begin{array}{cccc}
        1 & 0 & 0 & 0\\
        0 & \cos(\bQ\cdot\bs{r}_j) & \sin(\bQ\cdot\bs{r}_j) & 0 \\
        0 & -\sin(\bQ\cdot\bs{r}_j) & \cos(\bQ\cdot\bs{r}_j) & 0 \\
        0 & 0 & 0 & 1
    \end{array}
    \right), 
    \label{eq: spiral rotation matrix with charge}
\end{equation}
where $\bs{r}_j$ represents the coordinates of the lattice site $j$. As previously mentioned, the $\chit^{jj'}(t)$ are translationally invariant, so that their Fourier transform depends only on a single spatial momentum $\bq$. 

Within RPA, the susceptibilities are given by 
\begin{equation}
    \chit(q) = \chit_0(q)\left[\mathbb{1}-\Gamma_0\chit_0(q)\right]^{-1},
    \label{eq: chi RPA}
\end{equation}
where $\chit_0(q)$ is a matrix in the $a,b=0,1,2,3$ indices, $\mathbb{1}$ is the $4\times 4$ identity matrix, and $
\Gamma_0=2U\mathrm{diag}(-1,1,1,1)$. The bare susceptibilities on the real frequency axis are given by
\begin{equation}
     \begin{split}
         &\chit^{ab}_0(\bq,\omega) = \\
         &- \frac{1}{4} \int_\bk T \sum_{\nu_n} \tr \big[ 
        \sigma^a\,\widetilde{\mathbf{G}}_\bk(i\nu_n) \,
        \sigma^b\,\widetilde{\mathbf{G}}_{\bk+\bq}(i\nu_n+i\Omega_m) \big],
     \end{split}
\end{equation}
where $\Omega_m=2m\pi T$ ($m\in\mathbb{Z}$) denotes a bosonic Matsubara frequency and the substitution $i\Omega_m\to\omega+i0^+$ has been performed. Using \eqref{eq: spiral Gf comf}, one can perform the frequency sum, obtaining
\begin{equation}
     \chit^{ab}_0(\bq,\omega) = - \frac{1}{8}\sum_{\ell,\ell'=\pm}\int_\bk \mathcal{A}^{ab}_{\ell\ell'}(\bk,\bq)F_{\ell\ell'}(\bk,\bq,\omega),
     \label{eq: chi0 def}
\end{equation}
where we have defined 
\begin{equation}
    F_{\ell\ell'}(\bk,\bq,\omega)=\mathcal{P}\frac{f(E^\ell_\bk)-f(E^{\ell'}_{\bk+\bq})}{\omega+E^\ell_\bk-E^{\ell'}_{\bk+\bq}},
    \label{eq: Fll def}
\end{equation}
with $\mathcal{P}$ the principal value, and the coherence factors
\begin{equation}
    \mathcal{A}^{ab}_{\ell\ell'}(\bk,\bq)=\frac{1}{2}\Tr\left[\sigma^a u^\ell_\bk\sigma^b u^{\ell'}_{\bk+\bq}\right].
    \label{eq: coh fact def}
\end{equation}
Notice that in Eq.~\eqref{eq: Fll def} we have neglected the imaginary part of $1/(\omega+i0^++E^\ell_\bk-E^{\ell'}_{\bk+\bq})$, which is proportional to a $\delta$ function, and is responsible for the \emph{Landau damping} of the collective spin fluctuations~\cite{Bonetti2022}.   

The bare susceptibilities have a well defined sign under inversion of $\bq$ and $\omega$, and under the exchange of the indices $a$ and $b$. These symmetries can be derived from Eqs.~\eqref{eq: chi0 def}, \eqref{eq: Fll def}, and \eqref{eq: coh fact def} (see, for example, Ref.~\cite{Bonetti2022}) and are listed in Table~\ref{tab:symmetries}. Since neglecting the imaginary part of $F_{\ell\ell'}(\bk,\bq,\omega)$ makes the matrix $\chit_0^{ab}(q)$ hermitian, we deduce that those elements which are even (odd) under the exchange $a\leftrightarrow b$ are purely real (imaginary). At least within the RPA, the symmetry properties derived for the bare $\chit_0(q)$ hold also for the full susceptibilities $\chit(q)$.
\begin{table}[t]
\centering
\begin{tabular}{|c|c|c|c|c|}
        \hline
        $a,b$ & 0 & 1 & 2 & 3  \\
        \hline
        0 & $+,+,+$ & $+,+,+$ & $-,+,-$ & $-,-,+$ \\
        \hline
        1 & $+,+,+$ & $+,+,+$ & $-,+,-$ & $-,-,+$ \\
        \hline
        2 & $-,+,-$ & $-,+,-$ & $+,+,+$ & $+,-,-$ \\
        \hline
        3 & $-,-,+$ & $-,-,+$ & $+,-,-$ & $+,+,+$ \\
        \hline
\end{tabular}
\caption{Symmetries of the bare susceptibilities. The first two signs in each cell represent the sign change of $\chit_0^{ab}(q)$ by flipping the sign of $\bq$ and $\omega$, respectively. The third sign corresponds to the sign change of $\chit_0^{ab}(q)$ when $a$ and $b$ are interchanged.}
\label{tab:symmetries}
\end{table}

Goldstone modes associated with spontaneous symmetry breaking of the SU(2) spin symmetry appear in the susceptibilities. In particular, considering the limit $q\to 0$, all the off-diagonal elements of the type $\chit_0^{2a}(q)$ and $\chit_0^{a2}(q)$ vanish as they are odd in $\bq$ or $\omega$, while $\chit_0^{22}(q)$ takes the simple form
\begin{equation}
    \chit_0^{22}(0)=\int_\bk \frac{f(E^-_\bk)-f(E^+_\bk)}{4e_\bk}.
\end{equation}
The full susceptibility at $q=0$ therefore reads
\begin{equation}
    \chit^{22}(0)=\frac{\chit^{22}_0(0)}{1-2U\chit^{22}_0(0)}.
    \label{eq: chi22 q=0}
\end{equation}
We immediately see that when the gap equation~\eqref{eq: gap equation} is fulfilled, the denominator of~\eqref{eq: chi22 q=0} vanishes, signaling a gapless mode. This Goldstone mode represents fluctuations of the spins within the plane in which the magnetization lies. We therefore refer to it as \emph{in-plane} mode in the following. With a similar reasoning, one proves that in the limit $q\to\pm Q$ (with $Q=(\bQ,0)$), all the off diagonal elements of the type $\chit_0^{3a}(q)$ and $\chit_0^{a3}(q)$ vanish, leaving
\begin{equation}
    \chit^{33}(\pm Q)=\frac{\chit^{33}_0(\pm Q)}{1-2U\chit^{33}_0(\pm Q)}.
    \label{eq: chi33 q=pm Q}
\end{equation}
Noticing that $\chit_0^{33}(\pm Q)=\chit_0^{22}(0)$~\cite{Bonetti2022}, one sees that the denominator of \eqref{eq: chi33 q=pm Q} is vanishing. The susceptibility $\chit^{33}(q)$ therefore hosts two poles at $q=\pm Q$, which are due to gapless excitations of the spins outside the plane in which the magnetization lies. We refer to these modes as \emph{out-of-plane} modes. As already stated, in the spiral state three Goldstone modes appear, as \emph{all} three generators of the symmetry group SU(2) are broken. However, the two out-of-plane modes have degenerate dispersions, as it is expected for \emph{coplanar} order. 
\subsection{Small \texorpdfstring{$\bq$}{q} and \texorpdfstring{$\omega$}{omega} expansion of the susceptibilities}
\label{sec: small q and omega RPA}
We are now in the position to compute the right hand sides of Eqs.~\eqref{eq: spin stiff spiral def} and \eqref{eq: Z factors spiral def}. The small $\bq$ and $\omega$ expansion of the susceptibilities in the spiral state has already carried out in Ref.~\cite{Bonetti2022}, but for completeness we revisit it here in a slightly different form.  
\subsubsection{In-plane mode}
Using \eqref{eq: chi RPA}, the in-plane susceptibility can be conveniently written as 
\begin{equation}
    \chit^{22}(q)=\frac{\overline{\chi}_0^{22}(q)}{1-2U\overline{\chi}_0^{22}(q)}, 
    \label{eq: chi22 RPA}
\end{equation}
with
\begin{equation}
    \overline{\chi}_0^{22}(q)=\chit_0^{22}(q) + \sum_{a,b\in\{0,1,3\}}\chit_0^{2a}(q)\Gammat_{2}^{ab}(q)\chit_0^{b2}(q).
\end{equation}
$\Gammat_{2}(q)$ is given by
\begin{equation}
    \Gammat_{2}(q)=\left[\mathbb{1}_3-\Gamma_{0,2}\chit_{0,2}(q)\right]^{-1}\Gamma_{0,2},
\end{equation}
where $\Gamma_{0,2}^{ab}$ and $\chit_{0,2}^{ab}(q)$ are matrices obtained from $\Gamma_0^{ab}$ and $\chit_0^{ab}(q)$ removing the components where $a=2$ and/or $b=2$, and $\mathbb{1}_3$ denotes the $3\times3$ identity matrix.
For later convenience, we notice that for $q=0$, all the off-diagonal elements $\chit_0^{2a}(q)$ and $\chit_0^{a2}(q)$ vanish, so that $\Gammat_{2}(0)$ can be obtained from the full expression
\begin{equation}
    \Gammat(q)=\left[\mathbb{1}-\Gamma_{0}\chit_{0}(q)\right]^{-1}\Gamma_{0},
    \label{eq: Gamma(q) definition}
\end{equation}
selecting only the components in which the indices take the values 0,1, or 3. 

Setting $\omega=0$, the bare susceptibilities $\chit_0^{23}(q)$ and $\chit_0^{32}(q)$ vanish as they are odd in $\omega$. Moreover, in the limit $\bq\to\bzero$, $\chit_0^{2a}(\bq,0)$ and $\chit_0^{a2}(\bq,0)$, with $a=0,1$, are linear in $\bq$ as they are odd under $\bq\to-\bq$. The in-plane spin stiffness can be therefore written as 
\begin{equation}
    \begin{split}
        J^\smsqr_{\alpha\beta}=&
        -2\Delta^2\partial^2_{q_\alpha q_\beta}\overline{\chi}_0^{22}(0)\\
        =&-2\Delta^2\bigg[
        \partial^2_{q_\alpha q_\beta}\chit_0^{22}(0)\\&+
        2\sum_{a,b\in\{0,1\}}\partial_{q_\alpha}\chit_0^{2a}(0)\,\Gammat^{ab}(\bq\to\bzero,0)\,\partial_{q_\alpha}\chit_0^{b2}(0)
        \bigg],
    \end{split}
    \label{eq: J3 RPA}
\end{equation}
where we have used $\overline{\chi}_0^{22}(0)=\chit_0^{22}(0)=1/(2U)$, descending from the gap equation, and $\partial_{q_\alpha}f(0)$ is a shorthand for $\partial f(\bq,0)/\partial q_\alpha |_{\bq\to\bzero}$, and similarly for $\partial^2_{q_\alpha q_\beta}f(0)$.

In a similar way, if we set $\bq$ to $\bzero$ and consider the limit of small $\omega$, the terms where $a$ and/or $b$ are 0 or 1 vanish as $\chit_0^{2a}(q)$ and $\chit_0^{a2}(q)$ for a=0,1 are odd in $\bq$. On the other hand, $\chit_0^{23}(q)$ and $\chit_0^{32}(q)$ are linear in $\omega$ for small $\omega$. With these considerations, the in-plane dynamical susceptibility is given by
\begin{equation}
    \begin{split}
        \chi_\mathrm{dyn}^\smsqr=
        &2\Delta^2\partial^2_{\omega}\overline{\chi}_0^{22}(0)\\
        =&2\Delta^2\Big[
        \partial^2_{\omega}\chit_0^{22}(0)\\
        &+2\partial_{\omega}\chit_0^{23}(0)\,\Gammat^{33}(\bzero,\omega\to 0)\,\partial_{\omega}\chit_0^{32}(0)
        \Big],
    \end{split}
    \label{eq: chi3 RPA}
\end{equation}
where $\partial^n_{\omega}f(0)$ is a shorthand for $\partial^n f(\bzero,\omega)/\partial \omega^n |_{\omega\to 0}$, and $\Gammat^{33}(\bzero,\omega\to 0)$ can be cast in the simple form
\begin{equation}
    \Gammat^{33}(\bzero,\omega\to 0)=\frac{2U}{1-2U \chit_0^{33}(\bzero,\omega\to 0)}.
\end{equation}
\subsubsection{Out-of-plane modes}
Similarly to the in-plane mode, one can write the out-of-plane susceptibility in the form
\begin{equation}
    \chit^{33}(q)=\frac{\overline{\chi}_0^{33}(q)}{1-2U\overline{\chi}_0^{33}(q)}, 
    \label{eq: chi33 RPA}
\end{equation}
with
\begin{equation}
    \overline{\chi}_0^{33}(q)=\chit_0^{33}(q) + \sum_{a,b\in\{0,1,2\}}\chit_0^{3a}(q)\Gammat_{3}^{ab}(q)\chit_0^{b3}(q),
\end{equation}
where $\Gammat_{3}(q)$ is defined similarly to $\Gammat_{2}(q)$, removing the components that involve the index 3 instead of the index 2. We also notice that 
\begin{equation}
    \Gammat_{3}^{ab}(\bq,0)=\Gammat^{ab}(\bq,0),
\end{equation}
for $a$, $b=0,1,2$, because all the off-diagonal components $\chit_0^{3a}(q)$ and $\chit_0^{a3}(q)$ vanish for zero frequency. Using $\overline{\chi}_0^{33}(\pm Q)=\chit_0^{33}(\pm Q)=1/(2U)$, we obtain the out-of-plane spin stiffness
\begin{equation}
    J_{\alpha\beta}^\perp=
        -\Delta^2\partial^2_{q_\alpha q_\beta}\overline{\chi}_0^{33}(\pm Q)=
        -\Delta^2\partial^2_{q_\alpha q_\beta}\chit_0^{33}(\pm Q),
\end{equation}
where $\partial^2_{q_\alpha q_\beta} f(\pm Q)$ stands for $\partial^2f(\bq,0)/\partial q_\alpha\partial q_\beta |_{\bq\to\pm\bQ}$. 

In the limit $\omega\to 0$, all the $\chit_0^{3a}(q)$ and $\chit_0^{a3}(q)$, with $a=0,1,2$, are linear in $\omega$, and the dynamical susceptibility is given by
\begin{equation}
    \begin{split}
        \chi_\mathrm{dyn}^\perp&=\Delta^2 \partial^2_\omega \overline{\chi}_0^{33}(\pm Q)\\ 
        &=\Delta^2 \bigg[
        \partial^2_\omega \chit^{33}_0(\pm Q)\\
        &\phantom{=}+ 2\sum_{a,b\in\{0,1,2\}}\partial_\omega \chit_0^{3a}(\pm Q) \Gammat^{ab}(\pm Q) \partial_\omega \chit_0^{b3}(\pm Q)
        \bigg],
    \end{split}
\end{equation}
with $\partial^n_{\omega} f(\pm Q)$ a shorthand for $\partial^nf(\pm\bQ,\omega)/\partial \omega^n |_{\omega\to 0}$. We remark that for $\Gammat^{ab}(q)$ the limits $\bq\to\bQ$ and $\omega\to 0$ commute if $\bQ$ is not a high-symmetry wavevector, that is, if $E^\ell_{\bk+\bQ}\neq E^\ell_\bk$. 
\subsection{Gauge kernels}
\label{sec: gauge kernels}
To calculate the gauge kernels, that is, the left hand sides of Eqs.~\eqref{eq: spin stiff spiral def} and \eqref{eq: Z factors spiral def}, we couple our system to a SU(2) gauge field via a Peierls substitution in the quadratic part of action~\eqref{eq: Hubbard action}:
\begin{equation}
    \begin{split}
        \mathcal{S}_0[\psi,\psibar,A_\mu]=\int_0^\beta \!d\tau \sum_{jj'}&\psibar_j \Big[
        (\partial_\tau - A_{0,j} + \mu)\delta_{jj'}\\
        &+t_{jj'}e^{-\bs{r}_{jj'}\cdot(\boldsymbol{\nabla}-i \bs{A}_j)}
    \Big]\psi_j,
    \end{split}
    \label{eq: S0 coupled to SU(2) gauge field}
\end{equation}
where $e^{-\bs{r}_{jj'}\cdot\boldsymbol{\nabla}}$ is the translation operator from site $j$ to site $j'$, with $\bs{r}_{jj'}=\bs{r}_j-\bs{r}_{j'}$. Notice that under the transformation $\psi_j\to R_j\psi_j$, with $R_j\in\mathrm{SU(2)}$, the interacting part of the action~\eqref{eq: Hubbard action} is left unchanged, while the gauge field transforms according to \eqref{eq: SU(2) Amu transformation}. Since the gauge kernels correspond to correlators of two gauge fields, we expand \eqref{eq: S0 coupled to SU(2) gauge field} to second order in $A_\mu$. After a Fourier transformation one obtains
\begin{equation}
    \begin{split}
        \mathcal{S}_0[\psi,\psibar,A_\mu]=
        &-\int_k \psibar_k \left[i\nu_n+\mu-\eps_\bk\right]\psi_k\\
        &+\frac{1}{2}\int_{k,q}A_\mu^a(q) \gamma^\mu_{\bk}\, \psibar_{k+q}\sigma^a\psi_k\\
        &-\frac{1}{8}\int_{k,q,q'}A^a_\alpha(q-q')A^a_\beta(q') \gamma^{\alpha\beta}_\bk \psibar_{k+q}\psi_k,
    \end{split}
    \label{eq: S0 coupled with Amu SU(2) momentum space}
\end{equation}
where the first order coupling is given by $\gamma^\mu_\bk=(1,\boldsymbol{\nabla}_\bk\eps_\bk)$, and the second order one is $\gamma^{\alpha\beta}_\bk=\partial^2_{k_\alpha k_\beta}\eps_{\bk}$. Analyzing the coupling of the temporal component of the gauge field to the fermions in \eqref{eq: S0 coupled to SU(2) gauge field} and \eqref{eq: S0 coupled with Amu SU(2) momentum space}, we notice that the temporal components of the gauge kernel are nothing but the susceptibilities in the original (unrotated) spin basis
\begin{equation}
    K_{00}^{ab}(\bq,\bq',\omega)=\chi^{ab}(\bq,\bq',\omega),
\end{equation}
where $\omega$ is a real frequency. 

\begin{figure*}[t!]
    \centering
    \includegraphics[width=1.\textwidth]{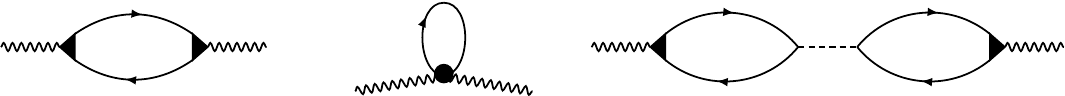}
    \caption{Diagrams contributing to the spin stiffnesses. The wavy line represents the external SU(2) gauge field, the solid lines the electronic Green's functions, the black triangles the paramagnetic vertex $\gamma^\mu_\bk \sigma^a$, the black circle the diamagnetic one $\gamma^{\alpha\beta}_\bk \sigma^0$, and the dashed line the effective interaction $\Gamma(\bq,\bq',\omega)$.}
    \label{fig: fig1}
\end{figure*}
The spatial components of the gauge kernel can be expressed in the general form (see Fig.~\ref{fig: fig1})
\begin{equation}
    \begin{split}
        K_{\alpha\beta}^{ab}(\bq,\bq',\omega)=
        &K_{\mathrm{para},\alpha\beta}^{ab}(\bq,\bq',\omega)
        +\delta_{ab}\,K_{\alpha\beta}^{\mathrm{dia}}\\
        +&\int_{\bq^{\prime\prime},\bq^{\prime\prime\prime}}\sum_{c,d}K_{\mathrm{para},\alpha 0}^{ac}(\bq,\bq^{\prime\prime},\omega)\times\\
        \times&\Gamma^{cd}(\bq^{\prime\prime},\bq^{\prime\prime\prime},\omega)K_{\mathrm{para},0\beta}^{db}(\bq^{\prime\prime\prime},\bq^{\prime},\omega),
    \end{split}
    \label{eq: gauge kernel formula}
\end{equation}
where $\Gamma(\bq',\bq^{\prime\prime},\omega)$ is the effective interaction~\eqref{eq: Gamma(q) definition} expressed in the unrotated basis. Within the RPA, the paramagnetic terms are given by
\begin{equation}
    \begin{split}
        &K_{\mathrm{para},\mu\nu}^{ab}(\bq,\bq',\omega)=\\
        &\hskip5mm-\frac{1}{4}\int_{\bk,\bk'}T\sum_{\nu_n}
    \gamma^\mu_\bk\gamma^\nu_{\bk'+\bq'}\tr\Big[\sigma^a\bs{G}_{\bk,\bk'}(i\nu_n)\sigma^b\times\\
    &\hskip20mm\times\bs{G}_{\bk'+\bq',\bk+\bq}(i\nu_n+i\Omega_m)\Big]\bigg\rvert_{i\Omega\to\omega+i0^+}.
    \end{split}
    \label{eq: paramagnetic contr Kernel}
\end{equation}
The Green's function in the unrotated basis takes the form
\begin{equation}
    \bs{G}_{\bk,\bk'}(i\nu_n)=\left(
    \begin{array}{cc}
        G_\bk(i\nu_n)\delta_{\bk,\bk'} & F_{\bk}(i\nu_n)\delta_{\bk,\bk'-\bQ} \\
        F_{\bk-\bQ}(i\nu_n)\delta_{\bk,\bk'+\bQ} & \overline{G}_{\bk-\bQ}(i\nu_n)\delta_{\bk,\bk'}
    \end{array}
    \right),
\end{equation}
where $\delta_{\bk,\bk'}$ is a shorthand for $(2\pi)^d\delta^d(\bk-\bk')$, and
\begin{subequations}
    \begin{align}
        &G_\bk(i\nu_n)=\frac{i\nu_n-\xi_{\bk+\bQ}}{(i\nu_n-\xi_{\bk})(i\nu_n-\xi_{\bk+\bQ})-\Delta^2},\\
        &\overline{G}_\bk(i\nu_n)=\frac{i\nu_n-\xi_{\bk}}{(i\nu_n-\xi_{\bk})(i\nu_n-\xi_{\bk+\bQ})-\Delta^2},\\
        &F_\bk(i\nu_n)=\frac{\Delta}{(i\nu_n-\xi_{\bk})(i\nu_n-\xi_{\bk+\bQ})-\Delta^2}.
    \end{align}
\end{subequations}
The diamagnetic term does not depend on $\bq$, $\bq'$ and $\omega$, and is proportional to the unit matrix in the gauge indices. It evaluates to 
\begin{equation}
    K_{\alpha\beta}^{\mathrm{dia}}=-\frac{1}{4}\int_{\bk,\bk'} T\sum_{\nu_n} (\partial^2_{k_\alpha k_\beta}\eps_\bk)\tr\left[\bs{G}_{\bk,\bk'}(i\nu_n)\right].
\end{equation}
We can now compute the spin stiffnesses and dynamical susceptibilities from the gauge kernels. 
\subsubsection{In-plane mode}
The in-plane spin stiffness is defined as
\begin{equation}
    J^\smsqr_{\alpha\beta}=-\lim_{\bq\to\bzero}K_{\alpha\beta}^{33}(\bq,0),
\end{equation}
where, similarly to Sec.~\ref{subsec: WI spiral}, we define as $K_{\mu\nu}(\bq,\omega)$ the prefactors of those components of the gauge kernels $K_{\mu\nu}(\bq,\bq',\omega)$ which are proportional to $\delta_{\bq,\bq'}$.
In addition to the bare term 
\begin{equation}
    J_{\alpha\beta}^{0,\smsqr}=-\lim_{\bq\to\bzero}K_{\alpha\beta}^{\mathrm{para},33}(\bq,0)-K_{\alpha\beta}^{\mathrm{dia}},
    \label{eq: J inplane bare}
\end{equation}
we find nonvanishing paramagnetic contributions that mix spatial and temporal components. They involve
\begin{subequations}\label{eq: off_diagonal kalphat_3a}
    \begin{align}
        &\lim_{\bq\to\bzero}K_{0\alpha}^{30}(\bq,\bq',0)=\kappa_\alpha^{30}(\bzero)\delta_{\bq',\bzero},\\
        &\lim_{\bq\to\bzero}K_{0\alpha}^{31}(\bq,\bq',0)=\kappa_\alpha^{31}(\bzero)\frac{\delta_{\bq',\bQ}+\delta_{\bq',-\bQ}}{2}, \label{eq: k31 def}\\
        &\lim_{\bq\to\bzero}K_{0\alpha}^{32}(\bq,\bq',0)=\kappa_\alpha^{32}(\bzero)\frac{\delta_{\bq',\bQ}-\delta_{\bq',-\bQ}}{2i}\label{eq: k32 def},
    \end{align}
\end{subequations}
where $\kappa_\alpha^{32}(\bzero)=\kappa_\alpha^{31}(\bzero)$. Noticing that for $a=0,1,2$, we have $\lim_{\bq\to\bzero}K^{3a}_{\alpha 0}(\bq,0)=\lim_{\bq\to\bzero}K^{a3}_{0\alpha}(\bq,0)$, and inserting this result into \eqref{eq: gauge kernel formula}, we obtain
\begin{equation}
    J_{\alpha\beta}^{\smsqr}=J_{\alpha\beta}^{0,\smsqr}-\sum_{a,b\in\{0,1\}}\kappa_\alpha^{3a}(\bzero)\Gammat^{ab}(\bq\to\bzero,0)\kappa_\beta^{3b}(\bzero),\label{eq: J3 gauge}
\end{equation}
where $\Gammat(\bq\to\bzero,0)$ is the effective interaction in the rotated spin basis, defined in Eq.~\eqref{eq: Gamma(q) definition}. Notice that the delta functions in Eq.~\eqref{eq: off_diagonal kalphat_3a} convert the unrotated $\Gamma$ to $\Gammat$ and, together with the equality $\kappa_\alpha^{32}(\bzero)=\kappa_\alpha^{31}(\bzero)$, they remove the terms where $a$ or $b$ equal 2 in the sum. 

The dynamical susceptibility is defined as
\begin{equation}
    \chi_\mathrm{dyn}^\smsqr=\lim_{\omega\to0}K_{00}^{33}(\bzero,\omega)=\lim_{\omega\to0}\chi^{33}(\bzero,\omega).
\end{equation}
From Eq.~\eqref{eq: spiral rotation matrix with charge} we deduce that
\begin{equation}
    \chi^{33}(\bq,\omega)=\chit^{33}(\bq,\omega).
\end{equation}
Remarking that for $\omega=0$ all the off-diagonal elements of the bare susceptibilities with one (and only one) of the two indices equal to 3 vanish, we obtain the RPA expression for $\chi_\mathrm{dyn}^\smsqr$
\begin{equation}
    \begin{split}
        \chi_\mathrm{dyn}^\smsqr = \lim_{\omega\to 0}
        \frac{\chit_0^{33}(\bzero,\omega)}{1-2U\chit_0^{33}(\bzero,\omega)}.
    \end{split}
    \label{eq: chi3 gauge}
\end{equation}
\subsubsection{Out-of-plane modes}
To compute the the out-of-plane stiffness, that is, 
\begin{equation}
    J_{\alpha\beta}^\perp=-\lim_{\bq\to\bzero}K_{\alpha\beta}^{22}(\bq,0),
\end{equation}
we find that all the paramagnetic contributions to the gauge kernel that mix temporal and spatial components vanish in the $\omega\to0$ and $\bq=\bq'\to\bzero$~\cite{notemixedkernelsoutofplane} limits. Moreover, the $\bq\to\bzero$ limit of the momentum diagonal paramagnetic contribution can be written as
\begin{equation}
    \begin{split}
        &\lim_{\bq\to\bzero}K_{\mathrm{para},\alpha\beta}^{22}(\bq,0)\\=
        &-\frac{1}{4}\int_{\bk,\bk'} T\sum_{\substack{\nu_n\\\zeta=\pm}}
        \gamma^\alpha_\bk\gamma^\beta_{\bk'}\tr\left[\sigma^\zeta\bs{G}_{\bk,\bk'}(i\nu_n)\sigma^{-\zeta}\bs{G}_{\bk',\bk}(i\nu_n)\right]\\
        =&-\frac{1}{2}\int_{\bk}T\sum_{\nu_n}\gamma^\alpha_\bk\gamma^\beta_{\bk}\,G_\bk(i\nu_n)\overline{G}_{\bk-\bQ}(i\nu_n),
    \end{split}
\end{equation}
where we have defined $\sigma^\pm=(\sigma^1\pm i\sigma^2)/2$. The out-of-plane spin stiffness is thus given by
\begin{equation}
    \begin{split}
        J^\perp_{\alpha\beta} = &-\frac{1}{2}\int_{\bk}T\sum_{\nu_n}\gamma^\alpha_\bk\gamma^\beta_{\bk}\,G_\bk(i\nu_n)\overline{G}_{\bk-\bQ}(i\nu_n)\\
        &-\frac{1}{4}\int_{\bk,\bk'} T\sum_{\nu_n} (\partial^2_{k_\alpha k_\beta}\eps_\bk)\tr\left[\bs{G}_{\bk,\bk'}(i\nu_n)\right]
    \end{split}
\end{equation}

Finally, we evaluate the dynamical susceptibility of the out-of-plane modes. This is defined as
\begin{equation}
    \chi_\mathrm{dyn}^\perp=\lim_{\omega\to 0}K_{00}^{22}(\bzero,\omega)=\lim_{\omega\to 0}\chi^{22}(\bzero,\omega).
\end{equation}
Applying transformation~\eqref{eq: rotated chis spiral RPA}, we can express the momentum-diagonal component of $\chi^{22}(\bq,\bq',\omega)$ in terms of the susceptibilities in the rotated basis as
\begin{equation}
    \begin{split}
        \chi^{22}(\bq,\omega)=\frac{1}{4}\sum_{\zeta=\pm}\Big[&
        \chit^{11}(\bq+\zeta\bQ,\omega)\\
        &+\chit^{22}(\bq+\zeta\bQ,\omega)\\
        &+2i\zeta\chit^{12}(\bq+\zeta\bQ,\omega)
        \Big],
        \label{eq: uniform chi22 spiral}
    \end{split}
\end{equation}
where we have used (see Sec.~\ref{sec: RPA}) $\chit^{12}(q)=-\chit^{21}(q)$. Sending $\bq$ to $\bzero$ in \eqref{eq: uniform chi22 spiral}, and using the symmetry properties of the susceptibilities for $\bq\to-\bq$ (see Table~\ref{tab:symmetries}), we obtain
\begin{equation}
    \begin{split}
        \chi^{22}(\bzero,\omega)&=\frac{1}{2}\left[\chit^{11}(\bQ,\omega)+\chit^{22}(\bQ,\omega)+2i\chit^{12}(\bQ,\omega)\right]\\
        &=2\chit^{-+}(\bQ,\omega),
    \end{split}
\end{equation}
with $\chit^{-+}(q)=\langle S^-(-q)S^+(q)\rangle$, and $S^\pm(q)=(S^1(q)\pm iS^2(q))/2$.
It is convenient to express $\chit^{-+}(\bQ,\omega)$ as 
\begin{equation}
    \begin{split}
        \chit^{-+}(\bQ,\omega)&=\chit^{-+}_0(\bQ,\omega)\\
        &\phantom{=}+\sum_{a,b\in\{0,1,2,3\}}\chit^{-a}_0(\bQ,\omega)\Gammat^{ab}(\bQ,\omega)\chit^{b+}_0(\bQ,\omega),
    \end{split}
    \label{eq: chit pm RPA}
\end{equation}
where we have defined
\begin{subequations}
    \begin{align}
        &\chit_0^{-a}(q)=\frac{1}{2}\left[\chit_0^{1a}(q)-i\chit_0^{2a}(q)\right],\\
        &\chit_0^{a+}(q)=\frac{1}{2}\left[\chit_0^{a1}(q)+i\chit_0^{a2}(q)\right].
    \end{align}
\end{subequations}
In the limit $\omega\to 0$, $\chit_0^{-3}(\bQ,\omega)$ and $\chit_0^{3+}(\bQ,\omega)$ vanish as they are odd in frequency (see Table~\ref{tab:symmetries}). We can now cast the dynamical susceptibility in the form
\begin{equation}
    \begin{split}
        \chi_\mathrm{dyn}^\perp&=2\chit^{-+}_0(Q)\\
        &\phantom{=}+2\sum_{a,b\in\{0,1,2\}}\chit^{-a}_0(Q)\Gammat^{ab}(Q)\chit^{b+}_0(Q),
    \end{split}
    \label{eq: chi22 gauge}
\end{equation}
or, equivalently,
\begin{equation}
    \begin{split}
        \chi_\mathrm{dyn}^\perp&=2\chit^{+-}_0(-Q)\\
        &\phantom{=}+2\sum_{a,b\in\{0,1,2\}}\chit^{+a}_0(-Q)\Gammat^{ab}(-Q)\chit^{b-}_0(-Q).
    \end{split}
\end{equation}
We remark that in the formulas above we have not specified in which order the limits $\bq\to\pm\bQ$ and $\omega\to0$ have to be taken as they commute. 
\subsection{Equivalence of RPA and gauge theory approaches}
In this Section, we finally prove that the expressions for the spin stiffnesses and dynamical susceptibilities obtained in Sec.~\ref{sec: small q and omega RPA} coincide with those of Sec.~\ref{sec: gauge kernels} via a direct evaluation.
\subsubsection{In-plane mode}
We start by computing the first term in Eq.~\eqref{eq: J3 RPA}. The second derivative of the 22-component of the bare susceptibility can be expressed as
\begin{equation}
    \begin{split}
        &-2\Delta^2\partial^2_{q_\alpha q_\beta}\chit_0^{22}(0)\\=
        &-\Delta^2\int_\bk \gamma_{\bk}^\alpha\gamma_{\bk+\bQ}^\beta\bigg[\frac{f(E^-_\bk)-f(E^+_\bk)}{4e_\bk^3}\\
        &\hskip3cm+\frac{f'(E^+_\bk)+f'(E^-_\bk)}{4e_\bk^2}\bigg],
    \end{split}
    \label{eq: J03 RPA}
\end{equation}
where $f'(x)=df/dx$ is the derivative of the Fermi function. On the other hand, the bare contribution to $J^\smsqr_{\alpha\beta}$ (Eq.~\eqref{eq: J inplane bare}) reads
\begin{equation}
    \begin{split}
        J^{0,\smsqr}_{\alpha\beta}=\frac{1}{4}\int_\bk T\sum_{\nu_n}\Big[
        &G_\bk(i\nu_n)^2\gamma_{\bk}^\alpha\gamma_{\bk}^\beta+\overline{G}_\bk(i\nu_n)^2\gamma_{\bk+\bQ}^\alpha\gamma_{\bk+\bQ}^\beta\\
        &-2F_\bk(i\nu_n)^2\gamma_{\bk}^\alpha\gamma_{\bk+\bQ}^\beta
        \Big]
        \\
        +\frac{1}{4}\int_\bk T\sum_{\nu_n} \Big[
        &G_\bk(i\nu_n)\gamma^{\alpha\beta}_\bk
        +\overline{G}_\bk(i\nu_n)\gamma^{\alpha\beta}_{\bk+\bQ}
        \Big].
    \end{split}
\end{equation}
%
The second (diamagnetic) term can be integrated by parts, giving
\begin{equation}
    \begin{split}
        -\frac{1}{4}\int_\bk T\sum_{\nu_n} \Big[
        &G^2_\bk(i\nu_n)\gamma_{\bk}^\alpha\gamma_{\bk}^\beta+\overline{G}^2_\bk(i\nu_n)\gamma_{\bk+\bQ}^\alpha\gamma_{\bk+\bQ}^\beta\\
        &+2F^2_\bk(i\nu_n)\gamma_{\bk}^\alpha\gamma_{\bk+\bQ}^\beta
        \Big],
    \end{split}
\end{equation}
where we have used the properties
\begin{subequations}
    \begin{align}
        &\partial_{k_\alpha}G_\bk(i\nu_n)=G^2_\bk(i\nu_n)\gamma^\alpha_\bk+F^2_\bk(i\nu_n)\gamma^\alpha_{\bk+\bQ},\\      
        &\partial_{k_\alpha}\overline{G}_\bk(i\nu_n)=\overline{G}^2_\bk(i\nu_n)\gamma^\alpha_{\bk+\bQ}+F^2_\bk(i\nu_n)\gamma^\alpha_{\bk}.
    \end{align}
    \label{eq: derivatives of G}
\end{subequations}
Summing up both terms, we obtain
\begin{equation}
    \begin{split}
        J_{0,\smsqr}^{\alpha\beta}=
    -\int_\bk T\sum_{\nu_n}\gamma_{\bk}^\alpha\gamma_{\bk+\bQ}^\beta &F^2_\bk(i\nu_n).
    \end{split}
    \label{eq: J03 explicit}
\end{equation}
Performing the Matsubara sum, we arrive at
\begin{equation}\label{eq: J0inplane}
    \begin{split}
        J^{0,\smsqr}_{\alpha\beta}=
        -\Delta^2\int_\bk \gamma_{\bk}^\alpha\gamma_{\bk+\bQ}^\beta\bigg[&\frac{f(E^-_\bk)-f(E^+_\bk)}{4e_\bk^3}\\
        +&\frac{f'(E^+_\bk)+f'(E^-_\bk)}{4e_\bk^2}\bigg],
    \end{split}
\end{equation}
which is the same result as in \eqref{eq: J03 RPA}. Furthermore, one can show that
\begin{subequations}
    \begin{align}
        &2i\Delta\partial_{q_\alpha}\chit_0^{20}(0)=-2i\Delta\partial_{q_\alpha}\chit_0^{02}(0)=\kappa^{30}_{\alpha}(\bzero),\\
        &2i\Delta\partial_{q_\alpha}\chit_0^{21}(0)=-2i\Delta\partial_{q_\alpha}\chit_0^{12}(0)=\kappa^{31}_{\alpha}(\bzero).
    \end{align}
    \label{eq: RPA vs gauge inplane J, offdiagonal}
\end{subequations}
Inserting results~\eqref{eq: J03 RPA}, \eqref{eq: J03 explicit}, and \eqref{eq: RPA vs gauge inplane J, offdiagonal} into \eqref{eq: J3 RPA} and \eqref{eq: J3 gauge}, we prove that these two expressions give the same result for the in-plane stiffness. Explicit expressions for $\kappa^{30}_{\alpha}(\bzero)$ and $\kappa^{31}_{\alpha}(\bzero)$ are given in Appendix~\ref{app: ka30 and ka31}.

If we now consider the dynamical susceptibility, it is straightforward to see that
\begin{equation}
    \begin{split}
        2\Delta^2\partial^2_\omega\chit_0^{22}(0)=&2i\Delta\partial_\omega\chit^{23}_0(0)=\lim_{\omega\to 0}\chit_0^{33}(\bzero,\omega)\\
        =&\Delta^2\int_\bk \frac{f(E^-_\bk)-f(E^+_\bk)}{4e^3_\bk},
    \end{split}
    \label{eq: RPA=gauge for chi3}
\end{equation}
which, if inserted into Eqs.~\eqref{eq: chi3 RPA} and \eqref{eq: chi3 gauge}, proves that the calculations of $\chi_\mathrm{dyn}^\smsqr$ via gauge kernels and via the low-energy expansion of the susceptibilities provide the same result.
\begin{widetext}
\subsubsection{Out-of-plane modes}
With the help of some lengthy algebra, one can compute the second momentum derivative of the bare susceptibility $\chit_0^{33}(q)$, obtaining
\begin{equation}
    \begin{split}
        -\Delta^2\partial^2_{q_\alpha q_\beta}\chit_0^{33}(Q)
        =&\frac{1}{8}\int_\bk\sum_{\ell,\ell'=\pm}\left(1-\ell\frac{h_\bk}{e_\bk}\right)\left(1+\ell\frac{h_{\bk+\bQ}}{e_{\bk+\bQ}}\right)\gamma^\alpha_{\bk+\bQ}\gamma^\beta_{\bk+\bQ}
        \frac{f(E^\ell_\bk)-f(E^{\ell'}_{\bk+\bQ})}{E^\ell_\bk-E^{\ell'}_{\bk+\bQ}}\\
        &-\frac{1}{8}\int_\bk \sum_{\ell=\pm}\left[\left(1-\ell\frac{h_\bk}{e_\bk}\right)^2\gamma^{\alpha}_{\bk+\bQ}+\frac{\Delta^2}{e_\bk^2}\gamma^{\alpha}_{\bk}\right]\gamma^\beta_{\bk+\bQ} f'(E^\ell_\bk)\\
        &-\frac{1}{8}\int_\bk \sum_{\ell=\pm}\left[\frac{\Delta^2}{e_\bk^2}(\gamma^{\alpha}_{\bk+\bQ}-\gamma^{\alpha}_{\bk})\right]\gamma^\beta_{\bk+\bQ} \frac{f(E^\ell_\bk)-f(E^{-\ell}_{\bk})}{E^\ell_\bk-E^{-\ell}_{\bk}}.
    \end{split}
\end{equation}
Similarly to what we have done for the in-plane mode, we integrate by parts the diamagnetic contribution to the gauge kernel. Its sum with the paramagnetic one gives
\begin{equation}
    \begin{split}
        -\lim_{\bq\to\bzero}K^{22}_{\alpha\beta}(\bq,\bq,0)=
        \frac{1}{2}\int_\bk T\sum_{\nu_n}\Big\{
        &\gamma^\alpha_{\bk+\bQ}\gamma^\beta_{\bk+\bQ}\overline{G}_\bk(i\nu_n)\left[G_{\bk+\bQ}(i\nu_n)-\overline{G}_\bk(i\nu_n)\right]\\
        -&\gamma^\alpha_{\bk}\gamma^\beta_{\bk+\bQ}F^2_\bk(i\nu_n)
        \Big\}.
    \end{split}
\end{equation}
Performing the Matsubara sums, one can prove the equivalence of the RPA and gauge theory approach for the calculation of $J_{\alpha\beta}^{\perp}$. 

Similarly, we obtain for the second frequency derivative of the bubble $\chit_0^{33}(q)$
\begin{equation}
    \begin{split}
        \Delta^2\partial^2_\omega\chit_0^{33}(Q)=-\frac{1}{8}\int_\bk\sum_{\ell,\ell'=\pm}
        \left(1-\ell\frac{h_\bk}{e_\bk}\right)\left(1+\ell\frac{h_{\bk+\bQ}}{e_{\bk+\bQ}}\right)
        \frac{f(E^\ell_\bk)-f(E^{\ell'}_{\bk+\bQ})}{E^\ell_\bk-E^{\ell'}_{\bk+\bQ}}=2\chit^{-+}_0(Q).
    \end{split}
    \label{eq: proof Z2, diagonal component}
\end{equation}
\end{widetext}
Furthermore, one can prove that
\begin{equation}
    \begin{split}
        \Delta\partial_\omega \chit_0^{3a}(Q)=&\Delta[\partial_\omega \chit_0^{a3}(Q)]^*\\
        =&\chit_0^{-a}(Q)=[\chit_0^{a+}(Q)]^*,
    \end{split}
    \label{eq: proof Z2, off diagonal components}
\end{equation}
for $a=0,1,2$. Inserting results \eqref{eq: proof Z2, diagonal component} and \eqref{eq: proof Z2, off diagonal components} into Eqs.~\eqref{eq: chi22 RPA} and \eqref{eq: chi22 gauge}, one sees that the RPA and gauge theory approaches are equivalent for the calculation of $\chi_\mathrm{dyn}^\perp$. In Appendix~\ref{app: chit0-a(Q)} we provide explicit expressions for the off diagonal bare susceptibilities $\chit^{-a}_0(Q)$.
\subsubsection{Remarks on more general models}
We remark that in the more general case of an interaction of the type
\begin{equation}
    \mathcal{S}_\mathrm{int}=\int_{k,k',q}\!U_{k,k'}(q)[\psibar_{k+q}\vec{\sigma}\psi_k]\cdot[\psibar_{k'-q}\vec{\sigma}\psi_{k'}],
\end{equation}
producing, in general, a $k$-dependent gap, the identities we have proven above do not hold anymore within the RPA, as additional terms in the derivative of the inverse susceptibilities emerge, containing expressions involving first and second derivatives of the gap with respect to the spatial momentum and/or frequency. In fact, in the case of nonlocal interactions, gauge invariance requires additional couplings to the gauge field in $\mathcal{S}_\mathrm{int}$, complicating our expressions for the gauge kernels. Similarly, even for action~\eqref{eq: Hubbard action}, approximations beyond the RPA produce in general a $k$-dependent $\Delta$, and vertex corrections in the kernels are required to obtain the same result as the one obtained expanding the susceptibilities. 
\subsection{N\'eel limit}
In this Section, we analyze the N\'eel limit, that is, $\bQ=(\pi/a_0,\dots,\pi/a_0)$. In this case, it is easy to see that, within the RPA, the bare susceptibilities in the rotated basis obey the identities
\begin{subequations}
    \begin{align}
        &\chit_0^{22}(\bq,\omega)=\chit_0^{33}(\bq+\bQ,\omega),\\
        &\chit_0^{20}(\bq,\omega)=\chit_0^{21}(\bq,\omega)=0,\\
        &\chit_0^{30}(\bq,\omega)=\chit_0^{31}(\bq,\omega)=0.
    \end{align}
\end{subequations}
Furthermore, we obtain for the mixed gauge kernels (see Appendix~\ref{app: ka30 and ka31})
\begin{equation}
    K^{ab}_{\mathrm{para},\alpha 0}(\bq,\bq',\omega)=K^{ab}_{\mathrm{para},0\alpha}(\bq,\bq',\omega)=0.
\end{equation}
We also notice that $K^{11}_{\alpha\beta}(\bq,\bq',0)$ and $K^{22}_{\alpha\beta}(\bq,\bq',0)$ have (different) momentum off-diagonal contributions for which $\bq'=\bq\pm 2\bQ$. If $\bQ=(\pi/a_0,\dots,\pi/a_0)$, these terms become diagonal in momentum, as $2\bQ= \bzero$, such that
\begin{subequations}
    \begin{align}
        &\lim_{\bq\to\bzero}K^{11}_{\alpha\beta}(\bq,0)=0,\\
        &K^{22}_{\alpha\beta}(\bq,0)=K^{33}_{\alpha\beta}(\bq,0).
    \end{align}
\end{subequations}

From the above relations, we can see that $J_{\alpha\beta}^\perp=J^\smsqr_{\alpha\beta}\equiv J_{\alpha\beta}$, and $\chi_\mathrm{dyn}^\perp=\chi_\mathrm{dyn}^\smsqr\equiv \chi_\mathrm{dyn}^\perp$, as expected for the N\'eel state.

From these considerations, we obtain for the spin stiffness
\begin{equation}
    \begin{split}
        J_{\alpha\beta}=&-\lim_{\bq\to\bzero}K^{22}_{\alpha\beta}(\bq,0)=-\lim_{\bq\to\bzero}K^{33}_{\alpha\beta}(\bq,0)\\
        &=-2\Delta^2\partial^2_{q_\alpha q_\beta}\chit_0^{22}(0)
        =-2\Delta^2\partial^2_{q_\alpha q_\beta}\chit_0^{33}(Q),
    \end{split}
\end{equation}
which implies that $J_{\alpha\beta}$ is given by Eq.~\eqref{eq: J0inplane}. If the underlying lattice is $C_4$-symmetric, the spin stiffness is isotropic in the N\'eel state, that is, $J_{\alpha\beta}=J\delta_{\alpha\beta}$. 
Similarly, for the dynamical susceptibility, we have
\begin{equation}
    \begin{split}
        \chi_\mathrm{dyn}^\perp=&\lim_{\omega\to 0}\chi^{22}(\bzero,\omega)=\lim_{\omega\to 0}\chi^{33}(\bzero,\omega)\\
        =&2\Delta^2\partial^2_\omega \chit_0^{22}(0)=2\Delta^2\partial^2_\omega \chit_0^{33}(Q),
    \end{split}
\end{equation}
which, combined with \eqref{eq: RPA=gauge for chi3}, implies
\begin{equation}
    \chi_\mathrm{dyn}^\perp=\lim_{\omega\to 0}\frac{\chit_0^{33}(\bzero,\omega)}{1-2U\chit_0^{33}(\bzero,\omega)},
\end{equation}
with $\chit_0^{33}(\bzero,\omega\to 0)$ given by Eq.~\eqref{eq: RPA=gauge for chi3}.

We notice that the dynamical susceptibility is obtained from the susceptibility by letting $\bq\to\bzero$ \emph{before} $\omega\to 0$. This order of the limits removes the intraband terms (that is, the $\ell=\ell'$ terms in Eq.~\eqref{eq: chi0 def}), which instead would yield a finite contribution to the \emph{uniform transverse susceptibility} $\chi^\perp\equiv\lim_{\bq\to\bzero}\chi^{22}(\bq,0)$. In the special case of an insulator at low temperature $T\ll\Delta$, the intraband contributions vanish and one has the identity $\chi_\mathrm{dyn}^\perp=\chi^\perp$, leading to the hydrodynamic relation for the spin wave velocity~\cite{Halperin1969} $c_s=\sqrt{J/\chi^\perp}$ (in an isotropic antiferromagnet). As noticed in Ref.~\cite{Sachdev1995}, in a doped antiferromagnet this hydrodynamic expression does not hold anymore, and one has to replace the uniform transverse susceptibility with the dynamical susceptibility. Since $J=0$ and $\chi_\mathrm{dyn}^\perp=0$ in the symmetric phase due to SU(2) gauge invariance, the expression $c_s=\sqrt{J/\chi_\mathrm{dyn}^\perp}$ yields a finite value $c_s$ at the critical point $\Delta\to 0$, provided that $J$ and $\chi_\mathrm{dyn}^\perp$ scale to zero with the same power of $\Delta$, as it happens within mean-field theory. Note that in the symmetric phase SU(2) gauge invariance does not pose any constraint on $\chi^\perp$, which is generally finite. 

In the simpler case of perfect nesting, that is, when $\xi_\bk=-\xi_{\bk+\bQ}$, corresponding to the half-filled particle-hole symmetric Hubbard model, and at zero temperature, expressions for $J$ and $\chi^\perp$ have been derived in Refs.~\cite{Schulz1995,Borejsza2004} for two spatial dimensions, and it is straightforward to check that our results reduce to these in this limit. Moreover,  Eqs.~31-34 in Ref.~\cite{Schulz1995} are similar to our Ward identities but no derivation is provided. 
\\
\section{Conclusion}
\label{sec: conclusion}
In conclusion, we have derived Ward identities for fermionic systems in which a gauge symmetry is globally broken. In particular, we have shown that the zero-energy and long-wavelength components of the gauge kernels are connected to the transverse susceptibilities of the order parameter by exact relations. We have analyzed several examples, namely a superconductor, a N\'eel antiferromagnet, and a spiral magnet. In the latter case, we have performed an explicit calculation of the transverse susceptibilities and of the gauge kernels within the random phase approximation and verified that the Ward identities are indeed fulfilled. Furthermore, we have considered the N\'eel limit $\bQ\to(\pi/a_0,\dots,\pi/a_0)$ and found that our RPA expressions for the spin stiffnesses and susceptibilities reduce to those previously obtained in Refs.~\cite{Schulz1995,Borejsza2004}. We have also shown that the hydrodynamic expression for the magnon velocity $c_s=\sqrt{J/\chi^\perp}$ does not hold in presence of gapless fermionic excitations, and it must be replaced by $c_s=\sqrt{J/\chi_\mathrm{dyn}^\perp}$. While $\chi^\perp$ is computed by taking the $\omega\to 0$ limit \emph{before} letting $\bq\to\bzero$ in the transverse susceptibility, $\chi_\mathrm{dyn}^\perp$ is obtained reversing the order of the limits. The equality $\chi_\mathrm{dyn}^\perp=\chi^\perp$ holds only for insulating antiferromagnets at low temperatures $T\ll\Delta$, as it is the case for the spin systems considered in Refs.~\cite{Halperin1969,Halperin1977}. 

Our findings find immediate application in the study of large distance properties of interacting fermion systems.
In fact, via a microscopic calculation, one can evaluate the coefficients of the effective actions listed in Sec.~\ref{sec: Ward Identities}, and study phenomena which are purely fluctuations-driven. Examples are the Berezinskii-Kosterlitz-Thouless transition in two-dimensional superconductors (see, for example, the microscopic calculation of Ref.~\cite{Metzner2019}) or strong magnetic fluctuations leading to the formation of a \emph{pseudogap} in low-dimensional electron systems~\cite{Borejsza2004,Scheurer2018}. Special care, however, must be taken when the fermions form Fermi surfaces, as the low-energy modes of the order parameter can decay in particle-hole pairs, producing a \emph{Landau damping}~\cite{Bonetti2022} that can affect the macroscopic and critical properties of the system~\cite{Millis1993,Sachdev1995}. An interesting extension of the present work would be to derive Ward identities for the damping of the Goldstone modes.
%
\section*{Acknowledgments}
I am grateful to L.~Debbeler, E.~K\"onig, J.~Mitscherling, M.~Puviani, J.~S\'ykora, and D.~Vilardi for stimulating discussions.
A particular thank goes to W.~Metzner for encouraging me to write this paper, enlightening discussions, and a careful and critical reading of the manuscript. 
\begin{widetext}
\appendix
\section{Expressions for \texorpdfstring{$\kappa_\alpha^{30}(\bzero)$}{ka30(0)} and \texorpdfstring{$\kappa_\alpha^{31}(\bzero)$}{ka31(0)}}
\label{app: ka30 and ka31}
In this appendix, we report explicit expressions for the off-diagonal paramagnetic contributions to the spin stiffness, namely $\kappa_\alpha^{30}(\bzero)$, and $\kappa_\alpha^{31}(\bzero)$. 

For $\kappa_\alpha^{30}(\bzero)$, we have, after making the trace in \eqref{eq: paramagnetic contr Kernel} explicit,
\begin{equation}
    \begin{split}
        \kappa_\alpha^{30}(\bzero)=\lim_{\bq\to\bzero}K_{\mathrm{para},\alpha 0}^{31}(\bq,\bq',0)=&-\frac{1}{4}\int_\bk T\sum_{\nu_n} \left\{\left[G^2_\bk(i\nu_n)+F^2_\bk(i\nu_n)\right]\gamma^\alpha_\bk-\left[\overline{G}^2_\bk(i\nu_n)+F^2_\bk(i\nu_n)\right]\gamma^\alpha_{\bk+\bQ}\right\}\delta_{\bq',\bzero}\\
        =&-\frac{1}{4}\int_\bk T\sum_{\nu_n}\left\{\partial_\bk\left[G_\bk-\overline{G}_\bk\right]+4F_\bk^2\,\partial_{k_\alpha}h_\bk\right\}\delta_{\bq',\bzero},
    \end{split}
\end{equation}
where we have made use of properties \eqref{eq: derivatives of G} in the last line. The first term vanishes when integrated by parts, while the Matsubara summation for the second yields
\begin{equation}
    \begin{split}
        \kappa_\alpha^{30}(\bzero)=
        -\frac{\Delta^2}{4}\int_\bk \left[\frac{f(E^-_\bk)-f(E^+_\bk)}{e^3_\bk}+\frac{f^\prime(E^+_\bk)+f^\prime(E^-_\bk)}{e_\bk^2}\right](\partial_{k_{\alpha}}h_{\bk}).
    \end{split}
\end{equation}
For $\kappa_\alpha^{31}(\bzero)$ we have
\begin{equation}
    \begin{split}
        \lim_{\bq\to\bzero}K_{\mathrm{para},\alpha 0}^{31}(\bq,\bq',0)=
        &-\frac{1}{4}\int_\bk T\sum_{\nu_n} \left[G_\bk(i\nu_n)F_\bk(i\nu_n)\gamma^\alpha_\bk-\overline{G}_\bk(i\nu_n)F_\bk(i\nu_n)\gamma^\alpha_{\bk+\bQ}\right]\left(\delta_{\bq',\bQ}+\delta_{\bq',-\bQ}\right).
    \end{split}
\end{equation}
Defining $\kappa_\alpha^{31}(\bzero)=2K_{\mathrm{para},\alpha 0}^{31}(\bzero,\bQ,0)$ (see Eq.~\eqref{eq: k31 def}), and performing the Matsubara sum, we obtain
\begin{equation}
    \begin{split}
    \kappa_\alpha^{31}(\bzero)=-\frac{\Delta^2}{4}
        \int_\bk\bigg\{ \left[\frac{h_\bk}{e_\bk}(\partial_{k_\alpha}g_\bk)+(\partial_{k_\alpha}h_\bk)\right]\frac{f'(E^+_\bk)}{e_\bk}
        +\Big[&\frac{h_\bk}{e_\bk}(\partial_{k_\alpha}g_\bk)-(\partial_{k_\alpha}h_\bk)\Big]\frac{f'(E^-_\bk)}{e_\bk}\\
        &+\frac{h_\bk}{e_\bk^2}(\partial_{k_\alpha}g_\bk) \frac{f(E^-_\bk)-f(E^+_\bk)}{e_\bk}\bigg\}.
    \end{split}
\end{equation}
Furthermore, it is easy to see that $K_{\alpha 0}^{31}(\bzero,\pm\bQ,0)=\mp iK_{\alpha 0}^{32}(\bzero,\pm\bQ,0)$, which, together with Eq.~\eqref{eq: k32 def} proves $\kappa_\alpha^{31}(\bzero)=\kappa^\alpha_{32}(\bzero)$. We remark that in the N\'eel limit both $\kappa_\alpha^{30}(\bzero)$ and $\kappa_\alpha^{31}(\bzero)$ vanish as their integrands are odd under $\bk\to\bk+\bQ$.
\section{Expressions for \texorpdfstring{$\chit_0^{-a}(Q)$}{chit0-a(Q)}}
\label{app: chit0-a(Q)}
We report here the RPA expressions for the off-diagonal bare susceptibilities $\chit_0^{-a}(Q)$, with $a=0,1,2$. They can all be obtained by computing the trace and the Matsubara summation in Eq.~\eqref{app: chit0-a(Q)}. We obtain
\begin{subequations}
    \begin{align}
        &\chit_0^{-0}(Q)=-\frac{1}{16}\int_\bk \sum_{\ell,\ell'=\pm}\left[\ell\frac{\Delta}{e_\bk}+\ell'\frac{\Delta}{e_{\bk+\bQ}}+\ell\ell'\frac{\Delta(h_{\bk+\bQ}-h_\bk)}{e_\bk e_{\bk+\bQ}}\right]F_{\ell\ell'}(\bk,\bQ,0),\\
        &\chit_0^{-1}(Q)=-\frac{1}{16}\int_\bk \sum_{\ell,\ell'=\pm}\left[1+\ell\frac{h_\bk}{e_\bk}-\ell'\frac{h_{\bk+\bQ}}{e_{\bk+\bQ}}-\ell\ell'\frac{h_\bk h_{\bk+\bQ}-\Delta^2}{e_\bk e_{\bk+\bQ}}\right]F_{\ell\ell'}(\bk,\bQ,0),\\
        &\chit_0^{-2}(Q)=+\frac{i}{16}\int_\bk \sum_{\ell,\ell'=\pm}\left[1+\ell\frac{h_\bk}{e_\bk}-\ell'\frac{h_{\bk+\bQ}}{e_{\bk+\bQ}}-\ell\ell'\frac{h_\bk h_{\bk+\bQ}+\Delta^2}{e_\bk e_{\bk+\bQ}}\right]F_{\ell\ell'}(\bk,\bQ,0).
    \end{align}
\end{subequations}
with $F_{\ell\ell'}(\bk,\bq,\omega)$ defined as in Eq.~\eqref{eq: Fll def}.
\end{widetext}
\bibliography{main.bib}

\begin{thebibliography}{66}%
\makeatletter
\providecommand \@ifxundefined [1]{%
 \@ifx{#1\undefined}
}%
\providecommand \@ifnum [1]{%
 \ifnum #1\expandafter \@firstoftwo
 \else \expandafter \@secondoftwo
 \fi
}%
\providecommand \@ifx [1]{%
 \ifx #1\expandafter \@firstoftwo
 \else \expandafter \@secondoftwo
 \fi
}%
\providecommand \natexlab [1]{#1}%
\providecommand \enquote  [1]{``#1''}%
\providecommand \bibnamefont  [1]{#1}%
\providecommand \bibfnamefont [1]{#1}%
\providecommand \citenamefont [1]{#1}%
\providecommand \href@noop [0]{\@secondoftwo}%
\providecommand \href [0]{\begingroup \@sanitize@url \@href}%
\providecommand \@href[1]{\@@startlink{#1}\@@href}%
\providecommand \@@href[1]{\endgroup#1\@@endlink}%
\providecommand \@sanitize@url [0]{\catcode `\\12\catcode `\$12\catcode `\&12\catcode `\#12\catcode `\^12\catcode `\_12\catcode `\%12\relax}%
\providecommand \@@startlink[1]{}%
\providecommand \@@endlink[0]{}%
\providecommand \url  [0]{\begingroup\@sanitize@url \@url }%
\providecommand \@url [1]{\endgroup\@href {#1}{\urlprefix }}%
\providecommand \urlprefix  [0]{URL }%
\providecommand \Eprint [0]{\href }%
\providecommand \doibase [0]{https://doi.org/}%
\providecommand \selectlanguage [0]{\@gobble}%
\providecommand \bibinfo  [0]{\@secondoftwo}%
\providecommand \bibfield  [0]{\@secondoftwo}%
\providecommand \translation [1]{[#1]}%
\providecommand \BibitemOpen [0]{}%
\providecommand \bibitemStop [0]{}%
\providecommand \bibitemNoStop [0]{.\EOS\space}%
\providecommand \EOS [0]{\spacefactor3000\relax}%
\providecommand \BibitemShut  [1]{\csname bibitem#1\endcsname}%
\let\auto@bib@innerbib\@empty
\bibitem [{\citenamefont {Goldstone}(1961)}]{Goldstone1961}%
  \BibitemOpen
  \bibfield  {author} {\bibinfo {author} {\bibfnamefont {J.}~\bibnamefont {Goldstone}},\ }\bibfield  {title} {\bibinfo {title} {{Field theories with "Superconductor" solutions}},\ }\href {https://doi.org/https://doi.org/10.1007/BF02812722} {\bibfield  {journal} {\bibinfo  {journal} {Il Nuovo Cimento}\ }\textbf {\bibinfo {volume} {19}},\ \bibinfo {pages} {154} (\bibinfo {year} {1961})}\BibitemShut {NoStop}%
\bibitem [{\citenamefont {Ward}(1950)}]{Ward1950}%
  \BibitemOpen
  \bibfield  {author} {\bibinfo {author} {\bibfnamefont {J.~C.}\ \bibnamefont {Ward}},\ }\bibfield  {title} {\bibinfo {title} {An identity in quantum electrodynamics},\ }\href {https://doi.org/10.1103/PhysRev.78.182} {\bibfield  {journal} {\bibinfo  {journal} {Phys. Rev.}\ }\textbf {\bibinfo {volume} {78}},\ \bibinfo {pages} {182} (\bibinfo {year} {1950})}\BibitemShut {NoStop}%
\bibitem [{\citenamefont {Takahashi}(1957)}]{Takahashi1957}%
  \BibitemOpen
  \bibfield  {author} {\bibinfo {author} {\bibfnamefont {Y.}~\bibnamefont {Takahashi}},\ }\bibfield  {title} {\bibinfo {title} {{On the generalized Ward identity}},\ }\href {https://doi.org/https://doi.org/10.1007/BF02832514} {\bibfield  {journal} {\bibinfo  {journal} {Il Nuovo Cimento}\ }\textbf {\bibinfo {volume} {6}},\ \bibinfo {pages} {371} (\bibinfo {year} {1957})}\BibitemShut {NoStop}%
\bibitem [{\citenamefont {Baym}\ and\ \citenamefont {Kadanoff}(1961)}]{Baym1961}%
  \BibitemOpen
  \bibfield  {author} {\bibinfo {author} {\bibfnamefont {G.}~\bibnamefont {Baym}}\ and\ \bibinfo {author} {\bibfnamefont {L.~P.}\ \bibnamefont {Kadanoff}},\ }\bibfield  {title} {\bibinfo {title} {{Conservation Laws and Correlation Functions}},\ }\href {https://doi.org/10.1103/PhysRev.124.287} {\bibfield  {journal} {\bibinfo  {journal} {Phys. Rev.}\ }\textbf {\bibinfo {volume} {124}},\ \bibinfo {pages} {287} (\bibinfo {year} {1961})}\BibitemShut {NoStop}%
\bibitem [{\citenamefont {Abrikosov}\ \emph {et~al.}(1965)\citenamefont {Abrikosov}, \citenamefont {Gorkov},\ and\ \citenamefont {Dzialoshinski}}]{Abrikosov1965}%
  \BibitemOpen
  \bibfield  {author} {\bibinfo {author} {\bibfnamefont {A.~A.}\ \bibnamefont {Abrikosov}}, \bibinfo {author} {\bibfnamefont {L.~P.}\ \bibnamefont {Gorkov}},\ and\ \bibinfo {author} {\bibfnamefont {I.~E.}\ \bibnamefont {Dzialoshinski}},\ }\href@noop {} {\emph {\bibinfo {title} {Methods of Quantum Field Theory in Statistical Physics}}}\ (\bibinfo  {publisher} {Pergamon},\ \bibinfo {address} {Elmsford, New York},\ \bibinfo {year} {1965})\BibitemShut {NoStop}%
\bibitem [{\citenamefont {Toyoda}(1987)}]{Toyoda1987}%
  \BibitemOpen
  \bibfield  {author} {\bibinfo {author} {\bibfnamefont {T.}~\bibnamefont {Toyoda}},\ }\bibfield  {title} {\bibinfo {title} {{Nonperturbative canonical formulation and Ward-Takahashi relations for quantum many-body systems at finite temperatures}},\ }\href {https://doi.org/https://doi.org/10.1016/0003-4916(87)90100-X} {\bibfield  {journal} {\bibinfo  {journal} {Ann.~Phys.}\ }\textbf {\bibinfo {volume} {173}},\ \bibinfo {pages} {226} (\bibinfo {year} {1987})}\BibitemShut {NoStop}%
\bibitem [{\citenamefont {Revzen}\ \emph {et~al.}(1989)\citenamefont {Revzen}, \citenamefont {Toyoda}, \citenamefont {Takahashi},\ and\ \citenamefont {Khanna}}]{Revzen1989}%
  \BibitemOpen
  \bibfield  {author} {\bibinfo {author} {\bibfnamefont {M.}~\bibnamefont {Revzen}}, \bibinfo {author} {\bibfnamefont {T.}~\bibnamefont {Toyoda}}, \bibinfo {author} {\bibfnamefont {Y.}~\bibnamefont {Takahashi}},\ and\ \bibinfo {author} {\bibfnamefont {F.~C.}\ \bibnamefont {Khanna}},\ }\bibfield  {title} {\bibinfo {title} {Baym-kadanoff criteria and the ward-takahashi relations in many-body theory},\ }\href {https://doi.org/10.1103/PhysRevB.40.769} {\bibfield  {journal} {\bibinfo  {journal} {Phys. Rev. B}\ }\textbf {\bibinfo {volume} {40}},\ \bibinfo {pages} {769} (\bibinfo {year} {1989})}\BibitemShut {NoStop}%
\bibitem [{\citenamefont {Bychkov}\ \emph {et~al.}(1966)\citenamefont {Bychkov}, \citenamefont {Gor'kov},\ and\ \citenamefont {Dzyaloshinskii}}]{Bychkov1966}%
  \BibitemOpen
  \bibfield  {author} {\bibinfo {author} {\bibfnamefont {Y.~A.}\ \bibnamefont {Bychkov}}, \bibinfo {author} {\bibfnamefont {L.~P.}\ \bibnamefont {Gor'kov}},\ and\ \bibinfo {author} {\bibfnamefont {I.}~\bibnamefont {Dzyaloshinskii}},\ }\bibfield  {title} {\bibinfo {title} {Possibility of superconductivity type phenomena in a one-dimensional system},\ }\href {http://jetp.ras.ru/cgi-bin/e/index/e/23/3/p489?a=list} {\bibfield  {journal} {\bibinfo  {journal} {Sov.~Phys.-JETP}\ }\textbf {\bibinfo {volume} {23}},\ \bibinfo {pages} {489} (\bibinfo {year} {1966})},\ \bibinfo {note} {[Zh.~Eksp.~Theor.~Fiz.~\textbf{50}, 738 (1966)]}\BibitemShut {NoStop}%
\bibitem [{\citenamefont {Hertz}\ and\ \citenamefont {Edwards}(1973)}]{Hertz1973}%
  \BibitemOpen
  \bibfield  {author} {\bibinfo {author} {\bibfnamefont {J.~A.}\ \bibnamefont {Hertz}}\ and\ \bibinfo {author} {\bibfnamefont {D.~M.}\ \bibnamefont {Edwards}},\ }\bibfield  {title} {\bibinfo {title} {{Electron-magnon interactions in itinerant ferromagnetism. I. Formal theory}},\ }\href {https://doi.org/10.1088/0305-4608/3/12/018} {\bibfield  {journal} {\bibinfo  {journal} {J.~Phys.~F3}\ }\textbf {\bibinfo {volume} {3}},\ \bibinfo {pages} {2174} (\bibinfo {year} {1973})}\BibitemShut {NoStop}%
\bibitem [{\citenamefont {Bickers}\ \emph {et~al.}(1989)\citenamefont {Bickers}, \citenamefont {Scalapino},\ and\ \citenamefont {White}}]{Bickers1989}%
  \BibitemOpen
  \bibfield  {author} {\bibinfo {author} {\bibfnamefont {N.~E.}\ \bibnamefont {Bickers}}, \bibinfo {author} {\bibfnamefont {D.~J.}\ \bibnamefont {Scalapino}},\ and\ \bibinfo {author} {\bibfnamefont {S.~R.}\ \bibnamefont {White}},\ }\bibfield  {title} {\bibinfo {title} {{Conserving Approximations for Strongly Correlated Electron Systems: Bethe-Salpeter Equation and Dynamics for the Two-Dimensional Hubbard Model}},\ }\href {https://doi.org/10.1103/PhysRevLett.62.961} {\bibfield  {journal} {\bibinfo  {journal} {Phys. Rev. Lett.}\ }\textbf {\bibinfo {volume} {62}},\ \bibinfo {pages} {961} (\bibinfo {year} {1989})}\BibitemShut {NoStop}%
\bibitem [{\citenamefont {Katanin}(2004)}]{Katanin2004}%
  \BibitemOpen
  \bibfield  {author} {\bibinfo {author} {\bibfnamefont {A.~A.}\ \bibnamefont {Katanin}},\ }\bibfield  {title} {\bibinfo {title} {{Fulfillment of Ward identities in the functional renormalization group approach}},\ }\href {https://doi.org/10.1103/PhysRevB.70.115109} {\bibfield  {journal} {\bibinfo  {journal} {Phys. Rev. B}\ }\textbf {\bibinfo {volume} {70}},\ \bibinfo {pages} {115109} (\bibinfo {year} {2004})}\BibitemShut {NoStop}%
\bibitem [{\citenamefont {Krien}\ \emph {et~al.}(2017)\citenamefont {Krien}, \citenamefont {van Loon}, \citenamefont {Hafermann}, \citenamefont {Otsuki}, \citenamefont {Katsnelson},\ and\ \citenamefont {Lichtenstein}}]{Krien2017}%
  \BibitemOpen
  \bibfield  {author} {\bibinfo {author} {\bibfnamefont {F.}~\bibnamefont {Krien}}, \bibinfo {author} {\bibfnamefont {E.~G. C.~P.}\ \bibnamefont {van Loon}}, \bibinfo {author} {\bibfnamefont {H.}~\bibnamefont {Hafermann}}, \bibinfo {author} {\bibfnamefont {J.}~\bibnamefont {Otsuki}}, \bibinfo {author} {\bibfnamefont {M.~I.}\ \bibnamefont {Katsnelson}},\ and\ \bibinfo {author} {\bibfnamefont {A.~I.}\ \bibnamefont {Lichtenstein}},\ }\bibfield  {title} {\bibinfo {title} {{Conservation in two-particle self-consistent extensions of dynamical mean-field theory}},\ }\href {https://doi.org/10.1103/PhysRevB.96.075155} {\bibfield  {journal} {\bibinfo  {journal} {Phys. Rev. B}\ }\textbf {\bibinfo {volume} {96}},\ \bibinfo {pages} {075155} (\bibinfo {year} {2017})}\BibitemShut {NoStop}%
\bibitem [{\citenamefont {Giuliani}\ and\ \citenamefont {Vignale}(2008)}]{GiulianiBook}%
  \BibitemOpen
  \bibfield  {author} {\bibinfo {author} {\bibfnamefont {G.}~\bibnamefont {Giuliani}}\ and\ \bibinfo {author} {\bibfnamefont {G.}~\bibnamefont {Vignale}},\ }\href@noop {} {\emph {\bibinfo {title} {{Quantum Theory of the Electron Liquid}}}}\ (\bibinfo  {publisher} {Cambridge University Press},\ \bibinfo {year} {2008})\BibitemShut {NoStop}%
\bibitem [{\citenamefont {Rostami}\ \emph {et~al.}(2017)\citenamefont {Rostami}, \citenamefont {Katsnelson},\ and\ \citenamefont {Polini}}]{Rostami2017}%
  \BibitemOpen
  \bibfield  {author} {\bibinfo {author} {\bibfnamefont {H.}~\bibnamefont {Rostami}}, \bibinfo {author} {\bibfnamefont {M.~I.}\ \bibnamefont {Katsnelson}},\ and\ \bibinfo {author} {\bibfnamefont {M.}~\bibnamefont {Polini}},\ }\bibfield  {title} {\bibinfo {title} {{Theory of plasmonic effects in nonlinear optics: The case of graphene}},\ }\href {https://doi.org/10.1103/PhysRevB.95.035416} {\bibfield  {journal} {\bibinfo  {journal} {Phys. Rev. B}\ }\textbf {\bibinfo {volume} {95}},\ \bibinfo {pages} {035416} (\bibinfo {year} {2017})}\BibitemShut {NoStop}%
\bibitem [{\citenamefont {Rostami}\ \emph {et~al.}(2021)\citenamefont {Rostami}, \citenamefont {Katsnelson}, \citenamefont {Vignale},\ and\ \citenamefont {Polini}}]{Rostami2021}%
  \BibitemOpen
  \bibfield  {author} {\bibinfo {author} {\bibfnamefont {H.}~\bibnamefont {Rostami}}, \bibinfo {author} {\bibfnamefont {M.~I.}\ \bibnamefont {Katsnelson}}, \bibinfo {author} {\bibfnamefont {G.}~\bibnamefont {Vignale}},\ and\ \bibinfo {author} {\bibfnamefont {M.}~\bibnamefont {Polini}},\ }\bibfield  {title} {\bibinfo {title} {{Gauge invariance and Ward identities in nonlinear response theory}},\ }\href {https://doi.org/https://doi.org/10.1016/j.aop.2021.168523} {\bibfield  {journal} {\bibinfo  {journal} {Ann.~Phys.}\ }\textbf {\bibinfo {volume} {431}},\ \bibinfo {pages} {168523} (\bibinfo {year} {2021})}\BibitemShut {NoStop}%
\bibitem [{\citenamefont {Edwards}\ and\ \citenamefont {Hertz}(1973)}]{Edwards1973}%
  \BibitemOpen
  \bibfield  {author} {\bibinfo {author} {\bibfnamefont {D.~M.}\ \bibnamefont {Edwards}}\ and\ \bibinfo {author} {\bibfnamefont {J.~A.}\ \bibnamefont {Hertz}},\ }\bibfield  {title} {\bibinfo {title} {{Electron-magnon interactions in itinerant ferromagnetism. {II}. Strong ferromagnetism}},\ }\href {https://doi.org/10.1088/0305-4608/3/12/019} {\bibfield  {journal} {\bibinfo  {journal} {J.~Phys.~F3}\ }\textbf {\bibinfo {volume} {3}},\ \bibinfo {pages} {2191} (\bibinfo {year} {1973})}\BibitemShut {NoStop}%
\bibitem [{\citenamefont {Schrieffer}(1999)}]{SchriefferBook}%
  \BibitemOpen
  \bibfield  {author} {\bibinfo {author} {\bibfnamefont {J.}~\bibnamefont {Schrieffer}},\ }\href@noop {} {\emph {\bibinfo {title} {{Theory Of Superconductivity}}}},\ Advanced Books Classics\ (\bibinfo  {publisher} {Avalon Publishing},\ \bibinfo {year} {1999})\BibitemShut {NoStop}%
\bibitem [{\citenamefont {Anderson}(1958{\natexlab{a}})}]{Anderson1958II}%
  \BibitemOpen
  \bibfield  {author} {\bibinfo {author} {\bibfnamefont {P.~W.}\ \bibnamefont {Anderson}},\ }\bibfield  {title} {\bibinfo {title} {{Coherent Excited States in the Theory of Superconductivity: Gauge Invariance and the Meissner Effect}},\ }\href {https://doi.org/10.1103/PhysRev.110.827} {\bibfield  {journal} {\bibinfo  {journal} {Phys. Rev.}\ }\textbf {\bibinfo {volume} {110}},\ \bibinfo {pages} {827} (\bibinfo {year} {1958}{\natexlab{a}})}\BibitemShut {NoStop}%
\bibitem [{\citenamefont {Anderson}(1958{\natexlab{b}})}]{Anderson1958}%
  \BibitemOpen
  \bibfield  {author} {\bibinfo {author} {\bibfnamefont {P.~W.}\ \bibnamefont {Anderson}},\ }\bibfield  {title} {\bibinfo {title} {{Random-Phase Approximation in the Theory of Superconductivity}},\ }\href {https://doi.org/10.1103/PhysRev.112.1900} {\bibfield  {journal} {\bibinfo  {journal} {Phys. Rev.}\ }\textbf {\bibinfo {volume} {112}},\ \bibinfo {pages} {1900} (\bibinfo {year} {1958}{\natexlab{b}})}\BibitemShut {NoStop}%
\bibitem [{\citenamefont {Nambu}(1960)}]{Nambu1960}%
  \BibitemOpen
  \bibfield  {author} {\bibinfo {author} {\bibfnamefont {Y.}~\bibnamefont {Nambu}},\ }\bibfield  {title} {\bibinfo {title} {{Quasi-Particles and Gauge Invariance in the Theory of Superconductivity}},\ }\href {https://doi.org/10.1103/PhysRev.117.648} {\bibfield  {journal} {\bibinfo  {journal} {Phys. Rev.}\ }\textbf {\bibinfo {volume} {117}},\ \bibinfo {pages} {648} (\bibinfo {year} {1960})}\BibitemShut {NoStop}%
\bibitem [{\citenamefont {Schulz}(1995)}]{Schulz1995}%
  \BibitemOpen
  \bibfield  {author} {\bibinfo {author} {\bibfnamefont {H.~J.}\ \bibnamefont {Schulz}},\ }\bibinfo {title} {{Functional Integrals for Correlated Electrons}},\ in\ \href {https://doi.org/10.1007/978-1-4899-1042-4_10} {\emph {\bibinfo {booktitle} {The Hubbard Model: Its Physics and Mathematical Physics}}},\ \bibinfo {editor} {edited by\ \bibinfo {editor} {\bibfnamefont {D.}~\bibnamefont {Baeriswyl}}, \bibinfo {editor} {\bibfnamefont {D.~K.}\ \bibnamefont {Campbell}}, \bibinfo {editor} {\bibfnamefont {J.~M.~P.}\ \bibnamefont {Carmelo}}, \bibinfo {editor} {\bibfnamefont {F.}~\bibnamefont {Guinea}},\ and\ \bibinfo {editor} {\bibfnamefont {E.}~\bibnamefont {Louis}}}\ (\bibinfo  {publisher} {Springer US},\ \bibinfo {address} {Boston, MA},\ \bibinfo {year} {1995})\ pp.\ \bibinfo {pages} {89--102}\BibitemShut {NoStop}%
\bibitem [{\citenamefont {Khalatinkov}(1965)}]{KhalatinkovBook}%
  \BibitemOpen
  \bibfield  {author} {\bibinfo {author} {\bibfnamefont {I.~M.}\ \bibnamefont {Khalatinkov}},\ }\href@noop {} {\emph {\bibinfo {title} {{An Introduction to the Theory of Superfluidity}}}}\ (\bibinfo  {publisher} {W.~A.~Benjamin, Inc.},\ \bibinfo {address} {New York},\ \bibinfo {year} {1965})\BibitemShut {NoStop}%
\bibitem [{\citenamefont {Halperin}\ and\ \citenamefont {Hohenberg}(1969)}]{Halperin1969}%
  \BibitemOpen
  \bibfield  {author} {\bibinfo {author} {\bibfnamefont {B.~I.}\ \bibnamefont {Halperin}}\ and\ \bibinfo {author} {\bibfnamefont {P.~C.}\ \bibnamefont {Hohenberg}},\ }\bibfield  {title} {\bibinfo {title} {{Hydrodynamic Theory of Spin Waves}},\ }\href {https://doi.org/10.1103/PhysRev.188.898} {\bibfield  {journal} {\bibinfo  {journal} {Phys. Rev.}\ }\textbf {\bibinfo {volume} {188}},\ \bibinfo {pages} {898} (\bibinfo {year} {1969})}\BibitemShut {NoStop}%
\bibitem [{\citenamefont {Halperin}\ and\ \citenamefont {Saslow}(1977)}]{Halperin1977}%
  \BibitemOpen
  \bibfield  {author} {\bibinfo {author} {\bibfnamefont {B.~I.}\ \bibnamefont {Halperin}}\ and\ \bibinfo {author} {\bibfnamefont {W.~M.}\ \bibnamefont {Saslow}},\ }\bibfield  {title} {\bibinfo {title} {{Hydrodynamic theory of spin waves in spin glasses and other systems with noncollinear spin orientations}},\ }\href {https://doi.org/10.1103/PhysRevB.16.2154} {\bibfield  {journal} {\bibinfo  {journal} {Phys. Rev. B}\ }\textbf {\bibinfo {volume} {16}},\ \bibinfo {pages} {2154} (\bibinfo {year} {1977})}\BibitemShut {NoStop}%
\bibitem [{\citenamefont {Haldane}(1983)}]{Haldane1983}%
  \BibitemOpen
  \bibfield  {author} {\bibinfo {author} {\bibfnamefont {F.}~\bibnamefont {Haldane}},\ }\bibfield  {title} {\bibinfo {title} {Continuum dynamics of the 1-d heisenberg antiferromagnet: Identification with the o(3) nonlinear sigma model},\ }\href {https://doi.org/https://doi.org/10.1016/0375-9601(83)90631-X} {\bibfield  {journal} {\bibinfo  {journal} {Physics Letters A}\ }\textbf {\bibinfo {volume} {93}},\ \bibinfo {pages} {464} (\bibinfo {year} {1983})}\BibitemShut {NoStop}%
\bibitem [{\citenamefont {Sachdev}\ \emph {et~al.}(1995)\citenamefont {Sachdev}, \citenamefont {Chubukov},\ and\ \citenamefont {Sokol}}]{Sachdev1995}%
  \BibitemOpen
  \bibfield  {author} {\bibinfo {author} {\bibfnamefont {S.}~\bibnamefont {Sachdev}}, \bibinfo {author} {\bibfnamefont {A.~V.}\ \bibnamefont {Chubukov}},\ and\ \bibinfo {author} {\bibfnamefont {A.}~\bibnamefont {Sokol}},\ }\bibfield  {title} {\bibinfo {title} {{Crossover and scaling in a nearly antiferromagnetic Fermi liquid in two dimensions}},\ }\href {https://doi.org/10.1103/PhysRevB.51.14874} {\bibfield  {journal} {\bibinfo  {journal} {Phys. Rev. B}\ }\textbf {\bibinfo {volume} {51}},\ \bibinfo {pages} {14874} (\bibinfo {year} {1995})}\BibitemShut {NoStop}%
\bibitem [{\citenamefont {Rastelli}\ \emph {et~al.}(1985)\citenamefont {Rastelli}, \citenamefont {Reatto},\ and\ \citenamefont {Tassi}}]{Rastelli1985}%
  \BibitemOpen
  \bibfield  {author} {\bibinfo {author} {\bibfnamefont {E.}~\bibnamefont {Rastelli}}, \bibinfo {author} {\bibfnamefont {L.}~\bibnamefont {Reatto}},\ and\ \bibinfo {author} {\bibfnamefont {A.}~\bibnamefont {Tassi}},\ }\bibfield  {title} {\bibinfo {title} {Quantum fluctuations in helimagnets},\ }\href {https://doi.org/10.1088/0022-3719/18/2/013} {\bibfield  {journal} {\bibinfo  {journal} {J. Phys. C}\ }\textbf {\bibinfo {volume} {18}},\ \bibinfo {pages} {353} (\bibinfo {year} {1985})}\BibitemShut {NoStop}%
\bibitem [{\citenamefont {Chandra}\ \emph {et~al.}(1990)\citenamefont {Chandra}, \citenamefont {Coleman},\ and\ \citenamefont {Larkin}}]{Chandra1990}%
  \BibitemOpen
  \bibfield  {author} {\bibinfo {author} {\bibfnamefont {P.}~\bibnamefont {Chandra}}, \bibinfo {author} {\bibfnamefont {P.}~\bibnamefont {Coleman}},\ and\ \bibinfo {author} {\bibfnamefont {A.~I.}\ \bibnamefont {Larkin}},\ }\bibfield  {title} {\bibinfo {title} {{A quantum fluids approach to frustrated Heisenberg models}},\ }\href {https://doi.org/10.1088/0953-8984/2/39/008} {\bibfield  {journal} {\bibinfo  {journal} {J. Phys. Condens. Matter}\ }\textbf {\bibinfo {volume} {2}},\ \bibinfo {pages} {7933} (\bibinfo {year} {1990})}\BibitemShut {NoStop}%
\bibitem [{\citenamefont {Shraiman}\ and\ \citenamefont {Siggia}(1992)}]{Shraiman1992}%
  \BibitemOpen
  \bibfield  {author} {\bibinfo {author} {\bibfnamefont {B.~I.}\ \bibnamefont {Shraiman}}\ and\ \bibinfo {author} {\bibfnamefont {E.~D.}\ \bibnamefont {Siggia}},\ }\bibfield  {title} {\bibinfo {title} {Excitation spectrum of the spiral state of a doped antiferromagnet},\ }\href {https://doi.org/10.1103/PhysRevB.46.8305} {\bibfield  {journal} {\bibinfo  {journal} {Phys. Rev. B}\ }\textbf {\bibinfo {volume} {46}},\ \bibinfo {pages} {8305} (\bibinfo {year} {1992})}\BibitemShut {NoStop}%
\bibitem [{\citenamefont {Kampf}(1996)}]{Kampf1996}%
  \BibitemOpen
  \bibfield  {author} {\bibinfo {author} {\bibfnamefont {A.~P.}\ \bibnamefont {Kampf}},\ }\bibfield  {title} {\bibinfo {title} {Collective excitations in itinerant spiral magnets},\ }\href {https://doi.org/10.1103/PhysRevB.53.747} {\bibfield  {journal} {\bibinfo  {journal} {Phys. Rev. B}\ }\textbf {\bibinfo {volume} {53}},\ \bibinfo {pages} {747} (\bibinfo {year} {1996})}\BibitemShut {NoStop}%
\bibitem [{\citenamefont {Shraiman}\ and\ \citenamefont {Siggia}(1989)}]{Shraiman1989}%
  \BibitemOpen
  \bibfield  {author} {\bibinfo {author} {\bibfnamefont {B.~I.}\ \bibnamefont {Shraiman}}\ and\ \bibinfo {author} {\bibfnamefont {E.~D.}\ \bibnamefont {Siggia}},\ }\bibfield  {title} {\bibinfo {title} {Spiral phase of a doped quantum antiferromagnet},\ }\href {https://doi.org/10.1103/PhysRevLett.62.1564} {\bibfield  {journal} {\bibinfo  {journal} {Phys. Rev. Lett.}\ }\textbf {\bibinfo {volume} {62}},\ \bibinfo {pages} {1564} (\bibinfo {year} {1989})}\BibitemShut {NoStop}%
\bibitem [{\citenamefont {Fr{\'{e}}sard}\ \emph {et~al.}(1991)\citenamefont {Fr{\'{e}}sard}, \citenamefont {Dzierzawa},\ and\ \citenamefont {Wölfle}}]{Fresard1991}%
  \BibitemOpen
  \bibfield  {author} {\bibinfo {author} {\bibfnamefont {R.}~\bibnamefont {Fr{\'{e}}sard}}, \bibinfo {author} {\bibfnamefont {M.}~\bibnamefont {Dzierzawa}},\ and\ \bibinfo {author} {\bibfnamefont {P.}~\bibnamefont {Wölfle}},\ }\bibfield  {title} {\bibinfo {title} {{Slave-Boson Approach to Spiral Magnetic Order in the Hubbard Model}},\ }\href {https://doi.org/10.1209/0295-5075/15/3/016} {\bibfield  {journal} {\bibinfo  {journal} {Europhys. Lett.}\ }\textbf {\bibinfo {volume} {15}},\ \bibinfo {pages} {325} (\bibinfo {year} {1991})}\BibitemShut {NoStop}%
\bibitem [{\citenamefont {Chubukov}\ and\ \citenamefont {Frenkel}(1992)}]{Chubukov1992}%
  \BibitemOpen
  \bibfield  {author} {\bibinfo {author} {\bibfnamefont {A.~V.}\ \bibnamefont {Chubukov}}\ and\ \bibinfo {author} {\bibfnamefont {D.~M.}\ \bibnamefont {Frenkel}},\ }\bibfield  {title} {\bibinfo {title} {{Renormalized perturbation theory of magnetic instabilities in the two-dimensional Hubbard model at small doping}},\ }\href {https://doi.org/10.1103/PhysRevB.46.11884} {\bibfield  {journal} {\bibinfo  {journal} {Phys. Rev. B}\ }\textbf {\bibinfo {volume} {46}},\ \bibinfo {pages} {11884} (\bibinfo {year} {1992})}\BibitemShut {NoStop}%
\bibitem [{\citenamefont {Chubukov}\ and\ \citenamefont {Musaelian}(1995)}]{Chubukov1995}%
  \BibitemOpen
  \bibfield  {author} {\bibinfo {author} {\bibfnamefont {A.~V.}\ \bibnamefont {Chubukov}}\ and\ \bibinfo {author} {\bibfnamefont {K.~A.}\ \bibnamefont {Musaelian}},\ }\bibfield  {title} {\bibinfo {title} {{Magnetic phases of the two-dimensional Hubbard model at low doping}},\ }\href {https://doi.org/10.1103/PhysRevB.51.12605} {\bibfield  {journal} {\bibinfo  {journal} {Phys. Rev. B}\ }\textbf {\bibinfo {volume} {51}},\ \bibinfo {pages} {12605} (\bibinfo {year} {1995})}\BibitemShut {NoStop}%
\bibitem [{\citenamefont {Kotov}\ and\ \citenamefont {Sushkov}(2004)}]{Kotov2004}%
  \BibitemOpen
  \bibfield  {author} {\bibinfo {author} {\bibfnamefont {V.~N.}\ \bibnamefont {Kotov}}\ and\ \bibinfo {author} {\bibfnamefont {O.~P.}\ \bibnamefont {Sushkov}},\ }\bibfield  {title} {\bibinfo {title} {{Stability of the spiral phase in the two-dimensional extended $t\text{\ensuremath{-}}J$ model}},\ }\href {https://doi.org/10.1103/PhysRevB.70.195105} {\bibfield  {journal} {\bibinfo  {journal} {Phys. Rev. B}\ }\textbf {\bibinfo {volume} {70}},\ \bibinfo {pages} {195105} (\bibinfo {year} {2004})}\BibitemShut {NoStop}%
\bibitem [{\citenamefont {Igoshev}\ \emph {et~al.}(2010)\citenamefont {Igoshev}, \citenamefont {Timirgazin}, \citenamefont {Katanin}, \citenamefont {Arzhnikov},\ and\ \citenamefont {Irkhin}}]{Igoshev2010}%
  \BibitemOpen
  \bibfield  {author} {\bibinfo {author} {\bibfnamefont {P.~A.}\ \bibnamefont {Igoshev}}, \bibinfo {author} {\bibfnamefont {M.~A.}\ \bibnamefont {Timirgazin}}, \bibinfo {author} {\bibfnamefont {A.~A.}\ \bibnamefont {Katanin}}, \bibinfo {author} {\bibfnamefont {A.~K.}\ \bibnamefont {Arzhnikov}},\ and\ \bibinfo {author} {\bibfnamefont {V.~Y.}\ \bibnamefont {Irkhin}},\ }\bibfield  {title} {\bibinfo {title} {{Incommensurate magnetic order and phase separation in the two-dimensional Hubbard model with nearest- and next-nearest-neighbor hopping}},\ }\href {https://doi.org/10.1103/PhysRevB.81.094407} {\bibfield  {journal} {\bibinfo  {journal} {Phys. Rev. B}\ }\textbf {\bibinfo {volume} {81}},\ \bibinfo {pages} {094407} (\bibinfo {year} {2010})}\BibitemShut {NoStop}%
\bibitem [{\citenamefont {Yamase}\ \emph {et~al.}(2016)\citenamefont {Yamase}, \citenamefont {Eberlein},\ and\ \citenamefont {Metzner}}]{Yamase2016}%
  \BibitemOpen
  \bibfield  {author} {\bibinfo {author} {\bibfnamefont {H.}~\bibnamefont {Yamase}}, \bibinfo {author} {\bibfnamefont {A.}~\bibnamefont {Eberlein}},\ and\ \bibinfo {author} {\bibfnamefont {W.}~\bibnamefont {Metzner}},\ }\bibfield  {title} {\bibinfo {title} {{Coexistence of Incommensurate Magnetism and Superconductivity in the Two-Dimensional Hubbard Model}},\ }\href {https://doi.org/10.1103/PhysRevLett.116.096402} {\bibfield  {journal} {\bibinfo  {journal} {Phys. Rev. Lett.}\ }\textbf {\bibinfo {volume} {116}},\ \bibinfo {pages} {096402} (\bibinfo {year} {2016})}\BibitemShut {NoStop}%
\bibitem [{\citenamefont {Eberlein}\ \emph {et~al.}(2016)\citenamefont {Eberlein}, \citenamefont {Metzner}, \citenamefont {Sachdev},\ and\ \citenamefont {Yamase}}]{Eberlein2016}%
  \BibitemOpen
  \bibfield  {author} {\bibinfo {author} {\bibfnamefont {A.}~\bibnamefont {Eberlein}}, \bibinfo {author} {\bibfnamefont {W.}~\bibnamefont {Metzner}}, \bibinfo {author} {\bibfnamefont {S.}~\bibnamefont {Sachdev}},\ and\ \bibinfo {author} {\bibfnamefont {H.}~\bibnamefont {Yamase}},\ }\bibfield  {title} {\bibinfo {title} {{Fermi Surface Reconstruction and Drop in the Hall Number due to Spiral Antiferromagnetism in High-${T}_{c}$ Cuprates}},\ }\href {https://doi.org/10.1103/PhysRevLett.117.187001} {\bibfield  {journal} {\bibinfo  {journal} {Phys. Rev. Lett.}\ }\textbf {\bibinfo {volume} {117}},\ \bibinfo {pages} {187001} (\bibinfo {year} {2016})}\BibitemShut {NoStop}%
\bibitem [{\citenamefont {Mitscherling}\ and\ \citenamefont {Metzner}(2018)}]{Mitscherling2018}%
  \BibitemOpen
  \bibfield  {author} {\bibinfo {author} {\bibfnamefont {J.}~\bibnamefont {Mitscherling}}\ and\ \bibinfo {author} {\bibfnamefont {W.}~\bibnamefont {Metzner}},\ }\bibfield  {title} {\bibinfo {title} {{Longitudinal conductivity and Hall coefficient in two-dimensional metals with spiral magnetic order}},\ }\href {https://doi.org/10.1103/PhysRevB.98.195126} {\bibfield  {journal} {\bibinfo  {journal} {Phys. Rev. B}\ }\textbf {\bibinfo {volume} {98}},\ \bibinfo {pages} {195126} (\bibinfo {year} {2018})}\BibitemShut {NoStop}%
\bibitem [{\citenamefont {Bonetti}\ \emph {et~al.}(2020)\citenamefont {Bonetti}, \citenamefont {Mitscherling}, \citenamefont {Vilardi},\ and\ \citenamefont {Metzner}}]{Bonetti2020}%
  \BibitemOpen
  \bibfield  {author} {\bibinfo {author} {\bibfnamefont {P.~M.}\ \bibnamefont {Bonetti}}, \bibinfo {author} {\bibfnamefont {J.}~\bibnamefont {Mitscherling}}, \bibinfo {author} {\bibfnamefont {D.}~\bibnamefont {Vilardi}},\ and\ \bibinfo {author} {\bibfnamefont {W.}~\bibnamefont {Metzner}},\ }\bibfield  {title} {\bibinfo {title} {{Charge carrier drop at the onset of pseudogap behavior in the two-dimensional Hubbard model}},\ }\href {https://doi.org/10.1103/PhysRevB.101.165142} {\bibfield  {journal} {\bibinfo  {journal} {Phys. Rev. B}\ }\textbf {\bibinfo {volume} {101}},\ \bibinfo {pages} {165142} (\bibinfo {year} {2020})}\BibitemShut {NoStop}%
\bibitem [{\citenamefont {Azaria}\ \emph {et~al.}(1990)\citenamefont {Azaria}, \citenamefont {Delamotte},\ and\ \citenamefont {Jolicoeur}}]{Azaria1990}%
  \BibitemOpen
  \bibfield  {author} {\bibinfo {author} {\bibfnamefont {P.}~\bibnamefont {Azaria}}, \bibinfo {author} {\bibfnamefont {B.}~\bibnamefont {Delamotte}},\ and\ \bibinfo {author} {\bibfnamefont {T.}~\bibnamefont {Jolicoeur}},\ }\bibfield  {title} {\bibinfo {title} {Nonuniversality in helical and canted-spin systems},\ }\href {https://doi.org/10.1103/PhysRevLett.64.3175} {\bibfield  {journal} {\bibinfo  {journal} {Phys. Rev. Lett.}\ }\textbf {\bibinfo {volume} {64}},\ \bibinfo {pages} {3175} (\bibinfo {year} {1990})}\BibitemShut {NoStop}%
\bibitem [{\citenamefont {Azaria}\ \emph {et~al.}(1992)\citenamefont {Azaria}, \citenamefont {Delamotte},\ and\ \citenamefont {Mouhanna}}]{Azaria1992}%
  \BibitemOpen
  \bibfield  {author} {\bibinfo {author} {\bibfnamefont {P.}~\bibnamefont {Azaria}}, \bibinfo {author} {\bibfnamefont {B.}~\bibnamefont {Delamotte}},\ and\ \bibinfo {author} {\bibfnamefont {D.}~\bibnamefont {Mouhanna}},\ }\bibfield  {title} {\bibinfo {title} {Low-temperature properties of two-dimensional frustrated quantum antiferromagnets},\ }\href {https://doi.org/10.1103/PhysRevLett.68.1762} {\bibfield  {journal} {\bibinfo  {journal} {Phys. Rev. Lett.}\ }\textbf {\bibinfo {volume} {68}},\ \bibinfo {pages} {1762} (\bibinfo {year} {1992})}\BibitemShut {NoStop}%
\bibitem [{\citenamefont {Azaria}\ \emph {et~al.}(1993{\natexlab{a}})\citenamefont {Azaria}, \citenamefont {Delamotte}, \citenamefont {Delduc},\ and\ \citenamefont {Jolicoeur}}]{Azaria1993}%
  \BibitemOpen
  \bibfield  {author} {\bibinfo {author} {\bibfnamefont {P.}~\bibnamefont {Azaria}}, \bibinfo {author} {\bibfnamefont {B.}~\bibnamefont {Delamotte}}, \bibinfo {author} {\bibfnamefont {F.}~\bibnamefont {Delduc}},\ and\ \bibinfo {author} {\bibfnamefont {T.}~\bibnamefont {Jolicoeur}},\ }\bibfield  {title} {\bibinfo {title} {{A renormalization-group study of helimagnets in $D = 2 + \varepsilon$ dimensions}},\ }\href {https://doi.org/https://doi.org/10.1016/0550-3213(93)90376-Z} {\bibfield  {journal} {\bibinfo  {journal} {Nuclear Physics B}\ }\textbf {\bibinfo {volume} {408}},\ \bibinfo {pages} {485} (\bibinfo {year} {1993}{\natexlab{a}})}\BibitemShut {NoStop}%
\bibitem [{\citenamefont {Azaria}\ \emph {et~al.}(1993{\natexlab{b}})\citenamefont {Azaria}, \citenamefont {Delamotte},\ and\ \citenamefont {Mouhanna}}]{Azaria1993_PRL}%
  \BibitemOpen
  \bibfield  {author} {\bibinfo {author} {\bibfnamefont {P.}~\bibnamefont {Azaria}}, \bibinfo {author} {\bibfnamefont {B.}~\bibnamefont {Delamotte}},\ and\ \bibinfo {author} {\bibfnamefont {D.}~\bibnamefont {Mouhanna}},\ }\bibfield  {title} {\bibinfo {title} {Spontaneous symmetry breaking in quantum frustrated antiferromagnets},\ }\href {https://doi.org/10.1103/PhysRevLett.70.2483} {\bibfield  {journal} {\bibinfo  {journal} {Phys. Rev. Lett.}\ }\textbf {\bibinfo {volume} {70}},\ \bibinfo {pages} {2483} (\bibinfo {year} {1993}{\natexlab{b}})}\BibitemShut {NoStop}%
\bibitem [{\citenamefont {Chubukov}\ \emph {et~al.}(1994{\natexlab{a}})\citenamefont {Chubukov}, \citenamefont {Senthil},\ and\ \citenamefont {Sachdev}}]{Sachdev1994}%
  \BibitemOpen
  \bibfield  {author} {\bibinfo {author} {\bibfnamefont {A.~V.}\ \bibnamefont {Chubukov}}, \bibinfo {author} {\bibfnamefont {T.}~\bibnamefont {Senthil}},\ and\ \bibinfo {author} {\bibfnamefont {S.}~\bibnamefont {Sachdev}},\ }\bibfield  {title} {\bibinfo {title} {{Universal magnetic properties of frustrated quantum antiferromagnets in two dimensions}},\ }\href {https://doi.org/10.1103/PhysRevLett.72.2089} {\bibfield  {journal} {\bibinfo  {journal} {Phys. Rev. Lett.}\ }\textbf {\bibinfo {volume} {72}},\ \bibinfo {pages} {2089} (\bibinfo {year} {1994}{\natexlab{a}})}\BibitemShut {NoStop}%
\bibitem [{\citenamefont {Chubukov}\ \emph {et~al.}(1994{\natexlab{b}})\citenamefont {Chubukov}, \citenamefont {Sachdev},\ and\ \citenamefont {Senthil}}]{Chubukov1994}%
  \BibitemOpen
  \bibfield  {author} {\bibinfo {author} {\bibfnamefont {A.~V.}\ \bibnamefont {Chubukov}}, \bibinfo {author} {\bibfnamefont {S.}~\bibnamefont {Sachdev}},\ and\ \bibinfo {author} {\bibfnamefont {T.}~\bibnamefont {Senthil}},\ }\bibfield  {title} {\bibinfo {title} {Quantum phase transitions in frustrated quantum antiferromagnets},\ }\href {https://doi.org/https://doi.org/10.1016/0550-3213(94)90023-X} {\bibfield  {journal} {\bibinfo  {journal} {Nuclear Physics B}\ }\textbf {\bibinfo {volume} {426}},\ \bibinfo {pages} {601} (\bibinfo {year} {1994}{\natexlab{b}})}\BibitemShut {NoStop}%
\bibitem [{\citenamefont {Azaria}\ \emph {et~al.}(1995)\citenamefont {Azaria}, \citenamefont {Lecheminant},\ and\ \citenamefont {Mouhanna}}]{Azaria1995}%
  \BibitemOpen
  \bibfield  {author} {\bibinfo {author} {\bibfnamefont {P.}~\bibnamefont {Azaria}}, \bibinfo {author} {\bibfnamefont {P.}~\bibnamefont {Lecheminant}},\ and\ \bibinfo {author} {\bibfnamefont {D.}~\bibnamefont {Mouhanna}},\ }\bibfield  {title} {\bibinfo {title} {The massive $\mathrm{CP}^{N-1}$ model for frustrated spin systems},\ }\href {https://doi.org/https://doi.org/10.1016/0550-3213(95)00514-S} {\bibfield  {journal} {\bibinfo  {journal} {Nuclear Physics B}\ }\textbf {\bibinfo {volume} {455}},\ \bibinfo {pages} {648} (\bibinfo {year} {1995})}\BibitemShut {NoStop}%
\bibitem [{\citenamefont {Borejsza}\ and\ \citenamefont {Dupuis}(2004)}]{Borejsza2004}%
  \BibitemOpen
  \bibfield  {author} {\bibinfo {author} {\bibfnamefont {K.}~\bibnamefont {Borejsza}}\ and\ \bibinfo {author} {\bibfnamefont {N.}~\bibnamefont {Dupuis}},\ }\bibfield  {title} {\bibinfo {title} {{Antiferromagnetism and single-particle properties in the two-dimensional half-filled Hubbard model: A nonlinear sigma model approach}},\ }\href {https://doi.org/10.1103/PhysRevB.69.085119} {\bibfield  {journal} {\bibinfo  {journal} {Phys. Rev. B}\ }\textbf {\bibinfo {volume} {69}},\ \bibinfo {pages} {085119} (\bibinfo {year} {2004})}\BibitemShut {NoStop}%
\bibitem [{\citenamefont {Elitzur}(1975)}]{Elitzur1975}%
  \BibitemOpen
  \bibfield  {author} {\bibinfo {author} {\bibfnamefont {S.}~\bibnamefont {Elitzur}},\ }\bibfield  {title} {\bibinfo {title} {{Impossibility of spontaneously breaking local symmetries}},\ }\href {https://doi.org/10.1103/PhysRevD.12.3978} {\bibfield  {journal} {\bibinfo  {journal} {Phys. Rev. D}\ }\textbf {\bibinfo {volume} {12}},\ \bibinfo {pages} {3978} (\bibinfo {year} {1975})}\BibitemShut {NoStop}%
\bibitem [{\citenamefont {Wen}(1989)}]{Wen1989}%
  \BibitemOpen
  \bibfield  {author} {\bibinfo {author} {\bibfnamefont {X.~G.}\ \bibnamefont {Wen}},\ }\bibfield  {title} {\bibinfo {title} {{Vacuum degeneracy of chiral spin states in compactified space}},\ }\href {https://doi.org/10.1103/PhysRevB.40.7387} {\bibfield  {journal} {\bibinfo  {journal} {Phys. Rev. B}\ }\textbf {\bibinfo {volume} {40}},\ \bibinfo {pages} {7387} (\bibinfo {year} {1989})}\BibitemShut {NoStop}%
\bibitem [{\citenamefont {Hansson}\ \emph {et~al.}(2004)\citenamefont {Hansson}, \citenamefont {Oganesyan},\ and\ \citenamefont {Sondhi}}]{Hansson2004}%
  \BibitemOpen
  \bibfield  {author} {\bibinfo {author} {\bibfnamefont {T.}~\bibnamefont {Hansson}}, \bibinfo {author} {\bibfnamefont {V.}~\bibnamefont {Oganesyan}},\ and\ \bibinfo {author} {\bibfnamefont {S.}~\bibnamefont {Sondhi}},\ }\bibfield  {title} {\bibinfo {title} {{Superconductors are topologically ordered}},\ }\href {https://doi.org/https://doi.org/10.1016/j.aop.2004.05.006} {\bibfield  {journal} {\bibinfo  {journal} {Ann.~Phys.}\ }\textbf {\bibinfo {volume} {313}},\ \bibinfo {pages} {497} (\bibinfo {year} {2004})}\BibitemShut {NoStop}%
\bibitem [{\citenamefont {Negele}\ and\ \citenamefont {Orland}(1998)}]{NegeleOrland}%
  \BibitemOpen
  \bibfield  {author} {\bibinfo {author} {\bibfnamefont {J.~W.}\ \bibnamefont {Negele}}\ and\ \bibinfo {author} {\bibfnamefont {H.}~\bibnamefont {Orland}},\ }\href@noop {} {\emph {\bibinfo {title} {{Quantum Many-Particle Systems}}}},\ edited by\ \bibinfo {editor} {\bibfnamefont {D.}~\bibnamefont {Pines}}\ (\bibinfo  {publisher} {Westview Press},\ \bibinfo {address} {Reading, Massachusetts},\ \bibinfo {year} {1998})\BibitemShut {NoStop}%
\bibitem [{not()}]{notedefstiffnesses}%
  \BibitemOpen
  \href@noop {} {}\bibinfo {note} {The spin stiffnesses and dynamical susceptibilities (or density-density uniform susceptibility, for the superconductor) can be equivalently defined as the coefficients of a low-energy expansion of the transverse susceptibilities. Here, we choose to define them from the gauge kernels and show that, within a conserving approximation, the two definitons are equivalent.}\BibitemShut {Stop}%
\bibitem [{\citenamefont {Berezinskii}(1971)}]{Berezinskii1971}%
  \BibitemOpen
  \bibfield  {author} {\bibinfo {author} {\bibfnamefont {V.~L.}\ \bibnamefont {Berezinskii}},\ }\bibfield  {title} {\bibinfo {title} {{Destruction of Long-range Order in One-dimensional and Two-dimensional Systems having a Continuous Symmetry Group I. Classical Systems}},\ }\href {http://www.jetp.ac.ru/cgi-bin/e/index/e/32/3/p493?a=list} {\bibfield  {journal} {\bibinfo  {journal} {Sov. Phys. JETP}\ }\textbf {\bibinfo {volume} {32}},\ \bibinfo {pages} {493} (\bibinfo {year} {1971})}\BibitemShut {NoStop}%
\bibitem [{\citenamefont {Kosterlitz}\ and\ \citenamefont {Thouless}(1973)}]{Kosterlitz1973}%
  \BibitemOpen
  \bibfield  {author} {\bibinfo {author} {\bibfnamefont {J.~M.}\ \bibnamefont {Kosterlitz}}\ and\ \bibinfo {author} {\bibfnamefont {D.~J.}\ \bibnamefont {Thouless}},\ }\bibfield  {title} {\bibinfo {title} {{Ordering, metastability and phase transitions in two-dimensional systems}},\ }\href {https://doi.org/10.1088/0022-3719/6/7/010} {\bibfield  {journal} {\bibinfo  {journal} {J. Phys. C}\ }\textbf {\bibinfo {volume} {6}},\ \bibinfo {pages} {1181} (\bibinfo {year} {1973})}\BibitemShut {NoStop}%
\bibitem [{Not()}]{NoteAFnoanomalousaverages}%
  \BibitemOpen
  \href@noop {} {}\bibinfo {note} {\hgl{Note that it is also possible to describe antiferromagnetism without the introduction of a sublattice magnetization, as shown in Ref.~\cite{Irkhin1986}.}}\BibitemShut {Stop}%
\bibitem [{\citenamefont {Wilson}\ \emph {et~al.}(2022)\citenamefont {Wilson}, \citenamefont {Curtis},\ and\ \citenamefont {Galitski}}]{Wilson2022}%
  \BibitemOpen
  \bibfield  {author} {\bibinfo {author} {\bibfnamefont {J.~H.}\ \bibnamefont {Wilson}}, \bibinfo {author} {\bibfnamefont {J.~B.}\ \bibnamefont {Curtis}},\ and\ \bibinfo {author} {\bibfnamefont {V.~M.}\ \bibnamefont {Galitski}},\ }\bibfield  {title} {\bibinfo {title} {{Analog spacetimes from nonrelativistic Goldstone modes in spinor condensates}},\ }\href {https://doi.org/10.1103/PhysRevA.105.043316} {\bibfield  {journal} {\bibinfo  {journal} {Phys. Rev. A}\ }\textbf {\bibinfo {volume} {105}},\ \bibinfo {pages} {043316} (\bibinfo {year} {2022})}\BibitemShut {NoStop}%
\bibitem [{\citenamefont {Bonetti}\ and\ \citenamefont {Metzner}(2022{\natexlab{a}})}]{Bonetti2022}%
  \BibitemOpen
  \bibfield  {author} {\bibinfo {author} {\bibfnamefont {P.~M.}\ \bibnamefont {Bonetti}}\ and\ \bibinfo {author} {\bibfnamefont {W.}~\bibnamefont {Metzner}},\ }\bibfield  {title} {\bibinfo {title} {{Spin stiffness, spectral weight, and Landau damping of magnons in metallic spiral magnets}},\ }\href {https://doi.org/10.1103/PhysRevB.105.134426} {\bibfield  {journal} {\bibinfo  {journal} {Phys. Rev. B}\ }\textbf {\bibinfo {volume} {105}},\ \bibinfo {pages} {134426} (\bibinfo {year} {2022}{\natexlab{a}})}\BibitemShut {NoStop}%
\bibitem [{\citenamefont {Auerbach}(1994)}]{AuerbachBook1994}%
  \BibitemOpen
  \bibfield  {author} {\bibinfo {author} {\bibfnamefont {A.}~\bibnamefont {Auerbach}},\ }\href@noop {} {\emph {\bibinfo {title} {{Interacting Electrons and Quantum Magnetism}}}}\ (\bibinfo  {publisher} {Springer-Verlag},\ \bibinfo {address} {New York},\ \bibinfo {year} {1994})\BibitemShut {NoStop}%
\bibitem [{not()}]{notemixedkernelsoutofplane}%
  \BibitemOpen
  \href@noop {} {}\bibinfo {note} {The terms $K^{\alpha 0}_{13}(\mathbf{0},\pm\mathbf{Q},0)$ and $K^{\alpha 0}_{23}(\mathbf{0},\pm\mathbf{Q},0)$ are zero only if $\mathbf{Q}$ is chosen such that it minimizes the free energy (see Eq.~\eqref{eq: MF free energy}). In fact, if this is not the case, they can be shown to be proportional to $\partial_{q_\alpha}\chit^0_{33}(\pm Q)$, which is finite for a generic $\mathbf{Q}$.}\BibitemShut {Stop}%
\bibitem [{\citenamefont {Metzner}\ and\ \citenamefont {Yamase}(2019)}]{Metzner2019}%
  \BibitemOpen
  \bibfield  {author} {\bibinfo {author} {\bibfnamefont {W.}~\bibnamefont {Metzner}}\ and\ \bibinfo {author} {\bibfnamefont {H.}~\bibnamefont {Yamase}},\ }\bibfield  {title} {\bibinfo {title} {{Phase stiffness in an antiferromagnetic superconductor}},\ }\href {https://doi.org/10.1103/PhysRevB.100.014504} {\bibfield  {journal} {\bibinfo  {journal} {Phys. Rev. B}\ }\textbf {\bibinfo {volume} {100}},\ \bibinfo {pages} {014504} (\bibinfo {year} {2019})}\BibitemShut {NoStop}%
\bibitem [{\citenamefont {Scheurer}\ \emph {et~al.}(2018)\citenamefont {Scheurer}, \citenamefont {Chatterjee}, \citenamefont {Wu}, \citenamefont {Ferrero}, \citenamefont {Georges},\ and\ \citenamefont {Sachdev}}]{Scheurer2018}%
  \BibitemOpen
  \bibfield  {author} {\bibinfo {author} {\bibfnamefont {M.~S.}\ \bibnamefont {Scheurer}}, \bibinfo {author} {\bibfnamefont {S.}~\bibnamefont {Chatterjee}}, \bibinfo {author} {\bibfnamefont {W.}~\bibnamefont {Wu}}, \bibinfo {author} {\bibfnamefont {M.}~\bibnamefont {Ferrero}}, \bibinfo {author} {\bibfnamefont {A.}~\bibnamefont {Georges}},\ and\ \bibinfo {author} {\bibfnamefont {S.}~\bibnamefont {Sachdev}},\ }\bibfield  {title} {\bibinfo {title} {{Topological order in the pseudogap metal}},\ }\href {https://doi.org/10.1073/pnas.1720580115} {\bibfield  {journal} {\bibinfo  {journal} {Proc. Natl. Acad. Sci. USA}\ }\textbf {\bibinfo {volume} {115}},\ \bibinfo {pages} {E3665} (\bibinfo {year} {2018})}\BibitemShut {NoStop}%
\bibitem [{\citenamefont {Millis}(1993)}]{Millis1993}%
  \BibitemOpen
  \bibfield  {author} {\bibinfo {author} {\bibfnamefont {A.~J.}\ \bibnamefont {Millis}},\ }\bibfield  {title} {\bibinfo {title} {Effect of a nonzero temperature on quantum critical points in itinerant fermion systems},\ }\href {https://doi.org/10.1103/PhysRevB.48.7183} {\bibfield  {journal} {\bibinfo  {journal} {Phys. Rev. B}\ }\textbf {\bibinfo {volume} {48}},\ \bibinfo {pages} {7183} (\bibinfo {year} {1993})}\BibitemShut {NoStop}%
\bibitem [{\citenamefont {Irkhin}\ and\ \citenamefont {Katsnelson}(1986)}]{Irkhin1986}%
  \BibitemOpen
  \bibfield  {author} {\bibinfo {author} {\bibfnamefont {V.~Y.}\ \bibnamefont {Irkhin}}\ and\ \bibinfo {author} {\bibfnamefont {M.~I.}\ \bibnamefont {Katsnelson}},\ }\bibfield  {title} {\bibinfo {title} {{On the description of the antiferromagnetism without anomalous averages}},\ }\href {https://doi.org/10.1007/BF01323431} {\bibfield  {journal} {\bibinfo  {journal} {Z.~Phys.~B}\ }\textbf {\bibinfo {volume} {62}},\ \bibinfo {pages} {201} (\bibinfo {year} {1986})}\BibitemShut {NoStop}%
\bibitem [{\citenamefont {Bonetti}\ and\ \citenamefont {Metzner}(2022{\natexlab{b}})}]{Bonetti2022_II}%
  \BibitemOpen
  \bibfield  {author} {\bibinfo {author} {\bibfnamefont {P.~M.}\ \bibnamefont {Bonetti}}\ and\ \bibinfo {author} {\bibfnamefont {W.}~\bibnamefont {Metzner}},\ }\bibfield  {title} {\bibinfo {title} {{SU(2) gauge theory of the pseudogap phase in the two-dimensional Hubbard model}},\ }\href {https://doi.org/10.1103/PhysRevB.106.205152} {\bibfield  {journal} {\bibinfo  {journal} {Phys. Rev. B}\ }\textbf {\bibinfo {volume} {106}},\ \bibinfo {pages} {205152} (\bibinfo {year} {2022}{\natexlab{b}})}\BibitemShut {NoStop}%
\bibitem [{\citenamefont {Goremykin}\ and\ \citenamefont {Katanin}(2024)}]{goremykin2024}%
  \BibitemOpen
  \bibfield  {author} {\bibinfo {author} {\bibfnamefont {I.~A.}\ \bibnamefont {Goremykin}}\ and\ \bibinfo {author} {\bibfnamefont {A.~A.}\ \bibnamefont {Katanin}},\ }\href@noop {} {\bibinfo {title} {{Antiferromagnetic and spin spiral correlations in the doped two-dimensional Hubbard model: gauge symmetry, Ward identities, and dynamical mean-field theory analysis}}} (\bibinfo {year} {2024}),\ \Eprint {https://arxiv.org/abs/2405.04277} {arXiv:2405.04277} \BibitemShut {NoStop}%
\end{thebibliography}%
\begin{widetext}
\vspace{1cm}
\begin{center}
    \large \textbf{Erratum: Local Ward identities for collective excitations in fermionic systems with spontaneously broken symmetries [Phys. Rev. B  106, 155105 (2022)]}
\end{center}
\vspace{1mm}
\setcounter{section}{0}
    We found a mistake in the original paper, namely the statement that the second derivative of the $\Gamma$-functional with respect to the gauge field returns the gauge kernel (Eq.~(10) of the original paper). 

    The Ward identities (WIs) presented in the original paper are nonetheless correct, provided that one does not interpret the object labeled as $K_{\mu\nu}$ as the gauge kernel but as the second derivative with respect to the gauge field of the generating functional $\Gamma$ computed at zero fields. The physical meaning, the usefulness, and the computation of such object are left for future research. 
    
    In the following, we present a revised derivation of the WIs for the \textit{actual} gauge kernel $K_{\mu\nu}$,  both for the case of a U(1) and of an SU(2) symmetry. 

    \section{U(1) symmetry}
    We consider a fermionic theory, described by the Grassmann fields $\psi$ and $\psibar$ and coupled to a U(1) gauge field $A_\mu$, and governed by the action $\mathcal{S}[\psi,\psibar,A_\mu]$. We define the U(1) gauge kernel as 
    \begin{equation}
        K_{\mu\nu}(x,x') \equiv \langle j_\mu(x) j_\nu(x')\rangle_c + \langle P_{\mu\nu}(x,x')\rangle\,,
    \end{equation}
    with $j_\mu(x) \equiv \frac{\delta\mathcal{S}}{\delta A_\mu(x)}\rvert_{A_\mu=0}$ the paramagnetic current operator, $P_{\mu\nu}(x,x')\equiv-\frac{\delta^2\mathcal{S}}{\delta A_\mu(x)\delta A_\nu(x')}\rvert_{A_\mu=0}$ the diamagnetic current operator, and where the (connected) average is taken by means of the path integral defined by $\mathcal{S}$. The gauge kernel can also be defined as the second derivative at zero source fields of the $\mathcal{G}$-generating functional of the original paper (Eq.~(1))
    \begin{equation}
        K_{\mu\nu}(x,x') \equiv - \frac{\delta^2\mathcal{G}[J,J^*,A_\mu]}{\delta A_\mu(x)\,\delta A_\nu (x')}\bigg|_{J=J^*=A_\mu=0}\,.
    \end{equation}
    The U(1) gauge kernel obeys the relation
    \begin{equation}\label{eq: WI1}
        \partial_\mu K_{\mu\nu}(x,x')=0\,,
    \end{equation}
    even in presence of spontaneous symmetry breaking. This can be readily derived from Eq.~(7) of the original paper. The above relation is somewhat obvious as it implies charge conservation, which must be fulfilled even in the superconducting state~\cite{SchriefferBook}. Fourier transforming Eq.~\eqref{eq: WI1}, one obtains
    \begin{equation}\label{eq: WI1 FT}
        i\omega K_{0\nu}(\bq,\omega) - iq_\alpha K_{\alpha\nu}(\bq,\omega) = 0\,,
    \end{equation}
    where $\alpha$ is only a spatial index and a sum over it is implied. Here and henceforth we use Einstein's convention for repeated indices, unless otherwise specified. Note that for the derivative in Eq.~\eqref{eq: WI1} to make sense, $A_\mu$ needs to be defined on a \textit{continuous} space-time, which implies, in case of a lattice system, that Eq.~\eqref{eq: WI1 FT} is valid only for those values of $|\bq|$ that are much smaller then the inverse lattice spacing. Setting $\nu=0$, $\bq=\bzero$, and $\nu=\beta$ (with $\beta$ also a spatial index), $\omega=0$, respectively, one gets 
    \begin{subequations}\label{eq: WI small q and omega}
        \begin{align}
            &K_{00}(\bq=\bzero,\omega)=0\,,\\
            &K_{\alpha\beta}(\bq,\omega=0)\simeq J_{\alpha\beta}(\bq)-\frac{[J_{\alpha\gamma}(\bq)q_\gamma] [J_{\beta\delta}(\bq)q_{\delta}]}{J_{\gamma\delta}(\bq)q_\gamma q_{\delta}}\,, \label{eq: small q Kmunu}
        \end{align}
    \end{subequations}
    where $J_{\alpha\beta}(\bq)$ a function approaching the superfluid stiffness for $\bq\to\bzero$ in the superconducting state. Eq.~\eqref{eq: small q Kmunu} is the most general form that $K_{\alpha\beta}(\bq,0)$ can take to obey $q_\alpha K_{\alpha\beta} (\bq,0)=0$. The above relations clearly show that Eqs.~(19) of the original paper cannot be correct if the correlator $K$ appearing there is the gauge kernel. While $\delta \mathcal{G}/\delta A_\mu=\delta \Gamma/\delta A_\mu$ descends from the properties of the Legendre transform, a similar relation does not apply to the second derivative of the $\mathcal{G}$- and $\Gamma$-functionals with respect to the gauge field. The correct form of Eq.~(10) is
    \begin{equation}\label{eq: d2Gamma dA dA}
        \frac{\delta^2\Gamma}{\delta A_\mu(x)\delta A_\nu(x')}\bigg|_{A_\mu=\phi=0} = - K_{\mu\nu}(x,x') +\int_{x^{\prime\prime},x^{\prime\prime\prime}}L_{\mu,a}(x,x^{\prime\prime})\chi^{-1}_{ab}(x^{\prime\prime},x^{\prime\prime\prime})L_{\nu,b}(x',x^{\prime\prime\prime})\,,
    \end{equation}
    with 
    \begin{equation}\label{eq: L-correlator}
        L_{\mu,a}(x,x') = -\frac{\delta^2\mathcal{G}}{\delta A_\mu(x) \delta J_a(x')}\bigg|_{A_\mu=J=0}.
    \end{equation}
    A proof of this relation is given in Appendix~\ref{app: derivation}.
    This proves that the $K$-correlator in Eq.~(10) of the original paper is not the gauge kernel. Thus, in Eqs.(15-20) of the original paper one should replace $K_{\mu\nu}(\bq,\omega)$ with (minus) the right hand side of Eq.~\eqref{eq: d2Gamma dA dA}.

    \ifx
    Using Eq.~(7) of the original paper and the property that the condensate $\varphi_0$ is given by $-\delta\mathcal{G}/\delta J_1|_{A_\mu=J=0}$, one obtains for the Fourier transform of $L_{\mu,2}$
    \begin{subequations}\label{eq: Lmu2}
        \begin{align}
            & L_{0,2}(\bq=\bzero,\omega) = \frac{2\varphi_0}{i\omega}, \\
            & L_{\alpha,2}(\bq,\omega=0) = i q_\alpha\frac{2\varphi_0}{|\bq|^2}\hskip5mm\text{for small }\bq.
        \end{align}
    \end{subequations}
    Similarly, one can prove that
    \begin{subequations}\label{eq: Lmu1}
        \begin{align}
            &\omega L_{0,1}(\bzero,\omega) = 0,\\
            &q_\alpha L_{\alpha,1}(\bzero,\omega) = 0.
        \end{align}
    \end{subequations}
    The above relations, combined with Eqs.~\eqref{eq: WI small q and omega} make Eqs.~(20) of the original paper a trivial statement. 
    \fi 
    %
    Eqs.~(20) of the original paper, despite being correct if one properly replaces $K_{\mu\nu}(\bq,\omega)$, are not particularly useful as it is not clear how to compute the second derivative of the $\Gamma$-functional with respect to the gauge field within a diagrammatic approach. In the following, we present a revised and more useful derivation of the Ward identities making use of the sole $\mathcal{G}$-functional. 

    We consider Eq.~(7) of the original paper using the parametrization in Eqs.~(8). We write $J_1(x)=h + \delta J_1(x)$, where $h$ is a uniform symmetry breaking field. Taking the derivative of Eq.~(7) of the original paper once with respect to $A_\mu$ and once with respect to $J_2$ and setting $A_\mu=J_2=\delta J_1=0$, we obtain
    \begin{subequations}
        \begin{align}
            &\partial_\mu K^h_{\mu\nu}(x,x') = -2h L^h_{\nu,2}(x',x),\\
            &\partial_\mu L^h_{\mu,2}(x,x') = -2h \chi^h_{22}(x,x') + 2\varphi_0^h\delta(x-x'),
        \end{align}
    \end{subequations}
    where $K^h_{\mu\nu}(x,x')$, $\chi_{ab}^h(x,x')\equiv-\frac{\delta^2\mathcal{G}}{\delta J_a(x)\delta J_b(x')}|_{\delta J_1=J_2=A_\mu=0}$, $\varphi_0^h\equiv-\frac{\delta\mathcal{G}}{\delta J_1(x)}|_{\delta J_1=J_2=A_\mu=0}$, and $L^h_{\mu,a}(x,x')$ are the gauge kernel, susceptibility, condensate fraction, and $L$-correlator, respectively (see Eq.~\eqref{eq: L-correlator}), computed in presence of a uniform explicit symmetry breaking field $h$. Combining the two equations above, we obtain
    \begin{equation}\label{eq: intermediate WI}
        \partial_\mu\partial'_\nu K^h_{\mu\nu}(x,x') = 4h^2\left( \chi^h_{22}(x,x') -\frac{\varphi_0^h}{h}\delta(x-x')\right),
    \end{equation}
    where $\partial_\nu'$ is a derivative over $x'$.
    
    Considering the global U(1) symmetry of the problem, one can obtain a relation similar to Eq.~(7) of the original paper:
    \begin{equation}
        \int_x \left[\frac{\delta\mathcal{G}}{\delta J_2(x)}J_1(x)-\frac{\delta\mathcal{G}}{\delta J_1(x)}J_2(x)\right]=0,
    \end{equation}
    where $\int_x$ is an integral over the spatio-temporal coordinates. From the above relation, one readily obtains
    \begin{equation}
        \chi_{22}^h(\bq=\bzero,\omega=0) = \frac{\varphi_0^h}{h}.
    \end{equation}
    This, combined with the properties $\chi_{22}^h(-\bq,\omega)=\chi_{22}^h(\bq,-\omega)=\chi_{22}^h(\bq,\omega)$, implies that the Fourier transform of the right hand side of Eq.~\eqref{eq: intermediate WI} is at least quadratic in frequency and/or momentum. We can thus assume the following small-$\bq$ and small-$\omega$ expansion of $\chi_{22}^h(\bq,\omega)$
    \begin{equation}\label{eq: small chi expansion}
        \chi_{22}^h(\bq,\omega) \simeq \frac{4(\varphi_0^h)^2}{-\chi_n^h \omega^2 + J_{\alpha\beta}^h q_\alpha q_\beta + 4h\varphi_0^h}.
    \end{equation}
    Taking the Fourier transform of Eq.~\eqref{eq: intermediate WI}, one gets
    \begin{subequations}
        \begin{align}
            &\lim_{\omega\to 0}K^h_{00}(\bzero,\omega) = -4h^2\,\frac{1}{2}\partial^2_\omega \chi^h_{22}(\bzero,\omega)\big |_{\omega\to0} = -\chi_n^h=-4(\varphi_0^h)^2\,\frac{1}{2}\partial_\omega^2\left(\frac{1}{\chi_{22}^h(\bzero,\omega)}\right)\bigg|_{\omega\to 0} ,\\
            &\lim_{\bq\to \bzero}K^h_{\alpha\beta}(\bq,0) = -4h^2\,\frac{1}{2} \partial^2_{q_\alpha q_\beta} \chi^h_{22}(\bq,0)\big |_{\bq\to\bzero} = J_{\alpha\beta}^h = 4(\varphi_0^h)^2\,\frac{1}{2}\partial^2_{q_\alpha q_\beta}\left(\frac{1}{\chi_{22}^h(\bq,0)}\right)\bigg|_{\bq\to \bzero}.
        \end{align}
    \end{subequations}
    Finally, taking the $h\to0$ limit, we obtain
    \begin{subequations}\label{eq: nice WI U(1)}
        \begin{align}
            & \chi_n = \lim_{h\to 0}\lim_{\omega\to 0}K^h_{00}(\bzero,\omega) = -4(\varphi_0)^2\,\frac{1}{2}\partial_\omega^2\left(\frac{1}{\chi_{22}(\bzero,\omega)}\right)\bigg|_{\omega\to 0},\label{eq: charge susc}\\
            &J_{\alpha\beta}=\lim_{h\to 0}\lim_{\bq\to \bzero}K^h_{\alpha\beta}(\bq,0) = 4(\varphi_0)^2\,\frac{1}{2}\partial^2_{q_\alpha q_\beta}\left(\frac{1}{\chi_{22}(\bq,0)}\right)\bigg|_{\bq\to \bzero}\label{eq: superfluid stiff},
        \end{align}
    \end{subequations}
    where $\chi_n$ and $J_{\alpha\beta}$ are the coefficients of the expansion~\eqref{eq: small chi expansion} obtained in absence of a symmetry breaking field. In particular, $J_{\alpha\beta}$ is the superfluid stiffness. Also, $\varphi_0$ and $\chi_{22}(\bq,\omega)$ are the condensate fraction and susceptibility obtained for zero $h$.

    The equations above are a more useful version of the Ward identities presented in the original paper. The calculation of $\lim_{h\to 0}K^h_{\mu\nu}$ can be performed in the same way as one computes the gauge kernel, introducing a symmetry breaking field, and sending it to zero at the end of the calculation. The equations above also clearly prove that the dynamical (Eq.~\eqref{eq: charge susc}) or static (Eq.~\eqref{eq: superfluid stiff}) limits do not commute with the $h\to 0$ limit in the gauge kernel. Inverting their order, one would obtain Eqs.~\eqref{eq: WI small q and omega} instead of those above.  
    %
    \section{SU(2) symmetry}
    Most of the considerations above also apply to the case of a spontaneously broken SU(2) symmetry, with minor modifications. The SU(2) gauge kernel has a similar definition to its U(1) counterpart. Also in this case, the SU(2) gauge kernel can only be obtained by taking the derivative of the $\mathcal{G}$-functional (and not from the $\Gamma$ functional).
    
    %
    %

    Relations similar to Eqs.~\eqref{eq: d2Gamma dA dA}, \eqref{eq: L-correlator}, can be obtained for a system with SU(2) symmetry, making, also in this case, Eq.~(40) of the original paper not useful for practical calculations. In the following, we derive WIs for the actual SU(2) gauge kernel.
    
    The analogue of Eq.~(7) is, in the SU(2)-symmetric case,
    \begin{equation} \label{eq: functional relation SU(2)}
        \partial_\mu \left(\frac{\delta\mathcal{G}}{\delta A_\mu^a(x)}\right)
        -\varepsilon^{a\ell m}\left[
        \frac{\delta\mathcal{G}}{\delta J^\ell(x)}J^m(x)
        +\frac{\delta\mathcal{G}}{\delta A_\mu^\ell(x)}A_\mu^m(x)
        \right]
        =0.
    \end{equation}
    We now set $J_a(x)=h_a(\mathbf{x}) + \delta J_a(x)$, where $h_a(\mathbf{x})$ is a symmetry breaking field. Its spatial dependence must take the same form as the spin condensate $S^h_a(\mathbf{x})=-\delta\mathcal{G}/\delta J_a(x)|_{A_\mu=\delta J=0}$. In essence, for every point $\mathbf{x}$, we have $S^h_a(\mathbf{x}) = (\varphi_0^h/h) h_a(\mathbf{x})$, where $h=\sqrt{h_a(\mathbf{x})h_a(\mathbf{x})}$ and $\varphi_0^h$ are constants.

    With some algebra, we can derive from Eq.~\eqref{eq: functional relation SU(2)} the following set of equations
    \begin{subequations}\label{eq: intermediate step SU(2)}
        \begin{align}
            &\partial_\mu K_{\mu\nu}^{h;ab}(x,x') = \varepsilon^{a\ell m} L^{h;b}_{\nu,\ell}(x',x) h_m(\mathbf{x}) + \varepsilon^{ab\ell} B_\nu^{h;\ell}(x)\delta(x-x'),\\
            &\partial_\mu L^{h;a}_{\mu,b}(x,x') = \varepsilon^{a\ell m}\chi^h_{\ell b}(x,x') h_m(\mathbf{x})-\varepsilon^{ab\ell}S^h_\ell(\mathbf{x})\delta(x-x'),\\
            &\partial_\mu B^{h;a}_\mu(x) = \varepsilon^{a\ell m}S^h_\ell(\mathbf{x}) h_m(\mathbf{x})=0\,,
        \end{align}
    \end{subequations}
    where the first (second) equation has been obtained deriving~\eqref{eq: functional relation SU(2)} with respect to $A_\nu(x')$ ($J_b(x')$) and setting $A_\mu=\delta J=0$. The third equation can be derived from~\eqref{eq: functional relation SU(2)} setting $A_\mu=\delta J=0$. Its right hand side is zero because we have assumed $S^h_a(\mathbf{x})$ and $h_a(\mathbf{x})$ to be parallel. We have also defined
    \begin{subequations}
        \begin{align}
            &L^{h;a}_{\mu,b}(x,x') = -\frac{\delta^2 \mathcal{G}}{\delta A^a_\mu(x)\delta J^b(x')}\bigg|_{A_\mu=\delta J=0}, \\
            &B^{h,a}_\mu(x) = -\frac{\delta\mathcal{G}}{\delta A_\mu^a(x)}\bigg|_{A_\mu=\delta J=0}.
        \end{align}
    \end{subequations}
    $K_{\mu\nu}^{h;ab}$ and $\chi_{ab}^h$ are the gauge kernel and spin susceptibility in presence of a symmetry breaking field. 

    Combining all Eqs.~\eqref{eq: intermediate step SU(2)}, we get
    \begin{equation}
        \begin{split}
            \partial_\mu\partial'_\nu K_{\mu\nu}^{h;ab}(x,x') &= \varepsilon^{a\ell m}\varepsilon^{b p r} \chi^h_{p\ell}(x,x') h_m(\mathbf{x}) h_r(\mathbf{x}') + (2\delta_{a\ell}\delta_{b m} - \delta_{a m}\delta_{b \ell} - \delta_{ab}\delta_{\ell m}) S^h_\ell(\mathbf{x}) h_m(\mathbf{x})\delta(x-x')\\
            &=\varepsilon^{a\ell m}\varepsilon^{b p r} \chi^h_{p\ell}(x,x') h_m(\mathbf{x}) h_r(\mathbf{x}') + h\,\varphi_0^h \left(\frac{h_a(\mathbf{x})h_b(\mathbf{x})}{h^2}-\delta_{ab}\right).
        \end{split}
    \end{equation}
    Similarly to the previous section, one can derive a functional identity for $\mathcal{G}$ descending from the \emph{global} SU(2) symmetry of the system, reading
    \begin{equation}
        \int_x \varepsilon^{a\ell m }\frac{\delta\mathcal{G}}{\delta J_\ell(x)}J_m(x) = 0\,.
    \end{equation}
    Taking a functional derivative with respect to $J_n(x')$, multiplying by $\varepsilon^{bnp} h_p(\mathbf{x}')$, summing over $n$ and $p$, setting $\delta J_a=A_\mu^a=0$, and integrating over $x'$, we obtain
    \begin{equation}\label{eq: global SU(2) WI}
        \int_x\int_{x'}\varepsilon^{a\ell m}\varepsilon^{b n p} \chi^h_{\ell n}(x,x') h_m(\mathbf{x})h_p(\mathbf{x}') = \frac{\varphi_0^h}{h} \int_x \left(\delta_{ab}\delta_{\ell m} - \delta_{a\ell}\delta_{b m}\right) h_\ell(\mathbf{x})h_m(\mathbf{x})\,.
    \end{equation}
    \subsection{Spiral order}
    In the case of spiral order, we have $h(\mathbf{x})=h(\cos(\bQ\cdot\mathbf{x}),\sin(\bQ\cdot\mathbf{x}),0)$ and $S^h(\mathbf{x})=\varphi_0^h(\cos(\bQ\cdot\mathbf{x}),\sin(\bQ\cdot\mathbf{x}),0)$. Eq.~\eqref{eq: global SU(2) WI} then gives
    \begin{subequations}
        \begin{align}
            &\chit_{22}^h(\bzero,0) = \frac{\varphi_0^h}{h},\\
            &\chit_{33}^h(\bQ,0) = \chit_{33}^h(-\bQ,0) = \frac{\varphi_0^h}{h},
        \end{align}
    \end{subequations}
    where $\chit^h_{ab}$ is the susceptibility in a \textit{rotated} spin basis, and it is connected to $\chi^h_{ab}$ with relations similar to Eqs.~(55) of the original paper. Setting $a=b=1,2,3$ in Eq.~\eqref{eq: intermediate step SU(2)}, Fourier transforming and using the above relations, one obtains
    \begin{subequations}
        \begin{align}
            &\lim_{\omega\to 0}K^{h;11}_{00}(\bzero,\omega) = \lim_{\omega\to 0}K^{h;22}_{00}(\bzero,\omega) = -\frac{h^2}{4} \partial_\omega^2 \chit^h_{33}(\bQ,\omega)|_{\omega\to 0},\\
            &\lim_{\bq\to \bzero}K^{h;11}_{\alpha\beta}(\bq,0) = \lim_{\bq\to \bzero}K^{h;22}_{\alpha\beta}(\bq,0) = -\frac{h^2}{4} \partial_{q_\alpha q_\beta}^2 \chit^h_{33}(\bq,0)|_{\bq\to \bQ},\\
            &\lim_{\omega\to 0}K^{h;33}_{00}(\bzero,\omega) = -\frac{h^2}{2} \partial_\omega^2 \chit^h_{22}(\bzero,\omega)|_{\omega\to 0},\\
            &\lim_{\bq\to \bzero}K^{h;33}_{\alpha\beta}(\bq,0) = -\frac{h^2}{2} \partial_{q_\alpha q_\beta}^2 \chit^h_{22}(\bq,0)|_{\bq\to \bzero},
        \end{align}
    \end{subequations}
    where, as in the original paper, $K_{\mu\nu}^{ab}(\bq,\omega)$ is the translational invariant component of the SU(2) gauge kernel. 

    Assuming the following forms for the susceptibilities, 
    \begin{subequations}
        \begin{align}
            &\chit^h_{22}(\bq,\omega) \simeq \frac{(\varphi_0^h)^2}{-\chi^{h;\smsqr}_\mathrm{dyn}\omega^2 + J^{h;\smsqr}_{\alpha\beta}q_\alpha q_\beta + h\varphi_0^h},\\
            &\chit^h_{33} (\bq,\omega) \simeq \sum_{\eta=\pm}\frac{(\varphi_0^h)^2/2}{-\chi^{h;\perp}_\mathrm{dyn}\omega^2 + J^{h;\perp}_{\alpha\beta}(q-\eta Q)_\alpha (q-\eta Q)_\beta + h\varphi_0^h /2},
        \end{align}
    \end{subequations}
    one can prove, following the steps performed in the U(1)-symmetric case, the final form of the Ward identities for a spiral magnet
    \begin{subequations}
        \begin{align}
            & \chi^{\perp}_\mathrm{dyn} = \lim_{h\to 0}\lim_{\omega\to 0}K^{h;11}_{00}(\bzero,\omega) = \lim_{h\to 0}\lim_{\omega\to 0}K^{h;22}_{00}(\bzero,\omega) = -\frac{(\varphi_0)^2}{2} \partial_\omega^2\left(\frac{1}{\chit^{33}(\bQ,\omega)}\right)\bigg|_{\omega\to 0}, \\
            &J_{\alpha\beta}^{\perp} = \lim_{h\to 0}\lim_{\bq\to \bzero}K^{h;11}_{\alpha\beta}(\bq,0) = \lim_{h\to 0}\lim_{\omega\to 0}K^{h;22}_{\alpha\beta}(\bq,0) = \frac{(\varphi_0)^2}{2} \partial_{q_\alpha q_\beta}^2\left(\frac{1}{\chit^{33}(\bq,0)}\right)\bigg|_{\bq\to \bQ}, \\
            & \chi^{\perp}_\mathrm{dyn} = \lim_{h\to 0}\lim_{\omega\to 0}K^{h;33}_{00}(\bzero,\omega) = -(\varphi_0)^2 \partial_\omega^2\left(\frac{1}{\chit^{22}(\bzero,\omega)}\right)\bigg|_{\omega\to 0}, \\
            &J_{\alpha\beta}^{\smsqr} = \lim_{h\to 0}\lim_{\bq\to \bzero}K^{h;33}_{\alpha\beta}(\bq,0) = (\varphi_0)^2 \partial_{q_\alpha q_\beta}^2\left(\frac{1}{\chit^{22}(\bq,0)}\right)\bigg|_{\bq\to \bzero}.
        \end{align}
    \end{subequations}
    Also in this case, the static or dynamic limit does not commute with the $h\to 0$ limit.
    \subsection{N\'eel order}
    The Ward identities in the case of N\'eel order can be straightforwardly derived following the steps performed in the previous subsection. Assuming $S^h(\mathbf{x})=\varphi_0^h(-1)^\mathbf{x}(1,0,0)$ and $h(\mathbf{x})=h(-1)^\mathbf{x}(1,0,0)$, they read as
    \begin{subequations}
        \begin{align}
            &\chi_\mathrm{dyn} = \lim_{h\to 0}\lim_{\omega\to 0}K^{h;22}_{00}(\bzero,\omega) = -(\varphi_0)^2 \partial_\omega^2
            \left(\frac{1}{\chi^{33}(\bQ,\omega)}\right)\bigg|_{\omega\to 0}, \\
            &J_{\alpha\beta} = \lim_{h\to 0}\lim_{\bq\to \bzero}K^{h;22}_{\alpha\beta}(\bq,0) =  (\varphi_0)^2 \partial_{q_\alpha q_\beta}^2\left(\frac{1}{\chi^{33}(\bq,0)}\right)\bigg|_{\bq\to \bQ},
        \end{align}
    \end{subequations}
    where now $\bQ$ takes the form $(\pi,\pi)$. The above equations remain valid upon exchanging the index "2" with "3".
    \section{Explicit calculation for a spiral magnet}
    Sec.~III of the original paper is fully correct, as the microscopic expressions for the spin stiffnesses and dynamical susceptibilities correspond to those derived in presence of a small symmetry breaking field that is sent to zero at the end of the calculation. 

    A misprint is present in Eq.~(A4) of the original paper. Its correct form is 
    \begin{equation}
    \begin{split}
    \kappa_\alpha^{31}(\bzero)=-\frac{\Delta}{4}
        \int_\bk\bigg\{ \left[\frac{h_\bk}{e_\bk}(\partial_{k_\alpha}g_\bk)+(\partial_{k_\alpha}h_\bk)\right]\frac{f'(E^+_\bk)}{e_\bk}
        +\Big[&\frac{h_\bk}{e_\bk}(\partial_{k_\alpha}g_\bk)-(\partial_{k_\alpha}h_\bk)\Big]\frac{f'(E^-_\bk)}{e_\bk}
        +\frac{h_\bk}{e_\bk^2}(\partial_{k_\alpha}g_\bk) \frac{f(E^-_\bk)-f(E^+_\bk)}{e_\bk}\bigg\}\,,
    \end{split}
    \end{equation}
    that is, compared to the original paper, the prefactor is $\Delta$ and not $\Delta^2$.
    \section{Note}
    In Ref.~\cite{Bonetti2022_II} the expressions for the spin stiffnesses and dynamical susceptibilities derived in the original manuscript have been used. Even though it was not explicitly stated, they have been derived applying a small symmetry breaking field to the system and sending it to zero \textit{after} performing the dynamical or static limit. The expressions in Ref.~\cite{Bonetti2022_II} are therefore correct within the (renormalization group improved) random phase approximation employed in the paper. 

    During the review process of this Erratum, Ref.~\cite{goremykin2024} appeared, which presents a similar derivation of the Ward identities above, and, as also discussed here, identifies the presence of an infinitesimal symmetry breaking field as crucial to obtain the correct formulas for the spin stiffnesses.
    \section*{Acknowledgments}
    I am thankful to W.~Metzner for noticing the error in the original paper. I am also very thankful to W.~Metzner, H.~Müller-Groeling and D.~Vilardi for enlightening discussions, a careful reading of the manuscript, and for correcting additional typos and misprints. 
    
    \appendix
    \section{Derivation of Eq.~\eqref{eq: d2Gamma dA dA}}
    \label{app: derivation}
    In this Appendix, we present a derivation of Eq.~\eqref{eq: d2Gamma dA dA}. The identity 
    \begin{equation}
        \frac{\delta\Gamma}{\delta A_\mu(x)}=\frac{\delta\mathcal{G}}{\delta A_\mu(x)}\,,
    \end{equation}
    descends directly from the properties of the Legendre transform that connects $\mathcal{G}$ and $\Gamma$. Taking a further derivative with respect to the gauge field in the above equation, we get
    \begin{equation}
        \frac{\delta^2\Gamma}{\delta A_\mu(x)\delta A_\nu(x')}=\frac{\delta^2\mathcal{G}}{\delta A_\mu(x)\delta A_\nu(x')} + \int_{x''}\frac{\delta^2\mathcal{G}}{\delta A_\mu(x)\delta J_a(x'')}\,\frac{\delta J_a(x'')}{\delta A_\nu(x')}\,.
    \end{equation}
    Setting the fields to zero, we obtain
    \begin{equation}\label{eq app: d2 gamma}
        \frac{\delta^2\Gamma}{\delta A_\mu(x)\delta A_\nu(x')}\bigg|_{A_\mu=\phi=0} = -K_{\mu\nu}(x,x') - \int_{x''} L_{\mu,a}(x,x'')\,\frac{\delta J_a(x'')}{\delta A_\nu(x')}\bigg|_{A_\mu=\phi=0}\,,
    \end{equation}
    with $L_{\mu,a}(x,x'')$ defined as in Eq.~\eqref{eq: L-correlator}.
    At this point, 
    %
    %
    
    one needs to express the source field $J_a(x)$ in terms of the "classical" field $\phi_a(x)$, which $\Gamma$ depends on. Since we only need its derivative with respect to the gauge field at zero sources, we are allowed to expand the relation connecting $\phi_a(x)$ and $J_a(x)$ to linear order in the fields:
    \begin{equation}
        \phi_a(x)\equiv-\frac{\delta\mathcal{G}}{\delta J_a(x)}\approx \varphi_{0,a} + \int_{x'} \left[\chi_{ab}(x,x')J_b(x') + L_{\mu,a}(x',x)A_\mu(x')\right]+\dots\,,
    \end{equation}
    with $\varphi_{0,a}\equiv- \frac{\delta\mathcal{G}}{\delta J_a(x)}|_{J=A_\mu=0}$ the condensate fraction. 
    Solving for $J_a(x)$, we obtain
    \begin{equation}\label{eq app: dJ dA}
        \frac{\delta J_a(x)}{\delta A_\mu(x')}\bigg|_{A_\mu=\phi=0} = -\int_{x''} \chi^{-1}_{ab}(x,x'') L_{\mu,b}(x',x'')\,,
    \end{equation}
    with the inverse susceptibility $\chi^{-1}_{ab}(x,x')$ obeying
    \begin{equation}
        \int_{x''}\chi_{ac}(x,x'')\chi^{-1}_{cb}(x'',x')=\delta(x-x')\delta_{ab}\,.
    \end{equation}
    Inserting Eq.~\eqref{eq app: dJ dA} into Eq.~\eqref{eq app: d2 gamma}, we obtain Eq.~\eqref{eq: d2Gamma dA dA}.
\end{widetext}
\end{document}